 \def \LongVersion {}
\ifdefined\LaTeXdiff
	\let\VersionWithComments\undefined
	
\fi

\ifdefined\LongVersion
	\newcommand{\generalFigsScaleFactor}{1}

	\newcommand{\ForLongVersion}[1]{#1}
	\newcommand{\ForShortVersion}[1]{}
\else
	\newcommand{\generalFigsScaleFactor}{.88}

	\newcommand{\ForLongVersion}[1]{}
	\newcommand{\ForShortVersion}[1]{#1}
\fi

\PassOptionsToPackage{table}{xcolor} %

\documentclass{lmcs}
\pdfoutput=1

\usepackage{lastpage}
\lmcsdoi{18}{1}{31}
\lmcsheading{}{\pageref{LastPage}}{}{}%
{Apr.~21,~2020}{Feb.~09,~2022}{}


\usepackage[utf8]{inputenc}

\usepackage{caption}

\usepackage{subcaption}
\usepackage{amsmath} %
\usepackage{amssymb} %

\ifdefined\VersionWithComments
	\usepackage{draftwatermark}
	\SetWatermarkText{draft}
	\SetWatermarkScale{15}
	\SetWatermarkColor[gray]{0.95}
\fi
\ifdefined\LaTeXdiff
	\usepackage{draftwatermark}
	\SetWatermarkText{diff}
	\SetWatermarkScale{15}
	\SetWatermarkColor[gray]{0.95}
\fi

\usepackage{multirow}
\usepackage{tikz}
\usetikzlibrary{automata, arrows, positioning}

\tikzstyle{PTA}=[auto, ->, >=stealth']
\tikzstyle{location}=[minimum size=12pt, circle, fill=blue!20, draw=black, text=black, inner sep=1.5pt, initial text={}] %
\tikzstyle{location2CM}=[location,thick,fill=green!30]
\tikzstyle{invariant}=[yshift=3, rectangle, draw=black, fill=white, text=black, inner sep=1pt]

\tikzstyle{decidable}=[green!80!black,-]
\tikzstyle{undecidable}=[red!80!black, dashed,-]
\tikzstyle{contribution}=[very thick]
\tikzstyle{legende}=[draw=none, font=\scriptsize]
\tikzstyle{unknown}=[black,dotted,thick]

\usepackage[ruled,vlined,linesnumbered]{algorithm2e}

\usepackage[
		colorlinks=true,
		citecolor=darkgreen,
		linkcolor=darkblue,
		urlcolor=darkpurple,
	]{hyperref}

\usepackage[capitalise,english,nameinlink]{cleveref}
\crefname{thm}{\text{Theorem}}{\text{Theorems}}
\crefname{cor}{\text{Corollary}}{\text{Corollaries}}
\crefname{lem}{\text{Lemma}}{\text{Lemmas}}
\crefname{prop}{\text{Proposition}}{\text{Propositions}}
\crefname{rem}{\text{Remark}}{\text{Remarks}}
\crefname{exa}{\text{Example}}{\text{Examples}}
\crefname{defi}{\text{Definition}}{\text{Definitions}}
\crefname{figure}{\text{Figure}}{\text{Figures}}

\newcommand{\init}{_0}

\newcommand{\A}{\ensuremath{\mathcal{A}}}

\newcommand{\Actions}{\Sigma}
\newcommand{\action}{\ensuremath{\sigma}}

\newcommand{\bounds}{\mathit{bounds}}
\newcommand{\boundinf}{\mathrm{inf}}
\newcommand{\boundsup}{\mathrm{sup}}
\newcommand{\C}{C}

\newcommand{\Clock}{X} %
\newcommand{\ClockCard}{H} %
\newcommand{\clock}{x} %
\newcommand{\clockval}{w} %
\newcommand{\ClocksZero}{\vec{0}}
\newcommand{\cms}{\ensuremath{\mathtt{q}}} %
\newcommand{\cmshalt}{\ensuremath{\cms_\textrm{halt}}} %
\newcommand{\cmspta}{\ensuremath{q}} %
\newcommand{\compOp}{\bowtie}
\newcommand{\locerror}{\ensuremath{\cmspta_{\mathtt{error}}}}
\newcommand{\lochalt}{\ensuremath{\cmspta_{\mathtt{halt}}}}
\newcommand{\lochaltprime}{\ensuremath{\cmspta'_{\mathtt{halt}}}}
\newcommand{\locsink}{\ensuremath{\loc_{\mathtt{sink}}}}
\newcommand{\edge}{e}
\newcommand{\Edges}{E}
\newcommand{\longuefleche}[1]{\stackrel{#1}{\mapsto}}
\newcommand{\longueflecheRel}{{\mapsto}}
\newcommand{\fleche}[1]{\stackrel{#1}{\rightarrow}}
\newcommand{\flecheRel}{{\rightarrow}}
\newcommand{\Fleche}[1]{\stackrel{#1}{\Rightarrow}}
\newcommand{\grandn}{{\mathbb N}}
\newcommand{\grandq}{{\mathbb Q}}
\newcommand{\grandqplus}{\grandq_{+}} %
\newcommand{\grandr}{{\mathbb R}}
\newcommand{\grandrplus}{\grandr_{+}} %
\newcommand{\grandz}{{\mathbb Z}}
\newcommand{\guard}{g}

\newcommand{\interval}{\mathcal{I}}
\newcommand{\invariant}{I}
\newcommand{\K}{K}

\newcommand{\loc}{\ell} %
\newcommand{\EGtoutsauflochaltprime}{\ensuremath{\mathrm{EG} (\Loc \setminus \{\lochaltprime{}\})}}
\newcommand{\locinit}{\loc\init}
\newcommand{\Loc}{L} %
\newcommand{\lterm}{\mathit{lt}}
\newcommand{\Param}{P} %
\newcommand{\param}{p} %
\newcommand{\ParamCard}{M} %
\newcommand{\pval}{v} %
\newcommand{\pvalinfsup}{\pval_{\boundinf/\boundsup}}

\newcommand{\plterm}{\mathit{plt}}
\newcommand{\varproblem}{\ensuremath{\phi}}
\newcommand{\Problem}{\ensuremath{\mathcal{P}}}
\newcommand{\Class}{\ensuremath{\mathcal{C}}}
\newcommand{\resets}{R}
\newcommand{\runlength}{m}
\newcommand{\sinit}{s\init} %

\newcommand{\somelocs}{T} %
\newcommand{\state}{\ensuremath{s}} %
\newcommand{\States}{S} %
\newcommand{\symbstate}{\ensuremath{\mathbf{s}}} %
\newcommand{\symbstateinit}{\symbstate\init} %
\newcommand{\Succ}{\mathsf{Succ}}
\newcommand{\timelapse}[1]{#1^\nearrow}
\newcommand{\timepast}[1]{#1^\swarrow}
\newcommand{\varrun}{\rho} %
\newcommand{\runLocs}{{\sf Locs}} %

\newcommand{\bounded}[2]{#1_{|#2}}
\newcommand{\boundinfin}[2]{\boundinf(#1,#2)}
\newcommand{\boundsupin}[2]{\boundsup(#1,#2)}
\newcommand{\ceil}[1]{\lceil #1 \rceil}
\newcommand{\floor}[1]{\lfloor #1 \rfloor}
\newcommand{\projectP}[1]{\ensuremath{#1{\downarrow_{\Param}}}}
\newcommand{\reset}[2]{\ensuremath{[#1]_{#2}}}
\newcommand{\valuate}[2]{\ensuremath{#2(#1)}}
\newcommand{\wv}[2]{#1|#2} %

\newcommand{\imitator}{\textsf{IMITATOR}}
\newcommand{\romeo}{\textsc{Roméo}}

\newcommand{\calA}{\ensuremath{\mathcal{A}}}
\newcommand{\calM}{\ensuremath{\mathcal{M}}}

\newcommand{\defProblem}[3]
{%
\noindent\fcolorbox{black}{white}{
	\begin{minipage}{.966\columnwidth}
		\textbf{#1 problem:}\\
		\textsc{Input}: #2\\
		\textsc{Problem}: #3
	\end{minipage}
}
	
	\smallskip
	
}

\ifdefined \VersionWithComments
	\usepackage{marginnote}
	\newcommand{\marginX}{\marginnote{\huge{\quad\quad\textbf{!}\quad\quad}}}
	\newcommand{\instructions}[1]{{\color{red}\marginX{}\textbf{Instructions: #1}}}
	\newcommand{\ea}[1]{\mbox{}{\color{blue}\marginX{}\textbf{[\'Etienne}: #1]}}
	\newcommand{\dl}[1]{\mbox{}{\color{green!50!black}\marginX{}\textbf{[Didier}: #1]}}
	\newcommand{\ohr}[1]{\mbox{}{\color{orange!90!black}\marginX{}\textbf{[Olivier}: #1]}}
	\newcommand{\reviewer}[2]{\mbox{}{\color{red}\marginX{}\textbf{[Reviewer #1}: ``#2'']}}
	\newcommand{\toutfaux}[1]{}
	\newcommand{\todo}[1]{\mbox{}{\color{red}{\textbf{TODO}\ifx#1\\\else:\ \fi #1}}} %
\else
	\newcommand{\instructions}[1]{}
	\newcommand{\ea}[1]{}
	\newcommand{\dl}[1]{}
	\newcommand{\ohr}[1]{}
	\newcommand{\reviewer}[2]{}
	\newcommand{\toutfaux}[1]{}
	\newcommand{\todo}[1]{}
\fi

\ifdefined \VersionWithComments
 	\definecolor{colorok}{RGB}{80,80,150}
\else
	\definecolor{colorok}{RGB}{0,0,0}
\fi

\newcommand{\eg}{\textcolor{colorok}{\textit{e.\,g.,}}\xspace}
\newcommand{\ie}{\textcolor{colorok}{\textit{i.\,e.,}}\xspace}

\newcommand\crefabbr[1]{%
\begingroup
	\crefname{thm}{\text{Th.}}{\text{Th.}}
	\crefname{corollary}{\text{Cor.}}{\text{Cor.}}
	\cref{#1}
\endgroup%
}

\newcommand{\colCellDec}[1]{\cellcolor{green!35}{\textbf{$\surd$#1}}}
\newcommand{\colCellDecNous}[1]{\cellcolor{green!70}\textbf{$\surd$#1}}
\newcommand{\colCellUndec}[1]{\cellcolor{red!35}{\emph{$\times$#1}}}
\newcommand{\colCellUndecNous}[1]{\cellcolor{red!70}{\emph{$\times$#1}}}
\newcommand{\cellOpenBefore}{\cellcolor{black!10}{open}}
\newcommand{\colCellOpen}{\cellcolor{yellow!70}}
\newcommand{\cellHeader}[1]{\cellcolor{blue!35}\textbf{#1}}

\begin{document}

\title[Reachability and liveness in parametric timed automata]{Reachability and liveness in parametric \texorpdfstring{\\}{} timed automata}
\titlecomment{{\lsuper*}%
	This manuscript is an extended version of~\cite{ALR16,ALime17}.
	}

\author[É.~André]{Étienne André\rsuper{a}}	%
\address{Université de Lorraine, CNRS, Inria, LORIA, F-54000 Nancy, France}	%
\urladdr{\url{https://www.loria.science/andre/}}  %

\author[D.~Lime]{Didier Lime\rsuper{b}}	%
\address{École Centrale de Nantes, Nantes Université, LS2N UMR CNRS 6004, Nantes, France}
\email{\{Didier.Lime,Olivier-h.Roux\}@ec-nantes.fr}  %
\urladdr{\url{https://pagesperso.ls2n.fr/~lime-d/}, \url{https://pagesperso.ls2n.fr/~roux-o/}}  %

\author[O.~H.~Roux]{Olivier H.~Roux\rsuper{b}}	%

\begin{abstract}
	We study timed systems in which some timing features are unknown parameters.
	Parametric timed automata (PTAs) are a classical formalism for such systems but for which most interesting problems are undecidable.
	Notably, the parametric reachability emptiness problem, \ie{} the emptiness of the parameter valuations set allowing to reach some given discrete state, is undecidable.
	Lower-bound/upper-bound parametric timed automata (L/U-PTAs) achieve decidability for reachability properties by enforcing a separation of parameters used as upper bounds in the automaton constraints, and those used as lower bounds.
	
	In this paper, we first study reachability.
	We exhibit a subclass of PTAs (namely integer-points PTAs) with bounded rational-valued parameters for which the parametric reachability emptiness problem is decidable.
	Using this class, we present further results improving the boundary between decidability and undecidability for PTAs and their subclasses such as L/U-PTAs.
	
	We then study liveness.
	We prove that:
	(1) deciding the existence of at least one parameter valuation for which there exists an infinite run in an L/U-PTA is \textsc{PSpace}-complete;
	(2) the existence of a parameter valuation such that the system has a deadlock is however undecidable;
	(3) \ForLongVersion{the problem of }the existence of a valuation for which a run remains in a given set of locations exhibits a very thin border between decidability and undecidability.
 \end{abstract}

 \maketitle

\ifdefined \VersionWithComments
	\textcolor{red}{\textbf{This is the version with comments. To disable comments, comment out line~3 in the \LaTeX{} source.}}
\fi

\section{Introduction}

Timed automata (TAs)~\cite{AD94} are a powerful formalism that extend finite-state automata with clocks (real-valued variables evolving linearly) that can be compared with integer constants in locations (``invariants'') and along transitions (``guards''); additionally, some clocks can be reset to~0 along transitions.
Many interesting problems for TAs (including the reachability of a location~\cite{AD94}) are decidable.
However, the classical definition of TAs is not tailored to verify systems only partially specified, especially when the value of some timing constants is not yet known.

We study here behaviors in the context of parametric timed systems, in which some timing features (\eg{} the duration of a task, a transmission delay in a network, the delay to trigger a watchdog, etc.) are not known and replaced by symbolic constants, called \emph{parameters}.
The objective of verification on such partially defined systems, is then to synthesize the possible valuations of parameters such that some properties are satisfied. 
\emph{Parametric timed automata} (PTAs)~\cite{AHV93} lift the limitation of timed automata by allowing the specification and the verification of systems where some of the timing constants are parametric.
PTAs extend TAs through the use of integer- or rational-valued parameters in place of timing constants in guards and invariants.
PTAs were used to model and verify a variety of case studies,
from hardware circuits to communication protocols (see, \eg{}~\cite{Andre19STTT} for a survey).

\paragraph{Problems of interest}
In this paper, we will be interested in several decision (and synthesis) problems, among which:
\begin{itemize}
	\item The EF-emptiness\footnote{%
		The EF and AF notations come from TCTL and denote reachability and unavoidability, respectively; they have been used in several works since~\cite{JLR15}.
	} problem (also called reachability-emptiness) asks: ``is the set of parameter valuations such that a given location is reachable empty?''
		Note that the emptiness refers here to the \emph{emptiness of the parameter valuations set} (and not, \eg{} of the language).
		
	\item The AF-emptiness problem (also called unavoidability-emptiness) asks: ``is the set of parameter valuations for which a given location is eventually reached for any run empty?''
\end{itemize}

\subsection{Related work}\label{ss:related}

\subsubsection{Decision problems for PTAs}
The expressive power of PTAs comes at the price of the undecidability of most interesting problems.
The EF-emptiness problem
	is undecidable in general~\cite{AHV93}.

\paragraph{Restricting the syntax}
	The undecidability of EF-emptiness still holds when the syntax is restricted:
	for example, EF-emptiness remains undecidable
		when parameters are bounded, \ie{} belong to a bounded domain (typically $[0,1]$)~\cite{Miller00},
		but also when only strict inequalities are used~\cite{Doyen07},
		or with a single integer-valued parameter~\cite{BBLS15}.

\paragraph{Restraining the number of variables}
	Bounding the number of parametric clocks and the number of parameters may yield decidability of EF-emptiness (see, \eg{}~\cite{AHV93,BBLS15,BO17}).
    It is known that the problem is undecidable with at least:
        three parametric clocks (\ie{} clocks compared with parameters) and one integer-valued parameter~\cite{BBLS15}
		or three parametric clocks and only one rational-valued parameter~\cite{Miller00},
		or only one parametric clock, three non-parametric clocks and one rational-valued parameter~\cite{Miller00}.

Conversely, some limitations on those numbers lead to decidability~\cite{AHV93,BO14,BBLS15,BO17,GH21}, though some open cases remain, notably for two clocks and at least two parameters: see~\cite{Andre19STTT} for a survey.

\paragraph{Logics and PTAs}
In~\cite{Wang96}, an approach is proposed for the verification of Parametric TCTL (PTCTL) formulas, where the problem is decidable.

In~\cite{BR07}, parameters are allowed both in the model and in the property (a PTCTL formula), and model checking is decidable with one parametric clock over discrete time, provided no equality appears in the formula.

In~\cite{BDR08}, the parametric model checking of a timed automaton against a parametric extension of TCTL is considered: the valuations set answering the parametric model checking problem is proved to be effectively computable, with some restrictions over the parametric TCTL syntax. %

In~\cite{Quaas14}, it is shown that the MTL model checking problem is undecidable for PTAs even with a single clock.

\subsubsection{Decision problems for L/U-PTAs}
In~\cite{HRSV02}, L/U-PTAs are introduced as a subclass of PTAs where each parameter is either always compared to a clock as a lower bound in guards and invariants, or always as an upper bound.

\paragraph{Decidability}
The EF-emptiness problem is decidable for L/U-PTAs~\cite{HRSV02}.
In~\cite{BlT09}, further results are proved for L/U-PTAs with integer-valued parameters:
	emptiness, finiteness and universality of the set of parameter valuations for which there exists an infinite accepting run are decidable.

\paragraph{Undecidability}
The first problems shown undecidable for L/U-PTAs are the \emph{constrained} EF-emptiness problem and constrained EF-universality problem~\cite{BlT09}:
here, ``constrained'' means that some parameters of the L/U-PTA can be constrained by an initial linear constraint, such as $\param_1 \leq 2 \times \param_2 + \param_3$.

In addition, the AF-emptiness problem is undecidable for L/U-PTAs~\cite{JLR15}.

The untimed language preservation problem (``given a parameter valuation, does there exist another valuation with the same untimed language?'') is undecidable for both PTAs and L/U-PTAs~\cite{ALM20}.

\paragraph{Intractability of synthesis}
It is shown in~\cite{JLR15} that the synthesis of the parameters reaching a given location in an L/U-PTA is intractable in practice, more specifically, it cannot be represented using a finite union of polyhedra.

\paragraph{L-PTAs and U-PTAs}
Two further subclasses have been defined in~\cite{BlT09}: L-PTAs and U-PTAs, where all parameters are always lower bounds and always upper bounds respectively.

On the positive side, exact synthesis algorithms are proposed for reachability properties in L-PTAs and U-PTAs over integer-valued parameters in~\cite{BlT09}.
On the negative side, it is shown in~\cite{ALR18FORMATS} that the full TCTL-emptiness problem (\ie{} the emptiness of the valuation set for which a particular TCTL formula is satisfied on the model) is undecidable even for U-PTAs.

\subsubsection{Parameter synthesis in practice}
Two main model checkers support parametric timed automata (or the similar formalism of parametric timed Petri nets, PTPNs~\cite{TLR09}): \imitator{}~\cite{Andre21} for PTAs, and \romeo{}~\cite{LRST09} for PTPNs.
In the general case, they both perform synthesis independently of the decidability results, \ie{} work in a ``best effort'', without any guarantee of termination.

\subsubsection{The power of integer points (integer parameter valuations)}
In~\cite{JLR15}, PTAs with bounded integer-valued parameters are considered.
The problem of exhibiting parameter valuations such that a given location is reachable or unavoidable becomes decidable, and two algorithms are provided that compute the exact such sets of integer valuations in a symbolic manner, \ie{} without performing an exhaustive enumeration.
In~\cite{ALR15}, it is shown that computing a parametric extrapolation of the integer hull of symbolic states allows one to synthesize (rational-valued) parameter valuations for bounded PTAs, guaranteeing the synthesis of at least all integer-valued valuations, but also sometimes most or even all rational-valued valuations.

\subsubsection{Summary of (un)decidability}
We give a summary of the known results prior to our contributions concerning EF-emptiness, EF-universality (``do all parameter valuations allow a given location to be reachable?'') and AF-emptiness in \cref{table:summary:decidability-before}.
We give from left to right the (un)decidability for bounded L/U-PTAs, L/U-PTAs, bounded PTAs, and PTAs.
Decidability is preceded by ``$\surd$'' (in green), whereas undecidability is preceded by ``$\times$'' (in red).

We give a second summary in \cref{table:summary:decidability:EG-before}, where EC-emptiness, ED-emptiness and EG-emptiness denote the emptiness of the parameter valuations set for which there exists an infinite run, there exists a maximal finite run (\ie{} deadlocked), and there exists a maximal (infinite or finite) run staying in a predefined set of locations, respectively.
In addition, ``bc'' and ``bo'' denote ``bounded with closed bounds'' (\ie{} of the form $[a, b]$) or ``bounded with open bounds'' (\ie{} for which at least one of the intervals has an open bound).
\begin{table}[tb!]
	\centering
	\setlength{\tabcolsep}{3pt} %
	\begin{tabular}{@{} | *{5}{c|}}
		\hline
		\cellHeader{Class} & \cellHeader{bL/U-PTAs} & \cellHeader{L/U-PTAs} & \cellHeader{bPTAs} & \cellHeader{PTAs} \\
		\hline
		EF-empt. & \cellOpenBefore{} & \colCellDec{\cite{HRSV02}} & \colCellUndec{\cite{Miller00}} & \colCellUndec{\cite{AHV93}} \\
		\hline
		EF-univ. & \cellOpenBefore{} & \colCellDec{\cite{BlT09}} & \cellOpenBefore{} & \cellOpenBefore{} \\
		\hline
		AF-empt. & \cellOpenBefore{} & \colCellUndec{\cite{JLR15}} & \cellOpenBefore{} & \colCellUndec{\cite{JLR15}} \\
		\hline
	\end{tabular}

	\caption{Decidability of reachability (and AF) problems for PTAs (prior to our work)} %
    \label{table:summary:decidability-before}
\end{table}
\begin{table}[tb!]
	\centering
	\begin{tabular}{@{} | *{5}{c|}}
		\hline
		\cellHeader{Class} & \cellHeader{PTAs} & \cellHeader{L/U-PTAs} & \cellHeader{bo L/U-PTAs} & \cellHeader{bc L/U-PTAs} \\
		\hline
        EC-emptiness & \cellOpenBefore{} & \cellOpenBefore{} & \cellOpenBefore{} & \cellOpenBefore{} \\
		\hline
		ED-emptiness & \cellOpenBefore{} & \cellOpenBefore{} & \cellOpenBefore{} & \cellOpenBefore{} \\
		\hline
		EG-emptiness & \cellOpenBefore{} & \cellOpenBefore{} & \cellOpenBefore{} & \cellOpenBefore{} \\
		\hline
		AF-emptiness & \colCellUndec{\cite{JLR15}} & \colCellUndec{\cite{JLR15}} & \cellOpenBefore{} & \cellOpenBefore{} \\
		\hline
	\end{tabular}

	\caption{Decidability of liveness problems for PTAs (prior to our work)} %
    \label{table:summary:decidability:EG-before}
\end{table}
\subsection{Contribution}
Following Lamport, properties of systems are often characterized as safety properties (``something bad will never happen'') and liveness properties (``something good will eventually happen'')~\cite{lamport-TSE-77}.
Safety generally reduces to reachability.
In this paper, we contribute to the study of both reachability and liveness for PTAs and their subclasses.

\subsubsection{Reachability}
L/U-PTAs is the only non-trivial\footnote{%
	The bounded integer PTAs of~\cite{JLR15} are arguably a trivial such subclass (even though the associated analysis techniques are not).
	}
	subclass of PTAs for which the EF-emptiness problem is decidable for an arbitrary number of clocks and parameters.
However, other results are disappointing for L/U-PTAs: undecidability of AF-emptiness, intractability of reachability-synthesis~\cite{JLR15}.
It is hence important to look for further subclasses of PTAs for which problems may be decidable.

\paragraph{Integer-point PTAs}
It is shown in~\cite{JLR15,ALR15} that integer points play a key role in decidability.
Hence, our first contribution here is to investigate integer-points PTAs (IP-PTAs), that are PTAs where each symbolic state contains at least one integer point (\ie{} an integer valuation of the clocks and the parameters).
We prove that the EF-emptiness problem is decidable for bounded IP-PTAs (\ie{} with a bounded parameter domain), even when parameters are rational-valued.
Although we show that it cannot be decided whether a bounded PTA is a (bounded) IP-PTA, we give two sufficient syntactic conditions: %
we show that bounded L/U-PTAs with non-strict inequalities are IP-PTAs and, more interestingly,
we introduce a new subclass of ``reset-PTAs'', that are also IP-PTAs, and for which, when bounded, the EF-emptiness problem is hence decidable too.
This class is only the second syntactic subclass of PTAs (after L/U-PTAs) for which this problem is decidable.

\paragraph{Decidability of PTAs}
Our second main contribution to reachability is to study several open problems for PTAs and several known subclasses (as well as the new class of IP-PTAs):
we study here the emptiness and universality of reachability (EF), as well as unavoidability emptiness (AF).
Emptiness is of utmost importance as, without decidability of the emptiness, exact synthesis is practically ruled out.
Universality checks whether all parameter valuations satisfy a property, which is important for applications where the designer has no power on some valuations; this is the case of networks, where some latencies (\eg{} the transmission time of some packets) may be totally arbitrary.

Among our results, we prove in particular that AF-emptiness is undecidable for both bounded IP-PTAs and bounded L/U-PTAs.
Overall, we significantly enhance the knowledge we have of decidability problems for PTAs and subclasses.

\subsubsection{Liveness} When a ``good'' behavior is not always eventually reached, it can be for two main reasons: either there is a deadlock (a state in which the system cannot evolve anymore), or there is a livelock (an infinite path never reaching the ``good'' behavior).
Both situations are captured by the CTL operator ``EG''~\cite{CES-86}.

With the notable exception of~\cite{JLR15,ALR18FORMATS}, and to some extent of~\cite{BlT09} which addresses the existence of cycles, all the works cited above focus on safety properties, through the basic problem of reachability. This is maybe not so surprising given that most results related to this simpler problem are already negative.

We nonetheless address here the problem of liveness in PTAs, and more precisely, with the negative result of~\cite{JLR15} on AF-emptiness in mind, we start from L/U-PTAs with rational-valued parameters and further refine both the model and the properties.
We prove (for the class of L/U-PTAs) that:
\begin{enumerate}
    \item deciding the existence of at least one parameter valuation for which there exists an infinite run (discrete cycle) in the automaton is \textsc{PSpace}-complete;
	\item deciding the existence of a parameter valuation such that the system has a deadlock is however undecidable;
	\item the problem of the existence of a valuation for which a run remains in a given set of locations exhibits a very thin border between decidability and undecidability: while this problem is decidable for L/U-PTAs with a bounded parameter domain with closed bounds, it becomes undecidable if either the assumption of boundedness or of closed bounds is lifted.
		This result confirms that L/U-PTAs stand at the border between decidability and undecidability.
\end{enumerate}

Differently from~\cite{BlT09}, we use here no accepting locations.
In addition, our parameters are not restricted to be integer-valued, and can be rational-valued.

\subsection{About this manuscript}
This manuscript is an extended version of~\cite{ALR16,ALime17}.
In addition to merging notations and preliminary concepts, and adding all proofs missing in~\cite{ALR16,ALime17}, we extended these manuscripts as follows:
\begin{itemize}
	\item we added more explanations on the technical parts;
    \item we modified the proof of \cref{theorem:EF-empty:undecidable} (our former version used a ``block'' statement in the 2-counter machine, as in~\cite{AHV93}, which some readers might find objectionable), impacting all subsequent proofs of \cref{section:reachability};
    \item we added the new Corollary~\ref{corollary:open-bounded-LUPTA-AF-undecidable};
    \item we added several summaries of results (in tables and graphics).
\end{itemize}

\subsection{Outline}
We recall the necessary preliminaries in \cref{section:preliminaries}.
We introduce integer-point PTAs (IP-PTAs) in \cref{section:IPPTA} and study their relationship with existing results.
We then study reachability problems in \cref{section:reachability} and liveness problems in \cref{section:liveness} for PTAs, L/U-PTAs and various other subclasses.
We summarize our results in \cref{section:summary} and discuss perspectives in \cref{section:conclusion}.

\section{Preliminaries}
\label{section:preliminaries}
\subsection{Clocks, parameters and constraints}

Let $\grandn$, $\grandz$, $\grandqplus$ and $\grandrplus$ denote the sets of non-negative integers, integers, non-negative rational numbers and non-negative real numbers respectively.
Let $\interval(\grandn)$ denote the set of non-necessarily closed intervals on~$\grandn$, \ie{} \ForLongVersion{the set of intervals }of the form $[a,b]$, $(a,b]$, $[a,b)$ or $(a,b)$ where $a,b\in \grandn$ and $a \leq b$.

Throughout this paper, we assume a set~$\Clock = \{ \clock_1, \dots, \clock_\ClockCard \} $ of \emph{clocks}, \ie{} real-valued variables that evolve at the same rate.
A clock valuation is a function
$\clockval : \Clock \rightarrow \grandrplus$.
We identify a clock valuation~$\clockval$ with the point $(\clockval(\clock_1), \dots, \clockval(\clock_{\ClockCard}))$ of $\grandrplus^\ClockCard$.
We write $\ClocksZero$ for the clock valuation that assigns $0$ to all clocks.
Given $d \in \grandrplus$, $\clockval + d$ denotes the valuation such that $(\clockval + d)(\clock) = \clockval(\clock) + d$, for all $\clock \in \Clock$.
Given $\resets \subseteq \Clock$, we define the \emph{reset} of a valuation~$\clockval$, denoted by $\reset{\clockval}{\resets}$, as follows: $\reset{\clockval}{\resets}(\clock) = 0$ if $\clock \in \resets$, and $\reset{\clockval}{\resets}(\clock)=\clockval(\clock)$ otherwise.

We assume a set~$\Param = \{ \param_1, \dots, \param_\ParamCard \} $ of \emph{parameters}, \ie{} unknown constants.
A parameter {\em valuation} $\pval$ is a function
$\pval : \Param \rightarrow \grandqplus$.
We identify a valuation~$\pval$ with the point $(\pval(\param_1), \dots, \pval(\param_{\ParamCard}))$ of $\grandqplus^\ParamCard$.
An \emph{integer} parameter valuation is \ForLongVersion{a valuation $\pval$ }such that $\forall \param\in \Param, \pval(\param)\in \grandn$. %
For short, we  also call \emph{integer points} such integer parameter valuations, in reference to the classical geometrical interpretation of sets of parameter valuations~\cite{JLR15}.

In the following, we assume %
	${\compOp} \in \{<, \leq, \geq, >\}$.
Throughout this paper, $\lterm$ denotes a linear term over $\Clock \cup \Param$ of the form $\sum_{1 \leq i \leq \ClockCard} \alpha_i \clock_i + \sum_{1 \leq j \leq \ParamCard} \beta_j \param_j + d$, with
	$\clock_i \in \Clock$,
	$\param_j \in \Param$,
	and
	$\alpha_i, \beta_j, d \in \grandz$.
A \emph{constraint}~$\C$ over $\Clock \cup \Param$ is a conjunction of inequalities of the form $\lterm \compOp 0$. 
Given a parameter valuation~$\pval$, $\valuate{\C}{\pval}$ denotes the constraint over~$\Clock$ obtained by replacing each parameter~$\param$ in~$\C$ with~$\pval(\param)$.
Likewise, given a clock valuation~$\clockval$, $\valuate{\valuate{\C}{\pval}}{\clockval}$ denotes the expression obtained by replacing each clock~$\clock$ in~$\valuate{\C}{\pval}$ with~$\clockval(\clock)$.
We say that %
$\pval$ \emph{satisfies}~$\C$,
denoted by $\pval \models \C$,
if the set of clock valuations satisfying~$\valuate{\C}{\pval}$ is nonempty.
Given a parameter valuation $\pval$ and a clock valuation $\clockval$, we denote by $\wv{\clockval}{\pval}$ the valuation over $\Clock\cup\Param$ such that 
for all clocks $\clock$, $\valuate{\clock}{\wv{\clockval}{\pval}}=\valuate{\clock}{\clockval}$
and 
for all parameters $\param$, $\valuate{\param}{\wv{\clockval}{\pval}}=\valuate{\param}{\pval}$.
We use the notation $\wv{\clockval}{\pval} \models \C$ to indicate that $\valuate{\valuate{\C}{\pval}}{\clockval}$ evaluates to true.
We say that $\C$ is \emph{satisfiable} if $\exists \clockval, \pval \text{ s.t.\ } \wv{\clockval}{\pval} \models \C$.

A \emph{(parametric) guard}~$\guard$ is a constraint over $\Clock \cup \Param$ defined by inequalities of the form
	$\clock \compOp \sum_{1 \leq j \leq \ParamCard} \beta_j \param_j + d$, with $\beta_j \in \{0, 1 \}$ and $d \in \grandz$.
Given an arbitrary order on $\Clock \cup \Param$, valuations can be seen as points in the real coordinate space of dimension $|\Clock\cup\Param|$, using $|A|$ to denote the cardinality of set~$A$.
Then the set of valuations satisfying constraint~$\C$ can be seen as a convex polyhedron in that vector space. In the sequel, we will often make the small abuse of using constraint and polyhedron indifferently.

\subsection{Parametric timed automata}
\subsubsection{Syntax}

Parametric timed automata (PTA) extend timed automata with parameters within guards and invariants in place of integer constants~\cite{AHV93}.

\begin{defi}\label{def:PTA}
	A PTA
	$\A$ is a tuple \mbox{$\A = (\Actions, \Loc, \locinit, \Clock, \Param, \invariant, \Edges)$}, where:
	\begin{enumerate}
		\item $\Actions$ is a finite set of actions,
		\item $\Loc$ is a finite set of locations,
		\item $\locinit \in \Loc$ is the initial location,
		\item $\Clock$ is a finite set of clocks,
		\item $\Param$ is a finite set of parameters,
		\item $\invariant$ is the invariant, assigning to every $\loc\in \Loc$ a parametric guard $\invariant(\loc)$,
		\item $\Edges$ is a finite set of edges  $\edge = (\loc,\guard,\action,\resets,\loc')$
		where
		$\loc,\loc'\in \Loc$ are the source and target locations, $\action \in \Actions$, $\resets\subseteq \Clock$ is a
		set of clocks to be reset, and
		$\guard$ (the transition guard) is a parametric guard.
	\end{enumerate}
\end{defi}

Given a parameter valuation $\pval$, we denote by $\valuate{\A}{\pval}$ the non-parametric timed automaton where all occurrences of a parameter~$\param_i$ have been replaced by~$\pval(\param_i)$.
In the following, we may denote as a \emph{timed automaton} any structure $\valuate{\A}{\pval}$, by assuming a rescaling of the constants:
	by multiplying all constants in $\valuate{\A}{\pval}$ by their least common denominator,
		we obtain an equivalent timed automaton (with integer constants, as in~\cite{AD94}).
\subsubsection{Concrete semantics}

Let us first recall the concrete semantics of TAs.

\begin{defi}[Concrete semantics of a TA]
	Given a PTA $\A = (\Actions, \Loc, \locinit, \Clock, \Param, \invariant, \Edges)$,
	and a parameter valuation~\(\pval\),
	the concrete semantics of $\valuate{\A}{\pval}$ is given by the timed transition system $(\States, \sinit, \flecheRel)$, with
	\begin{itemize}
		\item $\States = \{ (\loc, \clockval) \in \Loc \times \grandrplus^\ClockCard \mid \wv{\clockval}{\pval} \models \invariant(\loc) \}$ %
		\item $\sinit = (\locinit, \ClocksZero) $
		\item  $\flecheRel$ consists of the discrete and (continuous) delay transition relations:
				\begin{itemize}
			\item discrete transitions: $(\loc,\clockval) \fleche{\edge} (\loc',\clockval')$, %
				if $(\loc, \clockval) , (\loc',\clockval') \in \States$, there exists $\edge = (\loc,\guard,\action,\resets,\loc') \in \Edges$, $\clockval'= \reset{\clockval}{\resets}$, and $\wv{\clockval}{\pval} \models \guard$.
			\item delay transitions: $(\loc,\clockval) \fleche{d} (\loc, \clockval+d)$, with $d \in \grandrplus$, if $\forall d' \in [0, d], (\loc, \clockval+d') \in \States$.
		\end{itemize}
	\end{itemize}
\end{defi}

Moreover we write $(\loc, \clockval)\longuefleche{\edge} (\loc',\clockval')$ for a combination of a delay and discrete transition where
	$((\loc, \clockval), \edge, (\loc', \clockval')) \in \longueflecheRel$ if
		$\exists d, \clockval'' :  (\loc,\clockval) \fleche{d} (\loc,\clockval'') \fleche{\edge} (\loc',\clockval')$.

Given a TA~$\valuate{\A}{\pval}$ with concrete semantics $(\States, \sinit, \flecheRel)$,
we refer to the states of~$\States$ as the \emph{concrete states} of~$\valuate{\A}{\pval}$.
A (concrete) \emph{run} of~$\valuate{\A}{\pval}$ is a possibly infinite alternating sequence of concrete states of $\valuate{\A}{\pval}$ and edges starting from the initial concrete state $\sinit$ of the form 
$\sinit \longuefleche{\edge_0} \state_1\longuefleche {\edge_1} \cdots \longuefleche{\edge_{m-1}} \state_m \longuefleche{\edge_{m}} \cdots$, such that 
$\edge_i \in \Edges$ and $(\state_i , \edge_i , \state_{i+1}) \in \longueflecheRel$ for all $i = 0, 1, \dots$.
Given a state~$\state=(\loc, \clockval)$, we say that $\state$ is reachable (or that $\valuate{\A}{\pval}$ reaches $\state$) if $\state$ belongs to a run of $\valuate{\A}{\pval}$.
By extension, we say that $\loc$ is reachable in~$\valuate{\A}{\pval}$, if there exists a state $(l,\clockval)$ that is reachable.
For any run $\varrun$, we note $\runLocs$ the set of locations reached by~$\varrun$.
A maximal run is a run that is either infinite (\ie{} contains an infinite number of discrete transitions), or that cannot be extended by a discrete transition.
A maximal run is deadlocked if it is finite, \ie{} it contains a finite number of discrete transitions.
By extension, we say that a TA is deadlocked if it contains at least one deadlocked run.

\subsubsection{Symbolic semantics}\label{sss:symbolic}

We now recall the symbolic semantics of parametric timed automata (see \eg{}~\cite{HRSV02,ACEF09,JLR15}).

We first define the \emph{time elapsing} of a constraint~$\C$, denoted by $\timelapse{\C}$, as the constraint over $\Clock$ and $\Param$ obtained from~$\C$ by delaying all clocks by an arbitrary amount of time.
That is, $\timelapse{\C} = \{ \wv{\clockval'}{\pval} \mid \exists \clockval,d : \clockval \models \valuate{\C}{\pval} \land \forall \clock \in \Clock : \clockval'(\clock) = \clockval(\clock) + d, d \in \grandrplus \}$.
Dually, we define the \emph{past} of~$\C$, denoted by $\timepast{\C}$, as the constraint over $\Clock$ and $\Param$ obtained from~$\C$ by letting time pass backward by an arbitrary amount of time. %
That is, $\timepast{\C} = \{ \wv{\clockval'}{\pval} \mid \exists\clockval,d : \clockval \models \valuate{\C}{\pval} \land \forall \clock \in \Clock : \clockval'(\clock) + d = \clockval(\clock), d \in \grandrplus \}$.
Given $\resets \subseteq \Clock$, we define the \emph{reset} of~$\C$, denoted by $\reset{\C}{\resets}$, as the constraint obtained from~$\C$ by resetting the clocks in~$\resets$, and keeping the other clocks unchanged.
We denote by $\projectP{\C}$ the projection of~$\C$ onto~$\Param$, \ie{} obtained by existentially eliminating the clock variables (\eg{} using Fourier-Motzkin algorithm~\cite{schrijver-book-86}): $\pval\models\projectP{\C}$ iff $\exists \clockval: \wv{\clockval}{\pval}\models C$.
The (sets of clock valuations satisfying the) constraints generated by PTA can be represented by subsets of $\grandrplus^{|\Clock|}$ with a special form called \emph{parametric zone}~\cite{HRSV02}.
A parametric zone is a convex polyhedron over $\Clock \cup \Param$ in which all constraints on variables are of the form $\clock\compOp \plterm$ (parametric rectangular constraints), or $\clock_i -\clock_j \compOp \plterm$ (parametric diagonal constraints), where $\clock_i \in \Clock$, $\clock_j \in \Clock$ and $\plterm$ is a parametric linear term over $\Param$, \ie{} a linear term without clocks ($\alpha_i = 0$ for all $i$). The intersection of two parametric zones, the future of parametric zone, its past, and its reset for any subset of clocks, are again parametric zones~\cite{HRSV02,JLR15}.

\begin{defi}[Symbolic state]\label{def:symbolicstate}
A symbolic state is a pair $\symbstate = (\loc, \C)$ where $\loc \in \Loc$ is a location, and $\C$ its associated parametric zone.
\end{defi}

The initial symbolic state of~$\A$ is $\symbstateinit = \big(\locinit, \timelapse{(\{\ClocksZero\} \land \invariant(\loc_0))} \land \invariant(\loc_0) \big)$. From what precedes, and since $\{\ClocksZero\}$ and $\invariant(\loc_0)$ are parametric zones, the constraint in this initial symbolic state is a parametric zone.

The symbolic semantics relies on the $\Succ$ operation.
Given a symbolic state $\symbstate = (\loc, \C)$ and an edge $\edge = (\loc,\guard,\action,\resets,\loc')$,
$\Succ(\symbstate, \edge) = (\loc', \C')$, with $\C' = \timelapse{\big(\reset{(\C \land \guard)}{\resets} \land \invariant(\loc')\big )} \land \invariant(\loc')$.
The $\Succ$ operation is effectively computable, using polyhedral operations: note that the successor of a parametric zone~$\C$ is a parametric zone. %

A symbolic run of a PTA is a possibly infinite alternating sequence of symbolic states and edges starting from the initial symbolic state, of the form 
$\symbstate_0 \Fleche{\edge_0} \symbstate_1\Fleche {\edge_1} \cdots$,
such that for all $i = 0, 1, \dots$, we have $\edge_i \in \Edges$, and
$\symbstate_{i+1}=\Succ(\symbstate_i, \edge_i)$.
In the following, we simply refer to symbolic states belonging to a run of~$\A$ as symbolic states of~$\A$. %

Finally, we say that a symbolic state is \emph{reachable} if it belongs to a symbolic run.\label{newtext:reachable:symbolic}

\subsection{Subclasses of PTAs}

L/U-PTAs~\cite{HRSV02} constrain the use of parameters: parameters must be partitioned between lower-bound parameters (only compared to clocks in parametric guards as a lower bound) and upper-bound parameters. 
L/U-PTAs were notably studied in~\cite{HRSV02,BlT09,ALime17,ALR18FORMATS}.

\begin{defi}[L/U-PTA]\label{def:LUPTA} %
	An L/U-PTA is a PTA where the set of parameters is partitioned into lower-bound parameters and upper-bound parameters,
	where an upper-bound (resp.\ lower-bound) parameter~$\param_i$ is such that, 
    for every guard or invariant constraint $\clock \compOp \sum_{1 \leq j \leq \ParamCard} \beta_j \param_j + d$, we have: $\beta_i=1$ implies ${\compOp} \in \{ \leq, < \}$ (resp.\ ${\compOp} \in \{ \geq, > \}$).
\end{defi}

Recall from our definition of guard that $\beta_i$ can only be 0 or~1.

L/U-PTAs enjoy a well-known monotonicity property recalled in the following lemma (that corresponds to a reformulation of~\cite[Prop~4.2]{HRSV02}), stating that increasing upper-bound parameters or decreasing lower-bound parameters can only add behaviors.

\begin{lem}\label{lemma:HRSV02:prop4.2}
	Let~$\A$ be an L/U-PTA and~$\pval$ be a parameter valuation.
	Let $\pval'$ be a valuation such that
	for each upper-bound parameter~$\param^+$, $\pval'(\param^+) \geq \pval(\param^+)$
	and
	for each lower-bound parameter~$\param^-$, $\pval'(\param^-) \leq \pval(\param^-)$.
	Then any run of~$\valuate{\A}{\pval}$ is a run of $\valuate{\A}{\pval'}$. 
\end{lem}
\begin{exa}
	The PTA (fragment) in \cref{fig:inc-LU_m2c} page~\pageref{fig:inc-LU_m2c} is an L/U-PTA, with upper-bound parameter~$a^+$ and lower-bound parameter~$a^-$.
\end{exa}

In this paper, we will also consider \emph{bounded} PTAs, \ie{} PTAs with a bounded parameter domain that assigns to each parameter an infimum and a supremum, both integers.

\begin{defi}[bounded PTA]
	A \emph{bounded PTA} is $\bounded{\A}{\bounds}$, where $\A$ is a PTA, and $\bounds : \Param \rightarrow \interval(\grandn)$ assigns to each parameter~$\param$ an interval
	$[\boundinf, \boundsup]$,
		$(\boundinf, \boundsup]$,
		$[\boundinf, \boundsup)$,
		or
		$(\boundinf, \boundsup)$,
	with $\boundinf, \boundsup \in \grandn$.
	We use $\boundinfin{\param}{\bounds}$ and $\boundsupin{\param}{\bounds}$ to denote the infimum and the supremum of~$\param$, respectively.
	Note that we rule out $\infty$ as a supremum.
	
	We say that a bounded PTA is a \emph{closed bounded PTA} if, for each parameter~$\param$, its ranging interval $\bounds(\param)$ is of the form $[\boundinf, \boundsup]$; otherwise it is an \emph{open bounded PTA}.
	
	We define similarly bounded L/U-PTAs.
\end{defi}

Whereas bounded PTAs are naturally a subclass of PTAs, we showed in~\cite{ALR16FORMATS} that, for some equivalence relations, bounded L/U-PTAs are \emph{incomparable} with L/U-PTAs. In particular, this can be the case when the equivalence relation refers to the equality of the sets of parameter valuations such that some reachability property is satisfied, because boundedness imposes constraints on parameters that cannot be expressed in an L/U-PTA (\eg{} enforcing upper bounds on upper-bound parameters)%
	: a consequence is that undecidability results for bounded L/U-PTAs cannot be automatically extended to L/U-PTAs; conversely, decidability results for L/U-PTAs cannot be automatically extended to bounded L/U-PTAs.

\subsection{Decision and synthesis problems}\label{ss:decision-problems}

\subsubsection{Decision problems}
In this article, we mainly focus on safety and liveness properties expressed as basic (non-nested) Computation Tree Logic (CTL,~\cite{CES-86}) properties. Given a TA $\A$ and a subset of its locations $\somelocs$: 
\begin{itemize}
    \item Reachability (EF): does there exist a concrete run $\varrun$ such that $\runLocs(\varrun)\cap\somelocs\neq\emptyset$?
    \item Safety (AG): for all concrete runs $\varrun$, does $\runLocs(\varrun)\subseteq\somelocs$?
    \item Unavoidability (AF): for all maximal concrete runs $\varrun$, does $\runLocs(\varrun)\cap\somelocs\neq\emptyset$?
    \item Preservability (EG):  does there exist a maximal concrete run $\varrun$ such that $\runLocs(\varrun)\subseteq\somelocs$?
\end{itemize}

We also consider two variants of EG:
\begin{itemize}
    \item Deadlock-existence (denoted here by ED): does there exist a finite maximal concrete run?
    \item Cycle-existence (denoted here by EC): does there exists an \emph{infinite} maximal concrete run?
\end{itemize}

Let \Problem{} be a given a class of decision problems (reachability, unavoidability, etc.). We define parametric variants as follows:

\smallskip

\defProblem
	{\Problem-emptiness}
	{A PTA~\A{} and an instance $\varproblem$ of \Problem{}}
	{Is the set of parameter valuations $\pval$ such that $\valuate{\A}{\pval}$ satisfies $\varproblem$ empty?}

\defProblem
	{\Problem-universality}
	{A PTA~\A{} and an instance $\varproblem$ of \Problem{}}
	{Are all parameter valuations $\pval$ such that $\valuate{\A}{\pval}$ satisfies $\varproblem$?}

Emptiness is the most basic parametric question: is the set of parameter valuations such that the property holds empty?
Universality gives a robustness quality to the property and permits to effectively abstract an infinite number of verifications with concrete values.

Note that EF and AG are dual properties, and so are AF and EG.
In our parametric setting this implies that, for instance, EF-emptiness and AG-universality are dual, and so are AF-emptiness and EG-universality.

Also note that ED-emptiness is equivalent to AC-universality, where AC-universality asks whether all \ForLongVersion{parameter }valuations are such that all maximal runs are infinite.
Conversely, EC-emptiness is equivalent to AD-universality (for all valuations, all maximal runs are finite).

\subsubsection{Synthesis and membership problems}\label{sss:synthesis}
In addition to the aforementioned decision problems, we define synthesis problems: 

\defProblem
	{\Problem-synthesis}
	{A PTA~\A{} and an instance $\varproblem$ of \Problem{}}
	{Compute the set of parameter valuations $\pval$ such that $\valuate{\A}{\pval}$ satisfies $\varproblem$}

Finally, we are also interested in classical membership decision problems:

\defProblem
	{\Class-membership}
	{A PTA~\A{} and a subset $\Class$ of the set of all PTAs}
	{Does $\A$ belong to $\Class$?}

    \subsection{Undecidable problems for two-counter machines}\label{ss:2CM}
    Most of the undecidability proofs we use work by reduction of a 2-counter machine problem to our problems; we therefore briefly recall here how such a machine works, and the two problems we consider.

A deterministic 2-counter machine~\cite{Minsky67} has two non-negative counters $C_1$ and $C_2$, a finite number of states and \ForLongVersion{a finite number of }transitions, which can be of the form:
	\begin{itemize}
		\item ``when in state $\cms_i$, increment $C_k$ and go to $\cms_j$'';
		\item ``when in state $\cms_i$, if $C_k=0$ then go to $\cms_k$, otherwise decrement $C_k$ and go to~$\cms_j$''.
	\end{itemize}

The machine starts in state $\cms_0$ with the counters set to~0.
The \emph{halting problem} consists in deciding whether some distinguished state called \cmshalt{} can be reached or not.
The \emph{boundedness problem} asks whether the counters stay bounded or not along the execution of the machine.
Both problems are known to be undecidable~\cite{Minsky67}.

\section{Integer-points parametric timed automata}
\label{section:IPPTA}
\begin{figure}[tb]
	\newcommand{\tikzscale}{.62}
	\newcommand{\scalefactor}{.9}
	
	\centering
		\scalebox{\scalefactor}{
		\begin{tikzpicture}[scale=\tikzscale,->, >=stealth', auto]
			\draw[contribution] (-1, 0) circle [radius=2.5];
			\node[legende, rotate=50] at (-2, 1) {bounded L/U};

			\draw[] (1, 0) circle [radius=2.5];
			\node[legende] at (2.8, 0.8) {L/U};

			\draw[contribution] (0, -2) ellipse (3 and 2.5);
			\node[legende] at (0, -4) {IP-PTA};

			\draw[] (0, -1) ellipse (2.5 and .75);
			\node[legende] at (0, -1) {closed L/U};

			\draw[-] (-1, 2.5) --++ (-5, 0);
			\draw[draw=none] (-1, 2.5) arc[radius = 2.5, start angle= 90, end angle= -65] node(arcfin){}; %
			\draw[-,contribution] (arcfin) --++ (-2.5, -1.35); %
			\draw[-] (arcfin) --++ (-5, -2.7); %
			\node[legende,align=center] at (-4.5, -3) {bounded\\PTAs};

			\draw[] (-6, -5) rectangle (4.5, 3);
			\node[legende] at (3.75, -4) {PTAs};
			\end{tikzpicture}
		}

	\caption{Subclasses of PTAs considered in this paper}
    \label{figure:subclasses}
\end{figure}
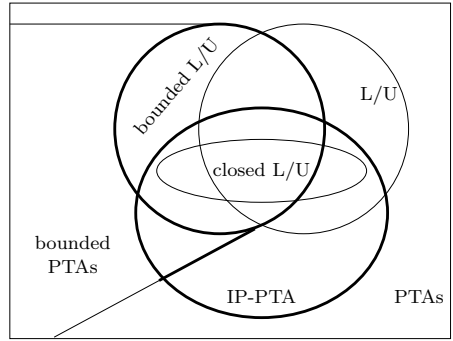
In this section, we introduce integer-points parametric timed automata (IP-PTAs for short), \ie{} a semantic subclass of PTAs in which any (reachable) symbolic state contains at least one integer point.

First, we compare in this section IP-PTAs with L/U-PTAs and show that the class of bounded IP-PTAs is strictly larger than bounded L/U-PTAs with non-strict inequalities. %
Then, our main result regarding this class will be to prove the decidability of the EF-emptiness problem for bounded IP-PTAs (\cref{ss:decidability}).
We will also show that synthesis is intractable in practice, and that the same holds for bounded L/U-PTAs (\cref{ss:synthesis}).
Finally, although we prove that the membership problem is undecidable for IP-PTAs, we will exhibit a syntactic sufficient condition, that provides a new subclass of PTAs for which the EF-emptiness problem is decidable (\cref{ss:membership}).

We give in \cref{figure:subclasses} a pictorial presentation of the subclasses (existing and introduced in this paper) and their relations---that will be proved in the remainder of this paper.
A similar graphical representation, this time for decidability results of various problems, will be given in \cref{figure:summary:decidability} in \cref{section:summary}.

\begin{defi}\label{definition:IP-PTA}
	A PTA~$\A$ is \ForLongVersion{said to be }an \emph{integer points PTA} (in short \emph{IP-PTA}) if, in any reachable symbolic state $(\loc, \C)$ of~$\A$, $\C$ contains at least one integer point.
\end{defi}
Let us now compare IP-PTAs and L/U-PTAs.
We first need the following lemma, stating that any reachable symbolic state of an L/U-PTA contains an integer parameter valuation.

\begin{lem}\label{lemma:LUPTA:IP}
	Let $(\loc, \C)$ be a reachable symbolic state of an L/U-PTA.
	Then $\projectP{\C}$ contains at least one integer point.
\end{lem}
\begin{proof}
    Consider a (non-empty) reachable symbolic state~$(\loc, \C)$ of an L/U-PTA.
	Let $\pval \models \projectP{\C}$.
    From the well-known monotonicity property of L/U-PTAs
		(recalled in Lemma~\ref{lemma:HRSV02:prop4.2})%
		, any parameter valuation such that the lower-bound parameters~$\param_i^-$ are lower or equal to~$\pval(\param_i^-)$ and upper-bound parameters~$\param_j^+$ are greater than or equal to~$\pval(\param_j^+)$ also belong to~$\projectP{\C}$.
	In particular, this is the case of the integer parameter valuation assigning~0 to all lower-bound parameters, and assigning to upper-bound parameters~$\param_j^+$ the smallest integer greater than or equal to~$\pval(\param_j^+)$.
\end{proof}

The previous lemma that ensures the presence of an integer parameter valuation in any symbolic state does not guarantee that an L/U-PTA is an IP-PTA, because clocks may have non-integer values.

\begin{prop}\label{proposition:IPPTA-LUPTA:incomp}
	The class of IP-PTAs is incomparable with the class of L/U-PTAs, in the sense that there exists an IP-PTA that is not an L/U-PTA and vice-versa.
\end{prop}
\begin{proof}
	\begin{itemize}
		\item%
			Consider an L/U-PTA with a transition guarded by $\clock > 0$ and resetting no clock, followed by a second location with invariant $\clock < 1$; then, necessarily, the symbolic state associated with this second location contains no integer point (as $\clock \in (0, 1)$ in that symbolic state).
			
		\item%
			It is easy to exhibit an IP-PTA that is not an L/U-PTA.
			This is for example the case of a simple PTA with only one location, one clock~$\clock$ and one parameter~$\param$ with a self-loop with guard $\clock = \param$ and resetting $\clock$.\qedhere
	\end{itemize}

\end{proof}

However, we can prove that any \emph{closed} L/U-PTA, \ie{} with only non-strict inequalities, 
is an IP-PTA.
In order to show that the class of closed L/U-PTAs is included in the class of IP-PTAs, we need the following lemma.

\begin{lem}\label{lemma:Didier}
	Let $\A$ be a PTA with only non-strict inequalities.
	Let $\symbstate = (\loc, \C)$ be a symbolic state of~$\A$.
	Then if $\projectP{\C}$ contains at least one integer parameter valuation, then $\C$ contains an integer point.
\end{lem}
\begin{proof}
    Since there is at least one integer parameter valuation $\pval$ in $\projectP{\C}$, then $\valuate{\C}{\pval}$ is not empty.
    Since $\pval$ is an integer valuation, $\valuate{\C}{\pval}$ is a zone of a timed automaton with integer constants, so the vertices of $\valuate{\C}{\pval}$ are integer points.
    Finally, there is at least one vertex in $\valuate{\C}{\pval}$ because all clocks are nonnegative (and hence are bounded from below by~$0$), and this vertex does belong to $\valuate{\C}{\pval}$ because it is topologically closed due to the non-strict constraints.
    So $\C$ contains at least one integer point.
\end{proof}
\begin{prop}\label{proposition:IPPTA-oLUPTA}
	The class of IP-PTAs is strictly larger than the class of closed L/U-PTAs.
\end{prop}
\begin{proof}
	From Lemmas~\ref{lemma:LUPTA:IP},~\ref{lemma:Didier}, and Proposition~\ref{proposition:IPPTA-LUPTA:incomp} ($\Leftarrow$).
\end{proof}

The previous result also holds for bounded PTAs:

\begin{cor}\label{proposition:bIPPTA-cbLUPTA}
	The class of bounded IP-PTAs is strictly larger than the class of closed bounded L/U-PTAs.
\end{cor}
\begin{proof}
	Lemma~\ref{lemma:LUPTA:IP} extends to bounded L/U-PTAs, since the bounds are integers (this would not hold otherwise).
	Then, the proof of Proposition~\ref{proposition:IPPTA-LUPTA:incomp} ($\Leftarrow$) holds with bounded IP-PTAs and closed bounded L/U-PTAs.
	Applying Lemma~\ref{lemma:Didier} concludes the proof.
\end{proof}
\begin{cor}\label{proposition:bIPPTA-bLUPTA}
	The class of bounded IP-PTAs is incomparable with the class of bounded L/U-PTAs.
	The class of bounded IP-PTAs is incomparable with the class of L/U-PTAs.
\end{cor}
\begin{proof}
	The proof of Proposition~\ref{proposition:IPPTA-LUPTA:incomp} can be applied with bounded PTAs on either side.
\end{proof}
\section{Reachability properties}\label{section:reachability}

In this section, we prove results for IP-PTAs.
While we present one key decidability result for this subclass (decidability of EF-emptiness), the subsequent undecidability results can then be lifted to more general subclasses, including general PTAs.

\paragraph{Outline}
We first provide a new proof for the undecidability of the EF-emptiness problem for general PTAs (\cref{ss:EF-emptiness}).
We then show that this proof can be modified to prove the undecidability of the EF-emptiness problem for closed bounded PTAs (\cref{ss:undecidability-EF-close-bounded}).
We then prove decidability of this problem for bounded IP-PTAs (\cref{ss:decidability}).
We then prove that EF-synthesis for IP-PTAs is intractable in practice (\cref{ss:synthesis}), and that the membership problem for IP-PTAs is undecidable (\cref{ss:membership}).
We finally prove the undecidability of EF-universality for IP-PTAs (\cref{ss:EF-universality}).

We will give a summary of the results proved in this section in \cref{table:summary:decidability} (in \cref{section:summary}).

\subsection{A new proof for the undecidability of EF-emptiness}\label{ss:EF-emptiness}

Historically, the EF-emptiness problem has been the most studied problem of those we have stated in \cref{ss:decision-problems}.
In the general setting of PTAs, several undecidability proofs are available, all based on reductions from the 2-counter machines halting problem~\cite{AHV93,Miller00,Doyen07,BBLS15,ALM20}.

We first propose yet another such construction that will be then modified to establish many of the results in this paper. Its main feature is that it works for a single rational parameter in $[0,1]$. While~\cite{Miller00} already proposes a similar result, we found it harder to extend to our purposes.
\begin{thm}
    The EF-emptiness problem is undecidable for bounded PTAs.
    \label{theorem:EF-empty:undecidable}
\end{thm}

\begin{proof}
We reduce from the halting problem of a 2-counter machine (2CM), which is undecidable~\cite{Minsky67}.
	\ForLongVersion{
	
	}%

	Given a machine $\calM$, we encode it as a PTA $\calA(\calM)$.
   	Let us describe this encoding in details, as we will modify it in the subsequent proofs.
	
	Each state $\cms_i$ of the machine is encoded as a location of the automaton, which we call~$\cmspta_i$.
	The counters are encoded using clocks $x$, $y$ and $z$ and one parameter~$a$, with the following relations with the values $c_1$ and $c_2$ of counters $C_1$ and $C_2$: when $x=0$, we have $y= 1-ac_1$ and $z=1-ac_2$.
	All three clocks are parametric, \ie{} are compared with~$a$ in some guard or invariant of the encoding.
    We will see that $a$ is a rational-valued bounded parameter, typically in~$[0,1]$ (although not bounding~$a$ has no impact on the proof).

	We initialize the clocks with the gadget in \cref{figure:EFemptiness:initial} (that also blocks the case where $a=0$).
	Note that, throughout the paper, we highlight in thick green style the locations of the PTA corresponding to a state of the 2CM (in contrast with other locations added in the encoding to maintain the matching between the clock values and the counter values).
	Since all clocks are initially~0, in \cref{figure:EFemptiness:initial} clearly, when in $\cmspta_0$ with $x=0$, we have $y=z=1$, which indeed corresponds to counter values~$0$.

		\begin{figure*}
			\begin{subfigure}[b]{.35\textwidth}
			\centering
			\scalebox{\generalFigsScaleFactor}{
				\begin{tikzpicture}[PTA, node distance=2.5cm]
				\node[location, initial] (l0) {$l_0$};
				\node[location, right of=l0] (l1) {$l_1$};
		
				\node[location2CM, right=of l1, xshift=-3em] (s0) {$\cmspta_0$};

				\path (l0) edge node[above, yshift=.5em]{\begin{tabular}{c}$x = a \land x > 0$\end{tabular}} (l1);
				\path (l1) edge node[above]{\begin{tabular}{c}$x = 1$ \\$x := 0$\end{tabular}} (s0);
				\end{tikzpicture}
			}
			
				\caption{Initial gadget}
				\label{figure:EFemptiness:initial}
			\end{subfigure}
			\hfill
			\begin{subfigure}[b]{.6\textwidth}
			\centering
			\scalebox{\generalFigsScaleFactor}{
			\begin{tikzpicture}[PTA, node distance=2cm]
				\node[location2CM] (si) {$\cmspta_i$};
				\node[location, xshift=-2em, right=of si] (l1) {$l_{i1}$};
				\node[location, above right=of l1] (l2) {$l_{i2}$};
				\node[location, below right=of l1] (l2p) {$l'_{i2}$};
				\node[location, below right=of l2] (l3) {$l_{i3}$};
				\node[location2CM, right=of l3, xshift=-2em] (sj) {$\cmspta_j$};

				\draw[->] (si) -- node {$x=0$} (l1);
				\draw[->] (l1) -- node {$\begin{array}{l}z=1\\ z:=0\end{array}$} (l2);
				\draw[->] (l1) -- node[swap] {$\begin{array}{l}y=a+1\\ y:=0\end{array}$} (l2p);
				\draw[->] (l2) -- node {$\begin{array}{l}y=a+1\\ y:=0\end{array}$} (l3);
				\draw[->] (l2p) -- node[swap] {$\begin{array}{l}z=1\\ z:=0\end{array}$} (l3);
				\draw[->] (l3) -- node {$\begin{array}{l}x=1\\ x:=0\end{array}$} (sj);
			\end{tikzpicture}
			}
			
			\caption{Increment gadget ($C_1$)}
			\label{fig:inc_m2c}
			\end{subfigure}
			\begin{subfigure}[b]{.9\textwidth}
			\centering
			\scalebox{\generalFigsScaleFactor}{
			\begin{tikzpicture}[PTA, node distance=2cm]
				\node[location2CM] (si) {$\cmspta_i$};
				\node[location, xshift=30, right=of si] (l1) {$l_{i1}$};
				\node[location, above right=of l1] (l2) {$l_{i2}$};
				\node[location, below right=of l1] (l2p) {$l'_{i2}$};
				\node[location, below right=of l2] (l3) {$l_{i3}$};
				\node[location2CM, right=of l3] (sj) {$\cmspta_j$};
				\node[location2CM, above=of si] (sk) {$\cmspta_k$};

				\draw[->] (si) -- node {$x=0$} node[below]{$y<1$} (l1);
				\draw[->] (l1) -- node {$\begin{array}{l}z=a + 1\\ z:=0\end{array}$} (l2);
				\draw[->] (l1) -- node[swap] {$\begin{array}{l}y=1\\ y:=0\end{array}$} (l2p);
				\draw[->] (l2) -- node {$\begin{array}{l}y=1\\ y:=0\end{array}$} (l3);
				\draw[->] (l2p) -- node[swap] {$\begin{array}{l}z=a+1\\ z:=0\end{array}$} (l3);
				\draw[->] (l3) -- node {$\begin{array}{c}x=a+1\\ x:=0\end{array}$} (sj);
				\draw[->] (si) -- node {$\begin{array}{l}x=0\\ y=1\end{array}$} (sk);
			\end{tikzpicture}
			}
			
			\caption{0-test and decrement gadget ($C_1$)}
			\label{fig:dec_m2c}
			\end{subfigure}
			\caption{EF-emptiness: gadgets}
		\end{figure*}
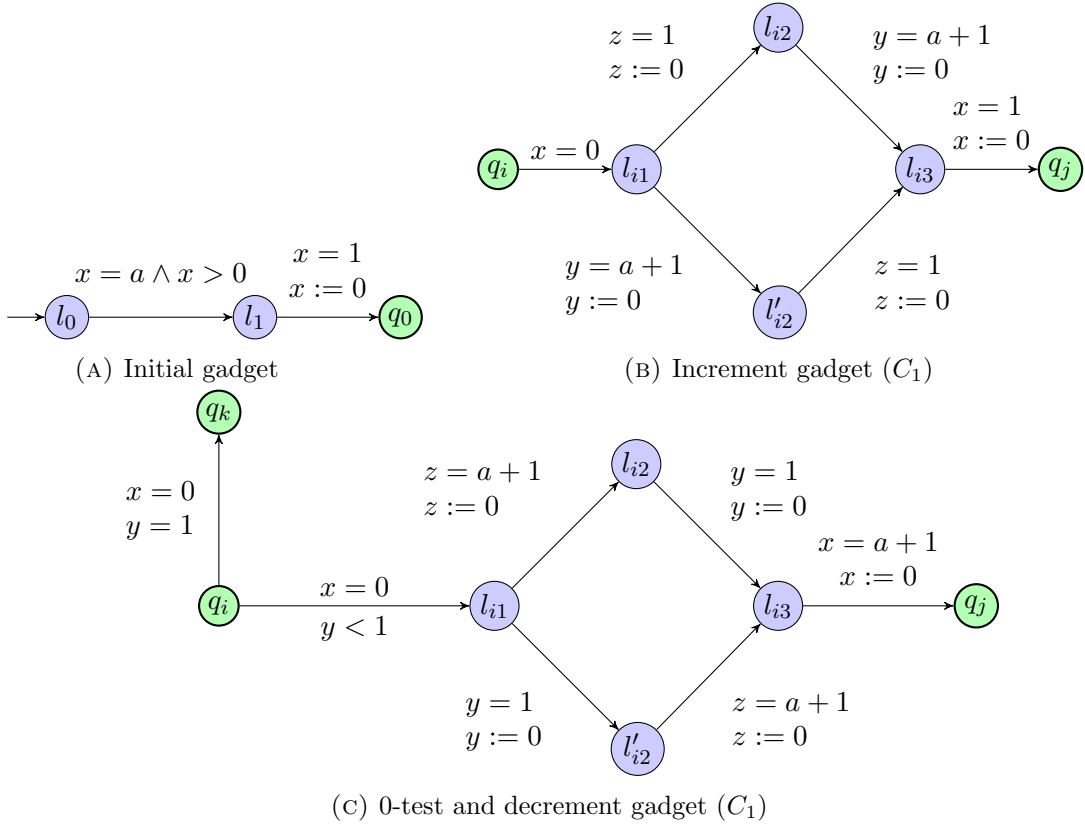

	We now present the gadget encoding the increment instruction of $C_1$ in \cref{fig:inc_m2c}.
	The transition from $\cmspta_i$ to $l_{i1}$ only serves to clearly indicate the entry in the increment gadget and is done in $0$ time unit.
	Since every edge is guarded by one equality, there are really only two paths that go through the gadget: one going through $l_{i2}$ and one through $l'_{i2}$. Let us begin with the former.
	We start from some encoding configuration: $x=0$, $y=1-ac_1$ and $z=1-ac_2$ in $\cmspta_i$ (and therefore the same in $l_{i1}$).
	We can enter $l_{i2}$ (after elapsing enough time) if $1-ac_2 \leq 1$, \ie{} $ac_2\geq 0$, which implies that $a\geq 0$, and when entering $l_{i2}$ we have $x=ac_2$, $y= 1 - a c_1 + a c_2$ and $z=0$.
	Then we can enter $l_{i3}$ if $1 - ac_1 + ac_2\leq 1 + a$, \ie{} $a(c_1 + 1) \geq ac_2$. When entering $l_{i3}$, we then have $x=a(c_1 + 1)$, $y=0$ and $z=a(c_1 + 1)-ac_2$.
	Finally, we can go to $\cmspta_j$ if $a(c_1 + 1)\leq 1$ and when entering $\cmspta_j$  we have $x=0$, $y=1-a(c_1 + 1)$ and $z=1-ac_2$, as expected.

	We now examine the second path.
	We can enter $l'_{i2}$ if $1-ac_1 \leq a+1$, \ie{} $a(c_1 + 1)\geq 0$, and when entering $l'_{i2}$ we have $x=a(c_1 + 1)$, $y=0$ and $z=1-ac_2+a(c_1 + 1)$. 
	Then we can go to $l_{i3}$ if $1 - ac_2 + a(c_1 + 1)\leq 1 + a$, \ie{} $a(c_1 + 1) \leq ac_2$. When entering $l_{i3}$, we then have $x=ac_2$, $y=ac_2-a(c_1 + 1)$ and $z=0$. 
	Finally, we can go to $\cmspta_j$ if $ac_2\leq 1$ and when entering $\cmspta_j$ we have $x=0$, $y=1-a(c_1 + 1)$ and $z=1-ac_2$, as expected.

	Remark that exactly one path can be taken depending on the respective order of $c_1 + 1$ and $c_2$, except when both are equal or $a=0$, in which cases both paths lead to the same configuration anyway (and the case $a=0$ is excluded by \cref{figure:EFemptiness:initial} anyway).

	Decrement is done similarly by replacing guards $y=a+1$ with $y=1$, and guards $x=1$ and $z=1$ with $x=a+1$ and $z=a+1$, respectively, as shown in \cref{fig:dec_m2c}.
    In addition, the 0-test is obtained by simply adding a transition from $\cmspta_i$ to~$\cmspta_k$ with guard $y=1 \land x=0$, which ensures that~$C_1 = 0$.
    Similarly, the guard from $\cmspta_i$ to $l_{i1}$ ensures that decrement is done only when the counter is not null.

	All those gadgets also work for $C_2$ by swapping $y$ and~$z$.
	
	The actions associated with the transitions do not matter; we can assume a single action~$\action$ on all transitions (omitted in all figures).

	We now prove that the machine halts iff there exists a parameter valuation~$\pval$ such that $\valuate{\A}{\pval}$ reaches location $\lochalt$.
	First note that if $a = 0$ the initial gadget cannot be passed, and so the machine does not halt.
	Assume $a > 0$.
	Consider two cases:
	\begin{enumerate}
		\item either the value of the counters is not bounded.
			Then, for any parameter valuation, at some point during an increment of, say, $C_1$ we will have $a(c_1 + 1)>1$ when taking the transition from $l_{i2}$ to $l_{i3}$ and the PTA will be blocked.
			Therefore, there exists no parameter valuation for which the PTA can reach $\lochalt$.
		\item or the value of the counters remains bounded.
			Let $c$ be their maximal value.
			Let us consider two subcases:
			\begin{enumerate}
				\item either the machine reaches $\cmshalt$: in that case, if $c=0$ and $0<a\leq 1$ or $c>0$ and $ca < 1$, then the PTA valuated with such parameter valuations correctly simulates the machine, yielding a (unique) run reaching location~$\lochalt$.
					The set of such valuations for~$a$ is certainly non-empty: $a=\frac{1}{2}$ belongs to it if $c=0$ and $a=\frac{1}{c}$ does otherwise.
				\item or the machine does not halt.
                    Then again, for a sufficiently small parameter valuation (\ie{} $a < 1$ if $c=0$ and $a \leq \frac{1}{c}$ otherwise), the machine is properly simulated, and since the machine does not halt, then the PTA never reaches $\lochalt$.
					For other values of $a$, the machine will block at some point in an increment gadget, because $a$ is not small enough and the guard to $\cmspta_j$ cannot be satisfied.
			\end{enumerate}
	\end{enumerate}
	Hence the machine halts iff there exists a parameter valuation~$\pval$ such that $\valuate{\A}{\pval}$ reaches $\lochalt$.
\end{proof}

\begin{rem}\label{remark:syntax2}
	In this paper, we allow guards and invariants of the form
		$\clock \compOp \sum_{1 \leq j \leq \ParamCard} \beta_j \param_j + d$,
	which is more restrictive than~\cite{BlT09} (that allows parametric coefficients different from 0 and 1, as well as diagonal constraints), but more permissive than~\cite{AHV93}, that only allows a syntax $\clock \compOp \param$.
		In fact, most papers in the literature define their own syntax (see~\cite{Andre19STTT} for a survey).
	We can adapt our proof to fit in the most restrictive syntax ($\clock \compOp \param$) as follows:
		transitions with ``$y=a+1$'' guards and ``$y:=0$'' reset can be equivalently replaced by one transition with a ``$y=1$'' guard and a reset of some additional clock~$w$, followed by a transition with a ``$w=a$'' guard and the ``$y:=0$'' reset (and similarly for $x$ and $z$ in the decrement gadget).
	This also allows the proof to work without complex parametric expressions in guards, using three additional clocks (we conjecture that a smarter encoding can be exhibited to factor these additional clocks, so as to use a single additional clock).
	A similar modification can be applied to all subsequent undecidability proofs.
\end{rem}
\subsection{Undecidability for closed  bounded PTAs}\label{ss:undecidability-EF-close-bounded}

Now, by reusing the previous proof, we can show that the EF-emptiness problem is undecidable for closed bounded PTAs.
This is an original result, as all existing results with bounded PTAs (\eg{}~\cite{Miller00,Doyen07,ALM20}) require (at least some) strict inequalities.
This result can also be seen as a complement to~\cite{Doyen07}, that proves the same result by using \emph{only} strict inequalities.

\newcommand{\theoremEFemptinessClosedBounded}{The EF-emptiness problem is undecidable for closed bounded PTAs.}
\begin{thm}\label{theorem:EF-empty:undecidable-closed-bounded}
	\theoremEFemptinessClosedBounded{}
\end{thm}
\begin{proof}
	First, recall that the (unique) parameter~$a$ in the proof of Theorem~\ref{theorem:EF-empty:undecidable} can be bounded by $[0,1]$, hence the PTA in the proof of Theorem~\ref{theorem:EF-empty:undecidable} is a bounded PTA.

	The encoding of the instructions of the 2-counter-machine used in the proof of Theorem~\ref{theorem:EF-empty:undecidable} is almost a closed PTA, as mostly non-strict inequalities are used, with two exceptions:
	\begin{enumerate}
		\item the initial gadget (\cref{figure:EFemptiness:initial}) requires a strict inequality to ensure $a > 0$;
		\item the 0-test and decrement gadget (\cref{fig:dec_m2c}) uses an inequality $y < 1$ to ensure the counter is not~0.
	\end{enumerate}
	We will remove these strict inequalities as follows:
	\begin{enumerate}
		\item we remove the guard between $\loc_0$ and $\loc_1$ in the initial gadget (\cref{figure:EFemptiness:initial}), that is we allow (so far) the valuation $a = 0$ that does not correctly simulate the machine;
		\item we remove the strict inequality $y < 1$ in the decrement gadget (\cref{fig:dec_m2c}) as follows.
		Instead of testing $y < 1$ when $x = 0$, it is equivalent to test $x > 0$ when $y = 1$ on the subsequent transitions.
		And testing $x > 0$ is equivalent to testing $x \geq a$ provided $a > 0$ (we will take care of this later on).
		The new gadget is given in \cref{figure:gadget-2CM-bounded-closed-PTA-decrement} and, provided $a > 0$, it is equivalent to the former gadget in \cref{fig:dec_m2c}.
	\end{enumerate}

	Now, with these new gadgets, the behavior of the encoding is the same as in the proof of Theorem~\ref{theorem:EF-empty:undecidable}, provided $a > 0$.
	Removing $a=0$ is necessary, as this valuation does not correctly encode the machine.
	We will therefore forbid $a = 0$ in a new final gadget, given in \cref{figure:gadget-2CM-bounded-closed-PTA-final}, with a transition from $\lochalt$ to a new location $\lochaltprime$ (via an intermediate location).
	This gadget allows us to ensure $a > 0$, without using strict inequalities; we believe this is the cornerstone of this proof of Theorem~\ref{theorem:EF-empty:undecidable-closed-bounded}.

		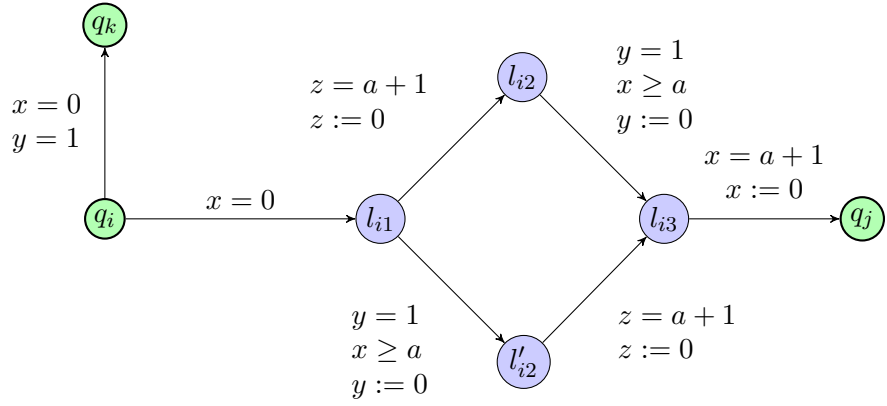
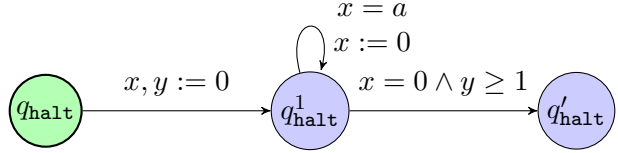
\begin{figure}
			\begin{subfigure}[b]{.9\textwidth}
			\centering
			\scalebox{\generalFigsScaleFactor}{
			\begin{tikzpicture}[PTA, node distance=2cm]
				\node[location2CM] (si) {$\cmspta_i$};
				\node[location, xshift=30, right=of si] (l1) {$l_{i1}$};
				\node[location, above right=of l1] (l2) {$l_{i2}$};
				\node[location, below right=of l1] (l2p) {$l'_{i2}$};
				\node[location, below right=of l2] (l3) {$l_{i3}$};
				\node[location2CM, right=of l3] (sj) {$\cmspta_j$};
				\node[location2CM, above=of si] (sk) {$\cmspta_k$};

				\draw[->] (si) -- node {$x=0$}  (l1);
				\draw[->] (l1) -- node {$\begin{array}{l}z=a + 1\\ z:=0\end{array}$} (l2);
				\draw[->] (l1) -- node[swap] {$\begin{array}{l}y=1\\x \geq a\\ y:=0\end{array}$} (l2p);
				\draw[->] (l2) -- node {$\begin{array}{l}y=1\\x \geq a\\ y:=0\end{array}$} (l3);
				\draw[->] (l2p) -- node[swap] {$\begin{array}{l}z=a+1\\ z:=0\end{array}$} (l3);
				\draw[->] (l3) -- node {$\begin{array}{c}x=a+1\\ x:=0\end{array}$} (sj);
				\draw[->] (si) -- node {$\begin{array}{l}x=0\\ y=1\end{array}$} (sk);
			\end{tikzpicture}
			}
			
			\caption{0-test and decrement gadget ($C_1$)}
			\label{figure:gadget-2CM-bounded-closed-PTA-decrement}
			\end{subfigure}
			\begin{subfigure}[b]{.9\textwidth}
			\centering
			\scalebox{\generalFigsScaleFactor}{
				\begin{tikzpicture}[PTA, node distance=2.5cm]
				\node[location2CM] (lhalt) {$\lochalt$};
				\node[location, right=of lhalt] (l1) {$\lochalt^1$};
				\node[location, right=of l1] (lhaltprime) {$\lochaltprime$};

				\path (lhalt) edge node[above]{\begin{tabular}{c}$x,y := 0$\end{tabular}} (l1);
				\path (l1) edge[loop above] node[right]{\begin{tabular}{c}$x = a$ \\$x := 0$\end{tabular}} (l1);
				\path (l1) edge node[above]{\begin{tabular}{c}$x = 0 \land y \geq 1$\end{tabular}} (lhaltprime);

				\end{tikzpicture}
			}
			
			\caption{Last transition of the 2CM encoding with non-strict inequalities}
			\label{figure:gadget-2CM-bounded-closed-PTA-final}
			\end{subfigure}

        \caption{Rewriting gadgets for EF-emptiness for closed bounded PTAs}
    \end{figure}

    Clearly, if $a=0$, taking the self-loop on $\lochalt^1$ will not allow time to elapse; and then there will be no way to leave $\lochalt^1$ with $x = 0$
		and $y \geq 1$; hence $\lochaltprime$ is not reachable for $ a = 0$.
    In contrast, if $0 < a \leq 1$, since $a$ is a rational number, then by taking an appropriate number of times the self-loop on $\lochalt^1$, we will eventually have $x = 0$, just after the last loop,
		and $y \geq 1$; hence $\lochaltprime$ will eventually be reached.
	To summarize:
	\begin{itemize}
		\item if $a=0$, the machine is not correctly encoded, but there is no way to reach $\lochaltprime$;
		\item if $0 < a \leq 1$, the machine is correctly encoded, and from Theorem~\ref{theorem:EF-empty:undecidable} we know that $\lochalt$ is reachable iff the machine halts.
		Since $\lochaltprime$ is reachable from $\lochalt$, then $\lochaltprime$ is reachable iff the machine halts.
	\end{itemize}
	Hence there exists a parameter valuation such that $\lochaltprime$ is reachable iff the machine halts.
	Finally, note that all our gadgets use non-strict inequalities, and therefore the built PTA is a closed bounded PTA.
\end{proof}
\subsection{A decidability result for bounded IP-PTAs}\label{ss:decidability}

Our main positive result in this section is that the EF-emptiness problem is decidable for bounded IP-PTAs.

\begin{thm}\label{theorem:EF-emptiness:decidable:Didier}
	The EF-emptiness problem is decidable (and \textsc{PSpace}-complete) for bounded IP-PTAs.
\end{thm}
\begin{proof}
	We first need to recall two lemmas relating symbolic and concrete runs (proved in~\cite{HRSV02,ACEF09}).

    Given a concrete (respectively symbolic) run $(\locinit, \vec{0}) \longuefleche{\edge_0} (\loc_1, \clockval_1) \longuefleche {\edge_1} \cdots \longuefleche{\edge_{m-1}} (\loc_m, \clockval_m)$ (respectively  $(\loc_0, \C_0) \Fleche{\edge_0} (\loc_1, \C_1) \Fleche {\edge_1} \cdots \Fleche{\edge_{m-1}} (\loc_m, \C_m)$),
we define the corresponding discrete sequence
as
	$\loc_0  \Fleche{\edge_0} \loc_1 \Fleche {\edge_1} \cdots \Fleche{\edge_{m-1}} \loc_m $.
Two runs (concrete or symbolic) are said to be equivalent if their associated discrete sequences are equal.

	\begin{lem} \label{lemma:reachability-condition} %
		Let $\A$ be a PTA, and~$\pval$ be a parameter valuation.
		Let $\varrun_s= \symbstate_0 \Fleche{\edge_0} \symbstate_1\Fleche {\edge_1} \cdots \Fleche{\edge_{m-1}} \symbstate_m$
        be a finite symbolic run of~$\A$ with $\symbstate_m = (\loc, \C)$.
		Then, there exists a finite concrete run 
        $\varrun_c = \sinit \longuefleche{\edge_0} \state_1\longuefleche {\edge_1} \cdots \longuefleche{\edge_{m-1}} \state_m$ 
        in the TA $\valuate{\A}{\pval}$ with $\state_m = (\loc, \clockval)$ (for some~$\clockval$) iff $\pval \models \projectP{\C}$.%
	\end{lem}
	\begin{lem} \label{lemma:reachability-condition:<=} %
		Let $\A$ be a PTA, and~$\pval$ be a parameter valuation.
		Let $\varrun_c = \sinit \longuefleche{\edge_0} \state_1\longuefleche {\edge_1} \cdots \longuefleche{\edge_{m-1}} \state_m$ be a finite concrete run of the TA $\valuate{\A}{\pval}$, with $\state_m=(\loc, \clockval)$.
		Then there exists a finite symbolic run $\varrun_s= \symbstate_0 \Fleche{\edge_0} \symbstate_1\Fleche {\edge_1} \cdots \Fleche{\edge_{m-1}} \symbstate_m$ in~$\A$, with $\symbstate_m=(\loc, \C)$, for some~$\C$ such that $\pval \models \projectP{\C}$.
	\end{lem}

	Let $\A$ be a bounded IP-PTA.
	EF-emptiness is false for $\A$ iff there exists a valuation~$\pval$ such that a run of $\valuate{\A}{\pval}$ reaches a location in some predefined set~$\somelocs$.
	Assume there exists a valuation~$\pval$ such that a run of $\valuate{\A}{\pval}$ reaches~$\loc$, with $\loc \in \somelocs$.
	From Lemma~\ref{lemma:reachability-condition:<=}, there exists a symbolic run of~$\A$ reaching a symbolic state $(\loc, \C)$, for some~$\C$.
	Since $\A$ is an IP-PTA, $\C$ contains at least one integer point.
	Hence there exists an integer parameter valuation $\pval' \models \projectP{\C}$; so from Lemma~\ref{lemma:reachability-condition}, there exists a concrete run of $\valuate{\A}{\pval'}$ reaching~$\loc$.
	This gives that EF-emptiness is false for $\A$ iff there exists an integer valuation~$\pval'$ such that a run of $\valuate{\A}{\pval'}$ reaches a location in~$\somelocs$.
	
    As a consequence, deciding whether some valuation permits to reach~$\loc$ reduces to deciding whether some  \emph{integer} valuation permits to do so, which, for bounded PTAs, is \textsc{PSpace}-complete~\cite{JLR15}.
\end{proof}

Since bounded IP-PTAs are incomparable with L/U-PTAs (for which the EF-emptiness problem is known to be decidable), and since L/U-PTAs are the only non-trivial subclass of PTAs for which this problem is known to be decidable, then Theorem~\ref{theorem:EF-emptiness:decidable:Didier} strictly extends the subclass of PTAs for which this problem is decidable.

In practice,~\cite{JLR15} proposes efficient symbolic algorithms to synthesize all the integer parameter valuations that permit to reach some given location, and thus to solve EF-emptiness for IP-PTAs.
\subsection{Intractability of the synthesis}\label{ss:synthesis}

Although the EF-emptiness problem is decidable for L/U-PTAs~\cite{HRSV02}, the synthesis seems to pose practical problems: it was shown in~\cite{JLR15} that the solution to the EF-synthesis problem for L/U-automata, if it can be computed, cannot be represented using any formalism for which emptiness of the intersection with equality constraints is decidable.
In particular, this rules out the possibility of computing the solution set as a finite union of polyhedra.

We reuse the intuition of this result and extend it to closed bounded L/U-PTAs.

\begin{thm}\label{theorem:intractability}
	If it can be computed, the solution to the EF-synthesis problem for closed bounded L/U-automata cannot be represented using any formalism for which emptiness of the intersection with equality constraints is decidable.
\end{thm}
\begin{proof}
	We reuse the idea of~\cite{BlT09} used for proving that constrained emptiness for infinite runs acceptance properties is undecidable, and reused in~\cite[Theorem 2]{JLR15}.
	Suppose that the solution to the EF-synthesis problem for closed bounded L/U-PTAs can be represented using a formalism for which emptiness of the intersection with equality constraints is decidable.
	Assume a closed bounded PTA~$\A$; for each parameter $\param_i$ of $\A$ that is used both as an upper bound and a lower bound, replace its occurrences as upper bounds by a fresh parameter $\param_i^u$ and its occurrences as lower bounds by a fresh parameter $\param_i^l$.
	We therefore obtain a closed bounded L/U-PTA.
	Assume we can derive a solution to the EF-synthesis problem for this closed bounded L/U-PTA, and let $\K$ be that solution.
	Then, by hypothesis, we can decide whether $\K \land \bigwedge_{i} \param_i^l = \param_i^u$ is empty or not; hence, we can solve the EF-emptiness for~$\A$, which contradicts the undecidability of EF-emptiness for closed bounded PTAs (from Theorem~\ref{theorem:EF-empty:undecidable-closed-bounded}).
\end{proof}

\begin{cor}\label{intractability-synthesis-IPPTA}
	If it can be computed, the solution to the EF-synthesis problem for IP-PTAs cannot be represented using any formalism for which emptiness of the intersection with equality constraints is decidable.
\end{cor}
\begin{proof}
	From the fact that a closed bounded L/U-PTA is an IP-PTA (Proposition~\ref{proposition:bIPPTA-cbLUPTA}).
\end{proof}

\subsection{Membership}\label{ss:membership}

We address here the problem of deciding whether a particular PTA is an IP-PTA.
We first show that it cannot be decided in general whether a PTA is a (bounded) IP-PTA.

\begin{thm}[undecidability of membership]\label{theorem:IPPTA-membership}
	It is undecidable whether a PTA is an IP-PTA, even when bounded.
\end{thm}
\begin{proof}
Let us consider again the PTA $\calA(\calM)$ encoding  the 	%
2-counter machine $\calM$ proposed in our proof of Theorem~\ref{theorem:EF-empty:undecidable-closed-bounded}.
	We make the following modification to the final gadget (\cref{figure:gadget-2CM-bounded-closed-PTA-final}): we forbid $a=1$ in the final location $\lochaltprime$, by adding $y > a$ on the final transition to~$\lochaltprime$.
	
	The PTA $\calA(\calM)$ has only one parameter~$a$ and all the symbolic states of $\calA(\calM)$ contain the integer value $a=0$ except the states corresponding to location~$\lochaltprime$.
  Since all constraints are non-strict (except on the final transition to~$\lochaltprime$), all reachable symbolic states except those associated with $\lochaltprime$ contain an integer point from Lemma~\ref{lemma:Didier}.
    Conversely, since $\lochaltprime$ can only be reached provided $0 < a < 1$, then the reachable symbolic states associated with $\lochaltprime$ contain \emph{no} integer point.

    Then the PTA ${\calA}({\calM})$ reaches the location $\lochaltprime$ if and only if ${\calA}({\calM})$ is not an IP-PTA.
	As a consequence, this PTA is an IP-PTA iff the 2-counter machine does not halt.
		Finally, note that this PTA can be bounded by imposing
        $0 \leq a \leq 1$, without any change in the reasoning above.
\end{proof}

Nevertheless, Proposition~\ref{proposition:IPPTA-oLUPTA} provides a sufficient syntactic membership condition, since any closed L/U-PTA is an IP-PTA.
In addition, we now define another new non-trivial set of restrictions leading to IP-PTAs:

\begin{defi}[Reset-PTA]
    A \emph{reset-PTA} is a PTA where:
    \begin{itemize}
        \item all guards and invariants are conjunctions of constraints of the form $x \leq p + k$, $x\geq p+k$, $x\leq k$, or $x\geq k$, with $x$ a clock, $p$ a parameter, and $k$ an integer;
        \item and all clocks are reset to~0 on any transition with a guard or a source location invariant in which a parameter appears. 
    \end{itemize}
\end{defi}

This kind of restriction is somewhat reminiscent of those enforced by \emph{initialized} hybrid automata~\cite{henzinger-JCSS-98} to obtain decidability. We now prove that reset-PTAs are IP-PTAs, which in turn means that the EF-emptiness problem is decidable for bounded reset-PTAs.
It is worth noting that, to the best our knowledge, bounded reset-PTAs and L/U-PTAs are the only non-trivial sets of syntactic restrictions of PTAs making the reachability emptiness problem decidable.

\begin{thm}\label{theorem:reset<=IP}
  Any reset-PTA is an IP-PTA.
\end{thm}
\begin{proof}
	We prove by induction that the symbolic states generated by reset-PTAs are zones with only non-strict constraints over the set of variables defined by the union of clocks and parameters. To simplify the proof a bit we omit invariants but including them would raise no theoretical difficulty. We then additionally prove as part of the induction that 
    there is no inequality involving two variables $x$ and $y$ in which $x$ and $y$ would not be of the same type (clock or parameter).

	The property clearly holds for the initial symbolic state: parameters are unconstrained, all clocks are equal and their common value is greater than or equal to~$0$.

	Now suppose this holds for some symbolic state and consider the successor of that symbolic state by some transition.
    Recall from the $\Succ$ definition in \cref{sss:symbolic} that this successor is computed by the following operations: intersection with the guard of the transition, reset of the clocks designated in the reset set of the transition, and finally time elapsing.%

	Due to the restriction on constraints, all guards are themselves zones, and it is well-known that the intersection of two zones is again a zone. Similarly, the reset operation on some of the variables of a zone again leads to a zone.
    The time elapsing of a proper subset of the variables in a zone however is not a zone in general (the same situation arises, \eg{} in stopwatch automata~\cite{CL00}).
	We therefore need to examine more closely the zone on which the time elapsing operates. Two cases arise:
	\begin{enumerate}
		\item the guard did not involve any parameter. Then, with the induction hypothesis, we still do not have any constraint between clocks and parameters after the intersection with the guard. A fortiori, we do not have any after the resets either. The time elapsing operation can be carried out by introducing a fresh non-negative variable $t$, performing variable substitutions $x\leftarrow x+t$ for each clock $x$, and finally eliminating $t$. This elimination can be done with the Fourier-Motzkin procedure (see \eg{}~\cite{schrijver-book-86}) which produces all the constraints not involving $t$ plus those obtained by writing all the combinations of a minorant of $t$ less or equal to a majorant of $t$. After the variable substitutions, $t$ does not appear in constraints between parameters, nor in diagonal clock constraints ($x-y\leq k$ gives $(x+t) - (y+t) \leq k$, \ie{} again $x-y\leq k$). Since there is no constraint between clocks and parameters, $t$ only appears in rectangular constraints that become of the form $y+t\leq k_1$ or $x+t\geq k_2$.
		Through the elimination procedure, this gives constraints of the form $y - x \leq k_1 - k_2$, and the expected result holds.
		
		\item the guard involves some parameters.
			Then before the reset we do have constraints between clocks and parameters. But then, from the definition of reset-PTAs, all clocks are reset along this transition, so these constraints are removed (as part of the elimination of clock variables) and replaced by constraints restricting the reintroduced clock variables to zero.
			Then, after the reset we do not have constraints between clocks and parameters anymore and the previous reasoning is again valid.
	\end{enumerate}

  We conclude the proof by noting that in any non-empty zone with integer coefficients all vertices are integer (see the discussion in~\cite{JLR15}). And following the proof of Lemma~\ref{lemma:Didier}, since all variables are non-negative, there is at least one such vertex, which does belong to the zone because all constraints are non-strict.
    This zone therefore contains at least an integer point.
\end{proof}

Recall that the synthesis is intractable for bounded IP-PTAs (from Corollary~\ref{intractability-synthesis-IPPTA}) and for bounded L/U-PTAs.
In contrast, and although studying reset-PTAs in detail goes beyond the scope of this work, we showed in~\cite{ALR21} that exact synthesis can be computed for bounded reset-PTAs as defined here, even when resets are done to (rational-valued) parameters.
\subsection{Undecidability of EF-Universality}\label{ss:EF-universality}

We show below that, unlike L/U-PTAs, the EF-universality problem is undecidable for IP-PTAs even bounded.
This result differentiates the classes of (bounded) L/U-PTAs and bounded IP-PTAs, and
	helps to understand better the boundary between decidability and undecidability for subclasses of PTAs.

\begin{thm}\label{theorem:bip-EFu}
The EF-universality problem is undecidable for bounded IP-PTAs with at least 3 parametric clocks and 2 rational-valued parameters.
\end{thm}
\begin{proof}
	We start from the encoding in our proofs of \cref{theorem:EF-empty:undecidable,theorem:EF-empty:undecidable-closed-bounded}.
	The main idea is, for all valuations of the parameter $a$ that are not small enough to properly encode the counters (\ie{} for some value $c$ of a counter, $1-ac<0$), to allow the PTA to directly go to a \locerror{} location.
	In order for our encoding to be an IP-PTA (in particular the symbolic states with location~\locerror{}), we add a new parameter~$b$, the value of which can be typically in~$[0,1]$.
	
	We then reduce the problem of knowing whether the counters of the machine grow unbounded along its execution, which is undecidable~\cite{Minsky67}, to the universality of the set of parameters that allow the encoding PTA to reach \locerror{}.

	Let us summarize our construction before going into details.
	\begin{enumerate}
		\item We reuse the increment gadget from \cref{theorem:EF-empty:undecidable,theorem:EF-empty:undecidable-closed-bounded} (\cref{fig:inc_m2c}) and extend it with a new location~$\locerror$ (details will follow);
		\item We reuse the 0-test and decrement gadget from \cref{theorem:EF-empty:undecidable-closed-bounded} (\cref{figure:gadget-2CM-bounded-closed-PTA-decrement}) as it is;
		\item We do not use the final gadget from $\lochalt$ to~$\lochaltprime$ (\cref{figure:gadget-2CM-bounded-closed-PTA-final}), \ie{} the last location is~$\lochalt$. (Recall that our target location is \locerror{} anyway.)
	\end{enumerate}

	Let us now go into details.
	First, we add a fresh location~\locerror{} to our PTA encoding, and we add two transitions from $l_0$ (the initial location of the PTA) to the \locerror{} location:
	\begin{itemize}
		\item one with guard $x=0 \land  x=a$, that can only be taken when $a=0$ and serves to ``eliminate'' this special case that does not correctly encode the counters.
		\item and one with guard $0 \leq x < 1 \land  x=b$, that can only be taken when $b \in [0, 1)$.\footnote{%
			This case may not be necessary in the proof; however, it makes the explanation simpler, as we can now discard from our reasoning valuations such that $b \in [0, 1)$.
		}
	\end{itemize}
	So, whenever $a = 0$ and $b \in [0, 1)$, the system can eventually reach \locerror{}.
	Hence, in the following, we only need to focus on the case where $a \in (0, 1]$ and $b = 1$.

	Let us now change the increment gadget of \cref{fig:inc_m2c} (when decrementing, there is no upper bound constraint on~$a$).
	More specifically, remark that, when incrementing $C_1$, the constraint that implies $a\leq \frac{1}{c_1 + 1}$ comes from the last transition in the path going through~$l_{i2}$.
	In the other path, $c_2$ is already greater than or equal to $c_1 + 1$ and therefore $a$ is already small enough to properly encode $c_1 + 1$ since it is small enough to encode~$c_2$.
    Our modified increment gadget is given in \cref{fig:incu_m2c}.
    
    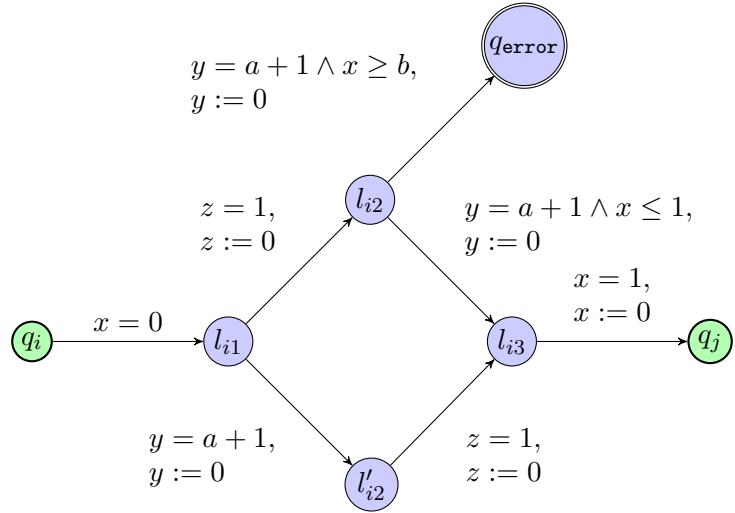
\begin{figure}
    \centering
\scalebox{\generalFigsScaleFactor}{
        \begin{tikzpicture}[PTA, node distance=2cm]
            \node[location2CM] (si) {$\cmspta_i$};
            \node[location, right=of si] (l1) {$l_{i1}$};
            \node[location, above right=of l1] (l2) {$l_{i2}$};
            \node[location, below right=of l1] (l2p) {$l'_{i2}$};
            \node[location, accepting, above right=of l2] (error) {\locerror{}};
            \node[location, below right=of l2] (l3) {$l_{i3}$};
            \node[location2CM, right=of l3] (sj) {$\cmspta_j$};
    
            \draw[->] (si) -- node {$x=0$} (l1);
            \draw[->] (l1) -- node {$\begin{array}{l}z=1,\\ z:=0\end{array}$} (l2);
            \draw[->] (l1) -- node[swap] {$\begin{array}{l}y=a+1,\\ y:=0\end{array}$} (l2p);
            \draw[->] (l2) -- node {$\begin{array}{l}y=a+1 \land x\leq 1,\\ y:=0\end{array}$} (l3);
            \draw[->] (l2) -- node {$\begin{array}{l}y=a+1 \land x \geq b,\\ y:=0\end{array}$} (error);
            \draw[->] (l2p) -- node[swap] {$\begin{array}{l}z=1,\\ z:=0\end{array}$} (l3);
            \draw[->] (l3) -- node {$\begin{array}{l}x=1,\\ x:=0\end{array}$} (sj);
        \end{tikzpicture}
        
        }
        \caption{EF-universality for bounded IP-PTAs: increment gadget}
        \label{fig:incu_m2c}
    \end{figure}

	In the transition from $l_{i2}$ to $l_{i3}$, if $a$ is not small enough then $x=a(c_1 + 1)$ will be greater than one and the final transition to $\cmspta_j$ cannot be taken.
	We therefore add a direct transition from $l_{i2}$ to $\locerror{}$ when $x \geq b$.
	Recall that we only care about the case where $b = 1$, hence this transition can be understood as $x \geq 1$; now the case where $x = 1$ is problematic, as the value of~$a$ is just small enough to encode the counters, and we still can reach \locerror{}, and the 2-counter machine is not properly encoded.
	However, since we are interested in universality, it suffices to take a valuation of~$a$ slightly smaller to properly encode the machine, as we explain more precisely below.

	We now prove that the counters of the machine grow unbounded along its execution iff for all values of $a$ and~$b$, the encoding PTA can reach \locerror{}.
	First recall that for $a=0$ or $b \in [0, 1)$, it is always possible to reach \locerror{} (from the initial state).
	When $a>0$ and $b = 1$, we have two cases:
	\begin{itemize}
		\item either the counters grow unbounded (say $C_1$ does), then whatever the value of $a>0$, at some point we have $ac_1 > 1$.
			More specifically, there is an increment of $C_1$ such that $ac_1 \leq 1$ and $a(c_1 + 1) > 1$, which also implies $a(c_1 + 1) \geq b$ (since $b = 1$).
			Then, when executing the corresponding increment gadget, \locerror{} can be reached from $l_{i2}$;
		\item or the counters stay bounded. Let $c$ be the maximal value of the counters.
            Recall that when entering $l_{i2}$ we have $x=ac_2$, $y= 1 - a c_1 + a c_2$ and $z=0$.
            Then, when $y=a+1$, we have $x=a(c_1+1)$. If $c_1+1=c$, then $x=ca$ is the largest value that $x$ can have in $l_{i2}$ when $y=a+1$.
		Observe that, due to the non-strict inequality $x \geq b$ in the guard from $l_{i2}$ to \locerror{}, one might still reach \locerror{} for a valuation of~$a$ such that $ca = 1$.
            Consequently, consider the parameter valuation $a =\frac{1}{c+1}$ and $b=1$. Then $ca<1=b$ and since $x$ is in $l_{i2}$ always at most equal to $ca$ when $y=a+1$, the guard to $\locerror{}$ is never true and the of valuations for which the automaton can reach \locerror{} is not universal.
	\end{itemize}

	It remains to show that the constructed PTA is an IP-PTA.
	With the exception of~$\locerror$, the result is clear: $a=0$ and $b = 0$ belongs to every reachable symbolic state, hence each symbolic state contains an integer parameter valuation, and hence from Lemma~\ref{lemma:Didier}, all symbolic states (except those with location~$\locerror$) contain at least one integer point.
    In addition, the two symbolic states taken by taking the two special transitions from the initial state to \locerror{} to handle $a = 0$ or $b \in [0, 1)$ also contain the integer point $x=y=a=b=0$.
	Now, let us consider the other symbolic states with location $\locerror$ (and reachable from some location $l_{i2}$ due to an increment).
	The projection onto the parameters of the associated constraint is $ 0  \leq b \leq 1 \land \frac{b}{i+1} \leq a \leq \frac{1}{i}$, with $i \in  \grandn$ denotes the current maximum valuation of the counter.
	Clearly, 
  $a = b = 0$ is a parametric integer point in this symbolic state; hence from Lemma~\ref{lemma:Didier} this symbolic state contains an integer point (in clocks and parameters dimensions).
	Hence this PTA is a (bounded) IP-PTA.
\end{proof}

\begin{cor}\label{corollary:EFu}
	The EF-universality problem is undecidable for IP-PTAs, for bounded PTAs, and for PTAs.
\end{cor}
\begin{proof}
	From Theorem~\ref{theorem:bip-EFu} and from the fact that a bounded IP-PTA is an IP-PTA, is a bounded PTA, and is a PTA.
\end{proof}

We now show that, in contrast to \emph{bounded} IP-PTAs, the EF-emptiness problem is undecidable for (unbounded) IP-PTAs.
This result emphasizes the fact that the boundedness of IP-PTAs is crucial to ensure decidability.

\begin{thm}\label{theorem:IPPTA-EF-undecidable}
	The EF-emptiness problem is undecidable for IP-PTAs.
\end{thm}
\begin{proof}
	The proof of the undecidability of the EF-emptiness problem for general PTAs in~\cite{AHV93} can be interpreted over integer parameter valuations.
	Any symbolic state contains at least one integer parameter valuation (the one that is large enough to correctly encode the value of the two counters), as well as all larger parameter valuations.
	Furthermore, since the proof only uses non-strict inequalities (in fact only equalities), from Lemma~\ref{lemma:Didier}, all symbolic states contain at least one integer point.
	Hence the PTA used in~\cite{AHV93} to encode the 2-counter machine is an IP-PTA.
\end{proof}

Finally, we show below (without surprise) that the EF-emptiness problem (shown to be decidable for L/U-PTAs~\cite{HRSV02}) and the EF-universality problem (shown to be decidable for integer-valued L/U-PTAs~\cite{BlT09}) are also decidable for bounded L/U-PTAs.
\begin{thm}\label{theorem:bL/UPTA-EF-decidable}
	The EF-emptiness and EF-universality problems are decidable for bounded L/U-PTAs.
\end{thm}
\begin{proof}
	In~\cite{HRSV02,BlT09}, it is shown that decreasing a lower-bound parameter $\param_i^-$ or increasing an upper-bound parameter $\param_j^+$ in an L/U-PTA~$\A$ can only add behaviors.
	Hence, deciding EF-emptiness can be done by testing the reachability of the location in the TA obtained from~$\A$ by instantiating all $\param_i^-$s with~$0$ and all $\param_j^+$s with~$\infty$.
	(Recall that testing the reachability of a location in a TA is decidable~\cite{AD94}.)
    For a bounded L/U-PTA, this can be done in a similar manner, by testing the reachability of the location in the TA obtained from~$\A$ by instantiating all $\param_i^-$s with their minimal value and all $\param_j^+$s with their maximal value in the (closed) bounded parameter domain.
	
	EF-universality can be solved similarly, except that $\param_i^-$s are replaced with their upper bound and $\param_j^+$s are replaced with their lower bound.
\end{proof}
\section{Liveness properties}\label{section:liveness}

In this section, we consider liveness properties.
More specifically, we consider the problems of the emptiness of the parameter valuations set
	for which all concrete runs eventually reach a given location (\cref{ss:AF}),
	for which there exists an infinite concrete run (\cref{section:EC-emptiness}),
	for which there exists a deadlock (\cref{section:ED-emptiness}),
	and for which a concrete run remains in a given set of locations (\cref{section:EG-emptiness}).

\subsection{Undecidability of AF-emptiness}\label{ss:AF}

It is known that AF-emptiness is undecidable for L/U-PTAs~\cite{JLR15}; reusing the encoding of the 2-counter machine proposed in our proof of Theorem~\ref{theorem:EF-empty:undecidable}, we now show that this result holds even for bounded L/U-PTAs.

\begin{thm}\label{theorem:bounded-LUPTA-AF-undecidable}
    The AF-emptiness problem is undecidable for (closed) bounded L/U-PTAs with at least 3 clocks and 2 rational-valued parameters.
\end{thm}
\begin{proof}
Let us consider the PTA $\calA(\calM)$ encoding the 2-counter machine $\calM$ proposed in our proof of 
Theorem~\ref{theorem:EF-empty:undecidable-closed-bounded}.
	The PTA $\calA(\calM)$ has only one parameter~$a$ which is used both as an upper bound and a lower bound.
	In our modified encoding, we replace~$a$ with two fresh parameters $a^-$ and~$a^+$.
	Then, in both the increment gadget (\cref{fig:inc_m2c}) and the 0-test and decrement gadget (\cref{figure:gadget-2CM-bounded-closed-PTA-decrement}), we replace the guard $y=1+a$ by a guard $1 + a^-\leq y$ and an invariant $y \leq 1 + a^+ $; we do the same for the guard $z=1+a$.
	We give the modified increment gadget in \cref{fig:inc-LU_m2c}. %

	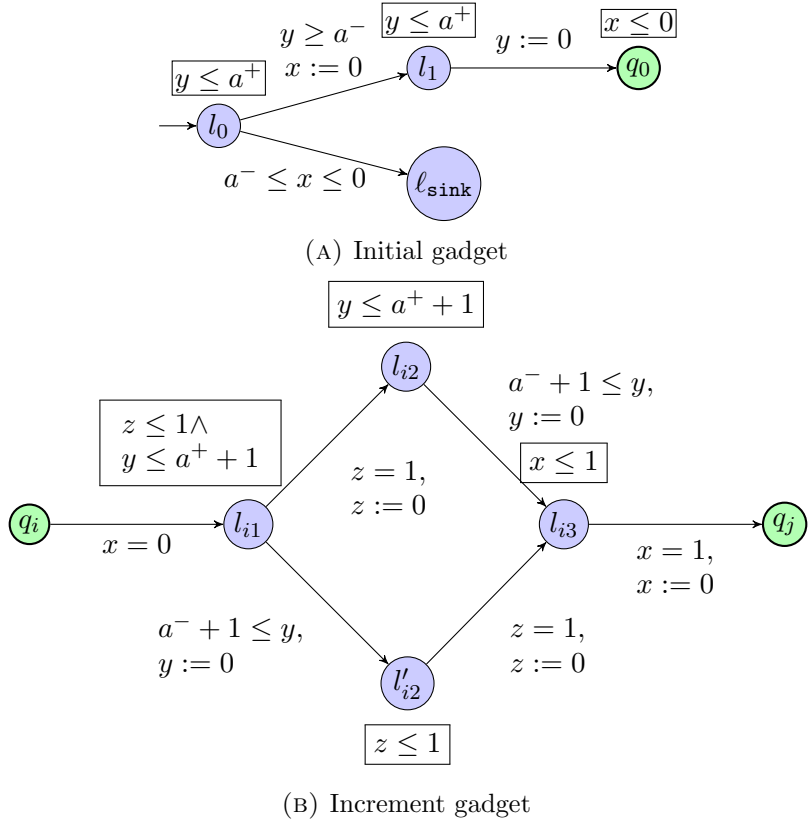
\begin{figure}

		\begin{subfigure}[b]{\textwidth}
			\centering
			\scalebox{\generalFigsScaleFactor}{
			\begin{tikzpicture}[node distance=2.2cm, PTA]
			\node[location, initial] (i0) {$l_0$};
			\node [invariant,above] at (i0.north) {$y\leq a^+$};
	
			\node[location, right=of i0, yshift=+2em] (i1) {$l_1$};
			\node [invariant,above] at (i1.north) {$y\leq a^+$};

			\node[location2CM, right=of i1] (l0) {$\cmspta_0$};
			\node [invariant,above] at (l0.north) {$x\leq 0$};
	
			\node[location, right=of i0, yshift=-2em] (caca) {$\locsink$};

			\path (i0) edge node[above]{\begin{tabular}{c}$y\geq a^-$\\ $x := 0$\end{tabular}} (i1);
			\path (i1) edge node[above]{\begin{tabular}{c}$y := 0$\end{tabular}} (l0);
	
			\path (i0) edge node[below, xshift=-1em]{\begin{tabular}{c}$ a^- \leq x \leq 0$\end{tabular}} (caca);
			\end{tikzpicture}
		}
		
			\caption{Initial gadget}
			\label{figure:AF-LUPTA:eqparam}
		\end{subfigure}
		\begin{subfigure}[b]{\textwidth}
		\centering
		\scalebox{\generalFigsScaleFactor}{
			\begin{tikzpicture}[PTA, node distance=2.3cm]
				\node[location2CM] (si) {$\cmspta_i$};
				\node[location, right=of si] (l1) {$l_{i1}$};
				\node[location, above right=of l1] (l2) {$l_{i2}$};
				\node[location, below right=of l1] (l2p) {$l'_{i2}$};
				\node[location, below right=of l2] (l3) {$l_{i3}$};
				\node[location2CM, right=of l3] (sj) {$\cmspta_j$};
		
			\node[above=+.1em of l1, xshift=-2em] {\fbox{$\begin{array}{l}z \leq 1 \wedge\\ y\leq a^+ + 1\end{array}$}};
			\node[above=0.05cm of l2] {\fbox{$y\leq a^+ + 1$}};
			\node[above =+0.08cm of l3] {\fbox{$x \leq 1$}};  
			\node[below=0.05cm of l2p] {\fbox{$z\leq 1$}};  
		
				\draw[->] (si) -- node[swap] {$x=0$} (l1);
				\draw[->] (l1) -- node[swap] {$\begin{array}{l}z=1,\\ z:=0\end{array}$} (l2);
				\draw[->] (l1) -- node[swap] {$\begin{array}{l}a^- + 1\leq y ,\\ y:=0\end{array}$} (l2p);
				\draw[->] (l2) -- node {$\begin{array}{l}a^- + 1\leq y ,\\ y:=0\end{array}$} (l3);
				\draw[->] (l2p) -- node[swap] {$\begin{array}{l}z=1,\\ z:=0\end{array}$} (l3);
				\draw[->] (l3) -- node[swap] {$\begin{array}{l}x=1,\\ x:=0\end{array}$} (sj);
			\end{tikzpicture}
		}
		
			\caption{Increment gadget}
			\label{fig:inc-LU_m2c}
		\end{subfigure}

    \caption{AF-emptiness for bounded L/U-PTAs}
    \end{figure}

	We initialize the parameters $a^-$ and $a^+$ with the gadget in \cref{figure:AF-LUPTA:eqparam} (adapted from~\cite{JLR15}) leading to the location~$\cmspta_0$.
	Clearly, starting from~$l_0$, we have ${\sf AF} (\cmspta_0)$ if and 
    only if $a^-=a^+>0$, because
	\begin{enumerate}
		\item if $a^-=0$ then it is possible to reach \locsink{} and therefore we do not have ${\sf AF} (\cmspta_0)$, and
		\item any run that reaches $l_1$ before $y$ is equal to $a^+$ can be extended by delaying a non-null amount of time into a run that will be blocked by the invariant of~$\cmspta_0$.
	\end{enumerate}
	So all runs should
	enter $l_1$ with $y=a^+$, which is the case if and only if $a^-=a^+$.
	We therefore obtain an L/U-automaton with $a^-=a^+$ and $a^+>0$.%

	Let us now go back to the increment gadget of \cref{fig:inc-LU_m2c}.
  As in the proof of Theorem~\ref{theorem:EF-empty:undecidable-closed-bounded}, when $a^-=a^+>0$, exactly one path can be taken depending on the respective order of $c_1 + 1$ and~$c_2$.
	The invariants allow to avoid infinite delays in locations and since $a^-=a^+$, no run can be blocked inside the gadget.
	The same reasoning can be made for the decrement and zero-testing gadgets.
	Moreover we can bound the PTA by $a^-, a^+ \in [0, 1]$ without loss of behavior.

    Hence we reduce the halting problem of 2-counter machine to the AF-emptiness problem for bounded L/U-PTAs: the machine halts iff there exists a value of $a^-=a^+>0$, such that the location $\lochalt{}$ is unavoidable in our bounded L/U-automaton.
\end{proof}
\begin{cor}\label{corollary:open-bounded-LUPTA-AF-undecidable}
	The AF-emptiness problem is undecidable for (closed) bounded L/U-PTAs with at least 3 clocks and 2 rational-valued parameters.
\end{cor}
\begin{proof}
	The proof of Theorem~\ref{theorem:bounded-LUPTA-AF-undecidable} works the same if we bound the PTA by setting $a^-, a^+ \in (0, 1]$.
\end{proof}
\newcommand{\enonceCorollaryAFundec}{%
	The AF-emptiness problem is undecidable for bounded IP-PTAs, for IP-PTAs and for bounded PTAs.
}
\begin{cor}\label{corollary:AF-undecidable}
	\enonceCorollaryAFundec{}
\end{cor}
\begin{proof}
	The AF-emptiness problem is undecidable for bounded L/U-PTAs (Theorem~\ref{theorem:bounded-LUPTA-AF-undecidable}), which immediately gives the undecidability for bounded PTAs.
	
  Furthermore, the PTA used in the proof of Theorem~\ref{theorem:bounded-LUPTA-AF-undecidable} only uses non-strict inequalities%
	; furthermore,
    $a^- = 0$ and $a^+ = 1$ is a parameter valuation solution of any symbolic state.
    Hence, from Lemma~\ref{lemma:Didier}, this PTA is a bounded IP-PTA, which gives the result for bounded IP-PTAs.
	As a consequence, the result also holds for general IP-PTAs.
\end{proof}
\subsection{Cycle-Existence-Emptiness}\label{section:EC-emptiness}
\begin{thm}\label{theorem:ECemptiness:cbLU}
  The cycle-existence-emptiness problem is decidable for closed bounded L/U-PTAs.
\end{thm}
\begin{proof}
	Recall that, thanks to the monotonicity property of L/U-PTAs (recalled in Lemma~\ref{lemma:HRSV02:prop4.2}), any run possible for a valuation~$\pval$ of the parameters is also possible for any valuation of the parameters for which the upper-bound (resp.\ lower-bound) parameters are larger (resp.\ smaller) than or equal to that of~$\pval$.
	
	Let $\bounded{\A}{\bounds}$ be a closed bounded L/U-PTA.
	Let $\pvalinfsup$ be the valuation such that, for each lower-bound parameter~$\param^-$, $\pvalinfsup(\param^-) = \boundinfin{\param^-}{\bounds}$ and, for each upper-bound parameter~$\param^+$, $\pvalinfsup(\param^+) = \boundsupin{\param^+}{\bounds}$.
	
	\begin{enumerate}
		\item If $\valuate{\A}{\pvalinfsup}$ contains an infinite run (which can be checked in \textsc{PSpace}~\cite{AD94}\ForLongVersion{, and can be performed efficiently in practice using, \eg{} the zone graph~\cite{BY03}}), then since $\bounded{\A}{\bounds}$ is closed, $\pvalinfsup$ belongs to~$\bounds$, and hence the set of \ForLongVersion{parameter }valuations that yield an infinite run is not empty.
		\item On the contrary, if $\valuate{\A}{\pvalinfsup}$ contains no infinite run, then from the monotonicity property of L/U-PTAs (Lemma~\ref{lemma:HRSV02:prop4.2}), no other valuation in~$\bounds$ gives a TA with an infinite run, as such a TA could only contain less runs.
	Hence the set of parameter valuations that yield an infinite run is empty.\qedhere
	\end{enumerate}

\end{proof}
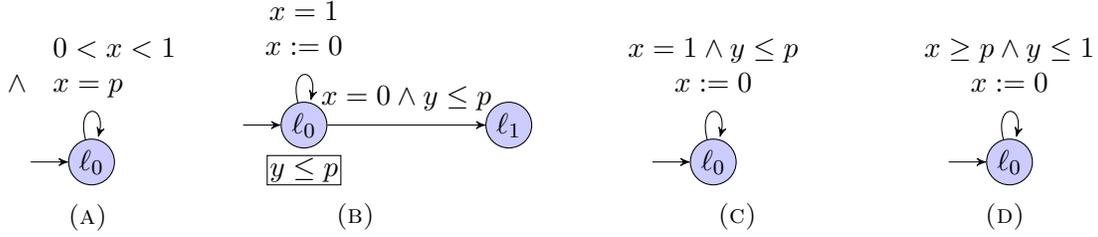
\begin{figure*}
	\begin{subfigure}[b]{.2\textwidth}
	\scalebox{\generalFigsScaleFactor}{
		\begin{tikzpicture}[PTA, node distance=2cm]
		\node[location, initial] (l0) {$\loc_0$};

		\path (l0) edge[loop above] node[above]{\begin{tabular}{l l} & $ 0 < x < 1 $\\$\land$ & $ x = \param$\end{tabular}} (l0);
		\end{tikzpicture}
	}

		\caption{}
		\label{figure:PTA:rational-but-not-integer}
	\end{subfigure}
	\begin{subfigure}[b]{.25\textwidth}
	\centering
	\scalebox{\generalFigsScaleFactor}{
		\begin{tikzpicture}[PTA, node distance=2.1cm]
		\node[location, initial] (l0) {$\loc_0$};
		\node [invariant,below,yshift=-.5em] at (l0.south) {$y \leq \param$};

		\node[location, right=of l0] (l1) {$\loc_1$};

		\path (l0) edge[loop above] node[above]{\begin{tabular}{c}$x = 1$ \\$x := 0$\end{tabular}} (l0);
		\path (l0) edge node[above]{\begin{tabular}{c}$x = 0 \land y \leq \param$\end{tabular}} (l1);
		\end{tikzpicture}
	}

		\caption{}
		\label{figure:example:infinite-run}
	\end{subfigure}
	\hfill
	\begin{subfigure}[b]{.25\textwidth}
	\scalebox{\generalFigsScaleFactor}{
		\begin{tikzpicture}[PTA, node distance=2cm]
		\node[location, initial] (l0) {$\loc_0$};

		\path (l0) edge[loop above] node[above]{\begin{tabular}{c}$x=1 \wedge y \leq \param$\\ $x:=0$\end{tabular}} (l0);
		\end{tikzpicture}
	}

		\caption{}
		\label{figure:counterexampleEG:LU-unbounded}
	\end{subfigure}
	\begin{subfigure}[b]{.20\textwidth}
	\scalebox{\generalFigsScaleFactor}{
		\begin{tikzpicture}[PTA, node distance=2cm]
		\node[location, initial] (l0) {$\loc_0$};

		\path (l0) edge[loop above] node[above]{\begin{tabular}{c}$x\geq\param\wedge y \leq 1$\\ $x:=0$\end{tabular}} (l0);
		\end{tikzpicture}
	}

		\caption{}
		\label{figure:counterexampleEG:LU-open}
	\end{subfigure}

	\caption{Examples of PTA (a) and L/U-PTAs (b--d)}
\end{figure*}

The above result cannot be used as such for non-bounded L/U-PTAs as a cycle that exists for an infinite parameter valuation may not exist for any finite parameter valuation: consider the L/U-PTA in \cref{figure:example:infinite-run}.
This L/U-PTA has an infinite run for $\param = \infty$, but for any parameter valuation (\ie{} different from~$\infty$), the number of self-loops in~$\loc_0$ is bounded by~$\param$, and hence finite.
However, extending to rational-valued parameters a result from~\cite{BlT09}, we can still prove decidability.

\begin{lem}\label{lemma:BlT09:Theorem8}
Given an L/U-PTA $\A$ and a subset of its locations~$\somelocs$, the problem of the existence of at least one parameter valuation~$\pval$ such that $\pval(\A)$ has a run passing infinitely often through~$\somelocs$ is \textsc{PSpace}-complete.
\end{lem}
\begin{proof}
	Let us prove that there exists a rational-valued valuation satisfying the property iff there exists an integer-valued valuation doing so.
	
	\begin{itemize}
		\item[$\Leftarrow$] Considering an integer valuation is also a rational-valued valuation, the result trivially holds.
		\item[$\Rightarrow$]
			Assume there exists a rational-valued parameter valuation~$\pval$ for which $\valuate{\A}{\pval}$ contains an infinite run passing infinitely often through locations of~$\somelocs$.
		Let $\pval'$ be the integer parameter valuation obtained from~$\pval$ as follows:
		\[\pval'(\param) = 
			\begin{cases}
			\pval(\param) & \text{if }\pval(\param) \in \grandn \\
			\ceil{\pval(\param)} & \text{if }\param\text{ is an upper-bound parameter}\\
			\floor{\pval(\param)} & \text{if }\param\text{ is a lower-bound parameter}\\
			\end{cases}
		\]
		
		From the monotonicity property of L/U-PTAs (Lemma~\ref{lemma:HRSV02:prop4.2}), if $\valuate{\A}{\pval}$ yields an infinite run passing infinitely often through locations of~$\somelocs$, then $\valuate{\A}{\pval'}$ does too.
	\end{itemize}
	Observe that this is not true for general PTAs:
    in \cref{figure:PTA:rational-but-not-integer}, there is an infinite run passing infinitely often through~$\loc_0$ iff $0 < \param < 1$; therefore, there exist rational-valued valuations satisfying the property, but no integer-valued valuation.

	Now, in~\cite[Theorem~8]{BlT09}, it is proved that the problem of the emptiness of the set of integer parameter valuations for which there exists an infinite run passing infinitely often through~$\somelocs$ is \textsc{PSpace}-complete.
	This concludes the proof.
\end{proof}
\begin{thm}\label{theorem:ECemptiness:LU}
  The cycle-existence-emptiness problem is \ForLongVersion{decidable and }\textsc{PSpace}-complete for L/U-PTAs.
\end{thm}
\begin{proof}
	Let $\A$ be an L/U-PTA.
	The set of parameter valuations for which $\A$ has an infinite run is empty iff the set of parameter valuations for which $\A$ has an infinite run passing infinitely often through~$\Loc$ (where $\Loc$ denotes all locations of~$\A$) is empty.
	Hence we can directly apply our intermediate Lemma~\ref{lemma:BlT09:Theorem8} to conclude that this problem is decidable and \textsc{PSpace}-complete.
\end{proof}

Without surprise\ForLongVersion{ (with the rule of thumb that any non-trivial problem for PTAs is undecidable)}, this problem becomes undecidable for general PTAs, even when bounded.
\begin{thm}\label{theorem:ECemptiness:PTA}
  The cycle-existence-emptiness problem is undecidable for (bounded) PTAs with at least 3~clocks and 1~parameter.
\end{thm}
\begin{proof}
	We reduce from the boundedness problem of a 2-counter machine, which is undecidable~\cite{Minsky67}.

    We start from the construction in the proof of \cref{theorem:EF-empty:undecidable} and add a self-loop (with no guard) on the location \lochalt{} (encoding the machine state \cmshalt{}), ensuring that whenever \lochalt{} is reachable then there exists an infinite run in the PTA.

    Then by examining exactly the same cases as before we see that the value of the counters remains bounded iff there exists a parameter valuation~$\pval$ such that $\valuate{\A}{\pval}$ yields an infinite run.
    Indeed the only two cases for which we do not have an infinite run correspond to
    \begin{enumerate}
    	\item $a=0$ (blocked by the initial gadget of \cref{figure:EFemptiness:initial}), or 
    	\item $a>0$ and the counters of the machine are unbounded.
    \end{enumerate}
    In the latter case, the PTA will block in the increment gadget when taking the transition from $l_{i2}$ to~$l_{i3}$.
\end{proof}

Finally note that the EC-emptiness problem for the class of open bounded L/U-PTAs (that does not fit in \cref{theorem:ECemptiness:cbLU,theorem:ECemptiness:LU}) remains an open problem.
We conjecture that this is decidable using techniques derived from the robustness results of~\cite{Sankur11} but the adaptation appears to require\ForLongVersion{ rather lengthy developments, with} techniques quite different from those presented here, and is thus left to future work.

\subsection{Deadlock-Existence-Emptiness}\label{section:ED-emptiness}
\begin{thm}\label{theorem:EDemptiness:cbLU}
  The deadlock-existence-emptiness problem is undecidable for closed bounded L/U-PTAs, with at least 3~clocks and 2~parameters.
\end{thm}
\newcommand{\preuveEDemptinesscbLU}{
\begin{proof}
	We will use a reduction from the halting problem of a 2-counter machine.
    Let us consider the encoding used in the proof of \cref{theorem:ECemptiness:PTA}, that we transform into an L/U-PTA by replacing any comparison of a clock with~$a$ (say $x = a$) into $x \leq a^+ \land x \geq a^-$, where $a^-$ (resp.\ $a^+$) is a lower-bound (resp.\ upper-bound) parameter. The crux of the proof is in the original enforcement of constraints in the encoding (in particular with location $\lochaltprime$) such that the deadlock property ensures that $a^- = a^+$.

	We give the modified increment gadget in \cref{figure:EDemptiness-bLUPTA:increment} (the decrement gadget is modified in a similar fashion).
	We replace the initial gadget (\cref{figure:EFemptiness:initial}) with the new one in \cref{figure:EDemptiness-bLUPTA:initial}.
	Before initializing the values of the counters, this gadget first ensures that $a^- \leq a^+$.

	\begin{figure}
		\begin{subfigure}[b]{.45\textwidth}
		\centering
		\scalebox{\generalFigsScaleFactor}{
			\begin{tikzpicture}[PTA, node distance=3cm]
			\node[location, initial] (i0) {$l_0$};

			\node[location, right=of i0] (i1) {$l_1$};

			\node[location2CM, right=of i1] (l0) {$\cmspta_0$};

			\path (i0) edge node[above]{\begin{tabular}{c}$a^- \leq x \leq a^+$\\$x,y,z := 0$\end{tabular}} (i1);
			\path (i1) edge node[above]{\begin{tabular}{c}$x = 1$\\$x := 0$\end{tabular}} (l0);

			\end{tikzpicture}
		}
        \caption{Initial gadget}
        \label{figure:EDemptiness-bLUPTA:initial}
		\end{subfigure}
		\hfill{}
		\begin{subfigure}[b]{.45\textwidth}
		\centering
		\scalebox{\generalFigsScaleFactor}{
			\begin{tikzpicture}[PTA, node distance=2cm]
			\node[location2CM] (lochalt) {\lochalt{}};

			\node[location, right=of lochalt] (lochaltprime) {\lochaltprime{}};

			\path (lochalt) edge[bend left] node[above]{$a^- \leq x < a^+$} (lochaltprime);
			\path (lochalt) edge[bend right] node[below]{$a^- \leq x \land x = 0$} (lochaltprime);
			\path (lochaltprime) edge[loop right] (lochaltprime);
	
			\end{tikzpicture}
		}
		
			\caption{Final gadget}
			\label{figure:EDemptiness-bLUPTA:final}
		\end{subfigure}
		\begin{subfigure}[b]{\textwidth}
	\centering
	
	\scalebox{\generalFigsScaleFactor}{
		\begin{tikzpicture}[PTA, auto, node distance=2.5cm]
			\node[location2CM] (si) {$\cmspta_i$};
			\node[location, right=of si] (l1) {$l_{i1}$};
			\node[location, above right=of l1] (l2) {$l_{i2}$};
			\node[location, below right=of l1] (l2p) {$l'_{i2}$};
			\node[location, below right=of l2] (l3) {$l_{i3}$};
			\node[location2CM, right=of l3] (sj) {$\cmspta_j$};

			\draw[->] (si) -- node {$x=0$} (l1);
			\draw[->] (l1) -- node {$\begin{array}{l}z=1\\ z:=0\end{array}$} (l2);
			\draw[->] (l1) -- node[swap] {$\begin{array}{l}a^- + 1 \leq y \leq a^+ + 1\\ y:=0\end{array}$} (l2p);
			\draw[->] (l2) -- node {$\begin{array}{l}a^- + 1 \leq y \leq a^+ + 1\\ y:=0\end{array}$} (l3);
			\draw[->] (l2p) -- node[swap] {$\begin{array}{l}z=1\\ z:=0\end{array}$} (l3);
			\draw[->] (l3) -- node {$\begin{array}{l}x=1\\ x:=0\end{array}$} (sj);
		\end{tikzpicture}
		}
	
		\caption{Increment gadget}
		\label{figure:EDemptiness-bLUPTA:increment}
		\end{subfigure}

	\caption{ED-emptiness for bounded L/U-PTAs}
    \end{figure}
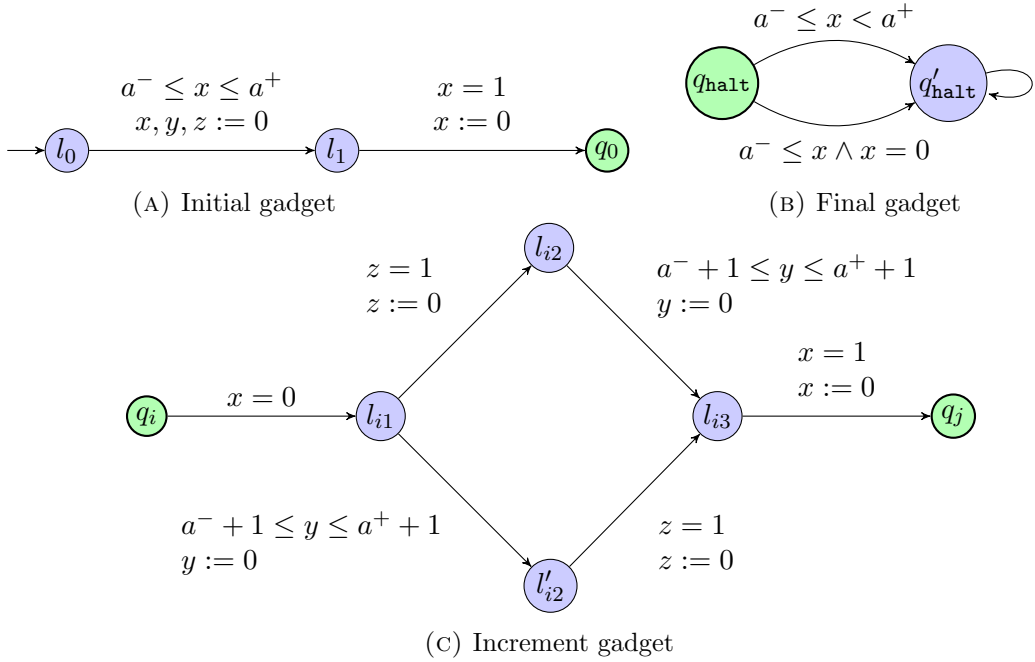

	We also add a new location~$\lochaltprime$ reachable from $\lochalt$ as shown in the final gadget in \cref{figure:EDemptiness-bLUPTA:final}.
    Finally, we add an unguarded transition (\ie{} a transition the guard of which is true) from any location of the encoding (including that of the initial gadget, but excluding~$\lochalt$) to location~$\lochaltprime$.
    That is, it is always possible to reach $\lochaltprime$ from any location without condition, except from~\lochalt{}. From that particular location, $\lochaltprime$ is reachable if and only if $a^- < a^+$ or $a^- = 0$.
    
	We assume the following bounds for the parameters: $a^-, a^+ \in [0,1]$.
	
	Let us show that there exists a parameter valuation for which the system contains at least one deadlock iff the 2-counter machine halts, which is undecidable~\cite{Minsky67}.
	Let us reason by cases on the valuations of $a^-$ and~$a^+$.
	\begin{enumerate}
		\item If $a^- > a^+$, the initial gadget cannot be passed, but thanks to the unguarded transitions to~$\lochaltprime$, all runs eventually end in $\lochaltprime$, from which the absence of deadlock is guaranteed by the unguarded self-loop.
		
		\item If $a^- < a^+$, the machine may not be properly simulated because some transitions do not occur at the right time and some run could reach \lochalt{} while the machine does not halt.
		Let us consider a run in the TA obtained with such a parameter valuation.
		\begin{enumerate}
			\item either this run is infinite and remains in the machine (\eg{} it loops infinitely through the increment, decrement and 0-test gadgets of our encoding). Then there is no deadlock.
            \item or this run would block in a gadget%
				; in that case, thanks to the unguarded transitions to~$\lochaltprime$, this run can go to~\lochaltprime{}, from which it is deadlock-free.
			\item or this run reaches \lochalt{} (recall that the value of~$x$ is necessarily~0 when entering~\lochalt{}); from there, thanks to the upper transition in \cref{figure:EDemptiness-bLUPTA:final}, it can reach~\lochaltprime{}, from which it is again deadlock-free.
		\end{enumerate}
		
		\item If $a^- = a^+ = 0$, the machine may again not be properly simulated: again we could reach \lochalt{} while the machine does not halt.
		The situation is similar to the previous case ($a^- < a^+$) except that in~\lochalt{} a run has to take the lower transition in \cref{figure:EDemptiness-bLUPTA:final} to reach~\lochaltprime{}, from which it is again deadlock-free.
		
		\item If $a^- = a^+ > 0$:
		\begin{enumerate}
			\item Either the machine does not halt:
			\begin{enumerate}
				\item …and the counters remain bounded: for some parameter valuations small enough to encode the value of the counters (typically $a^- = a^+ \leq \frac{1}{c}$, where $c$ is the maximum value of both $C_1$ and $C_2$) then the PTA correctly simulates the infinite execution of the machine, and the system is deadlock-free.
				(Note that such valuations can also lead to~\lochaltprime{} anytime, but this is harmless since this location guarantees the absence of deadlocks.)
				For other valuations, at some point we have $a^- c_1 > 1$; more specifically, there is an increment of $C_1$ such that $a^- c_1 \leq 1$ and $a^-(c_1 + 1) > 1$.
				Hence, the run cannot continue in the encoding, but can reach \lochaltprime{}, from where the run is non-blocking.
				\item …and the counters are unbounded. Then whatever the value of $a^->0$, at some point we have $a^-c_1 > 1$.
				Then, when executing the corresponding increment gadget, \lochaltprime{} can be reached from $l_{i2}$, from where the run is non-blocking.
			\end{enumerate}
			Hence if the machine does not halt, the system is deadlock-free for all parameter valuations.
			
			\item Or the machine halts.
				In this case, if $c$ is the maximum value of both $C_1$ and $C_2$ over the (necessarily finite) halting execution of the machine, and if $c>0$, then for valuations such that $a^- = a^+ \leq \frac{1}{c}$, then there exists one run that correctly simulates the machine (beside plenty of runs that will go to \lochaltprime{} due to the unguarded transitions from all locations except~$\lochalt$); this run that correctly simulates the machine eventually reaches \lochalt{}.
				From \lochalt{}, for such valuations, the system is deadlocked: indeed, the transitions from $\lochalt$ to $\lochaltprime$ can only be taken if $a^- < a^+$ or $a^- = 0$.
				And there is no unguarded transition from~$\lochalt$ to~$\lochaltprime$, which is crucial for the correctness of our encoding. %
				The set of such valuations for which there exists a run that correctly simulates the machine is certainly non-empty: $a^- = a^+ =\frac{1}{c}$ belongs to it (if $c=0$ then we choose, \eg{} $a^- = a^+ = \frac{1}{2}$).
			Hence, if the 2-counter machine halts, there exist parameter valuations for which a run has no discrete successor, and hence the system is not deadlock-free.
		\end{enumerate}

	\end{enumerate}

	Hence the 2-counter machine halts iff the set of valuations for which the automaton has at least one deadlock is not empty.
\end{proof}
}
	\preuveEDemptinesscbLU{}

\newcommand{\corollaryEDemptinessLUPTA}{The deadlock-existence-emptiness problem is undecidable for open bounded L/U-PTAs, for L/U-PTAs, for bounded PTAs and for PTAs, with at least 3~clocks and 2~parameters.}
\begin{cor}\label{corollary:EDemptiness:LU-PTA}
	\corollaryEDemptinessLUPTA{}
\end{cor}
\begin{proof}
    Let us consider each formalism:
	\begin{description}
		\item[\textbf{open bounded L/U-PTAs}] In the above construction, we can assume, \eg{} $a^- \in (0,1]$, which does not impact the proof.
		\item[\textbf{L/U-PTAs}] The bounds on the parameters are not required in the above construction: for valuations larger than~1 (that necessarily do not simulate correctly the machine), a gadget may block, therefore leading to~\lochaltprime{}, from which the system is deadlock-free, hence without impacting the spirit of the proof.
		\item[\textbf{bounded PTAs}] From the fact that a bounded L/U-PTA is a bounded PTA.
		\item[\textbf{PTAs}] From the fact that an L/U-PTA is a PTA.
	\end{description}
	Observe that the number of parameters can be reduced to~1 for (possibly bounded) PTAs by merging $a^-$ and~$a^+$ into a single parameter~$a$.
\end{proof}
\subsection{EG-Emptiness}\label{section:EG-emptiness}

We finally prove in the following that the EG-emptiness problem is decidable for closed bounded L/U-PTAs (\cref{theorem:EGemptiness:cbLU}), and that lifting either closedness or boundedness leads to undecidability (\cref{theorem:EGemptiness:obLU,theorem:EGemptiness:LU}).

\begin{thm}\label{theorem:EGemptiness:cbLU}
  The EG-emptiness problem is decidable for closed bounded L/U-PTAs.
\end{thm}
\newcommand{\preuveEGemptinesscbLU}{

We will use Lemma~\ref{lemma:HRSV02:prop4.2} to deal with infinite paths but it is of no use for deadlocks: by decreasing lower-bounds or increasing upper-bounds, some deadlocks can actually be removed.
\begin{proof}
	Let $\bounded{\A}{\bounds}$ be a closed bounded L/U-PTA and $\somelocs$ be a subset of its locations.
	Since $\A$ is closed and bounded, for each parameter $\param$, $\bounds(\param)$ is a closed interval $[m^-(\param),m^+(\param)]$.

	\ForLongVersion{The basic monotonicity property of L/U-PTAs (see }Lemma~\ref{lemma:HRSV02:prop4.2}\ForLongVersion{)} ensures that the TA~$\valuate{\A}{\pvalinfsup}$, where~$\pvalinfsup$ is obtained by valuating lower-bound parameters $\param^-$ by  $m^-(\param)$ and upper-bound parameters $\param^+$ by $m^+(\param)$, includes all the runs that could be produced with other parameter valuations.
	Consequently, if there is an infinite path for some valuation, there is one for $\pvalinfsup$ (note that, as emphasized above, this is not true for deadlocks). 	
    
    In~$\valuate{\A}{\pvalinfsup}$, it is decidable to find an infinite path staying in $\somelocs$, or conclude that none exists:
        this can be encoded into the CTL formula $EG (\somelocs \land X G)$, to be verified on the (finite) region graph of $\valuate{\A}{\pvalinfsup}$~\cite{AD94}. Since the region equivalence is a time-abstract bisimulation~\cite{TY01}, this means for $\valuate{\A}{\pvalinfsup}$ ``there exists a path that remains in~$\somelocs$ and in which every state has a discrete successor (possibly after letting some time elapse) in~$\somelocs$''. That path therefore has an infinite number of discrete actions.
	If we do find such a path, we can then terminate by answering yes to the EG-emptiness problem.
	If we do not, then in~$\valuate{\A}{\pvalinfsup}$, all paths staying in $\somelocs$ are finite.
	If we keep only discrete actions and locations, which are in finite number, the resulting paths therefore form a finite tree.
	Let us recall again that, thanks to Lemma~\ref{lemma:HRSV02:prop4.2}, %
		all the discrete paths that stay in $\somelocs$ and can be obtained with any parameter valuation, belong to that tree.

	We can now explicitly compute the symbolic states (following the symbolic semantics recalled in \cref{sss:symbolic}) for all the paths in the finite tree (not only those that are maximal).
	Recall that each symbolic state~$\symbstate$ is a pair $(\loc, \C)$, where $\loc$ is a location and $\C$ a convex polyhedron representing all parameter valuations and clock valuations that can be reached by the given discrete path.
	In each of these polyhedra, we can explicitly check for the existence of a deadlock, in the line of~\cite{Andre16}:
	\begin{enumerate}
		\item remove all parts that are in the past of the guard of an outgoing transition in $\A$ (using operation $\timepast{\C}$), and that would satisfy the target location invariant;
		\item test for emptiness.
	\end{enumerate}

    If the result is not empty then there exists a point in the tested set which can be decomposed into a parameter valuation and clock valuation such that, by any time elapsing from the clock valuation, none of the guards can become true. We therefore have a deadlock. If the result is empty, by the same reasoning, we can take a transition (possibly by first letting some time elapse) from all states of $\C$, so none of them are deadlocked.
	Note that both operations can be performed using classical polyhedral operations.

	If we find a deadlock, then we can terminate and answer yes to the EG-emptiness problem.
	Otherwise, we can terminate and answer no, because we have checked all the potential discrete paths staying in $\somelocs$ for any parameter valuation.
\end{proof}
}
	\preuveEGemptinesscbLU{}

Note that this proof fails when the L/U-PTA is not bounded or closed.
In particular, the closedness plays a key role in the sense that we are able to test the valuation~$\pvalinfsup$.
Consider\ForLongVersion{ first} the L/U-PTA in \cref{figure:counterexampleEG:LU-unbounded}\ForLongVersion{ made of a single location and a single loop with guard $x=1 \wedge y \leq \param$ and a reset of $x$, where $x,y$ are clocks and $\param$ a parameter.
This is clearly an L/U-PTA}.
As $\param$ grows, there are more and more discrete behaviors, but there is no cycle for any parameter valuation. 
In~\cite{BlT09}, the authors provide a finite upper bound $N_{\A}$ for the upper-bound parameters such that if there exists a valuation such that the valuated L/U-PTA has an accepting run, then the valuation giving $0$ to lower bound parameters and $N_{\A}$ to upper-bound parameters also ensures the existence of an accepting run.
That bound used in this example would indeed prove the non-existence of a cycle for any parameter value, but it does not in turn allow us to derive a finite tree containing all the discrete behaviors, for any possible parameter value (a larger bound would still give more runs).

Similarly, now consider the L/U-PTA in \cref{figure:counterexampleEG:LU-open}.
If~$0$ is excluded from the domain of~$\param$, we have a behavior similar to the previous example: as $\param$ gets closer and closer to~$0$, we have more and more discrete behaviors.
And even if we could derive a lower bound \emph{à la}~\cite{BlT09} ensuring the non-existence of a cycle here, it would not give a finite tree of all the possible discrete behaviors, for any parameter value.

We can actually exhibit a very thin border between decidability and undecidability of L/U-PTAs by proving that, given a bounded L/U-PTA~$\bounded{\A}{\bounds}$ with a single open bound in~$\bounds$ or an unbounded L/U-PTA, the EG-emptiness problem becomes undecidable.

\begin{thm}\label{theorem:EGemptiness:obLU}
	The EG-emptiness problem is undecidable for open bounded L/U-PTAs, with at least 4 clocks and 4 parameters.
\end{thm}
\begin{proof}
	We will use a reduction from the halting problem of a 2-counter machine.

	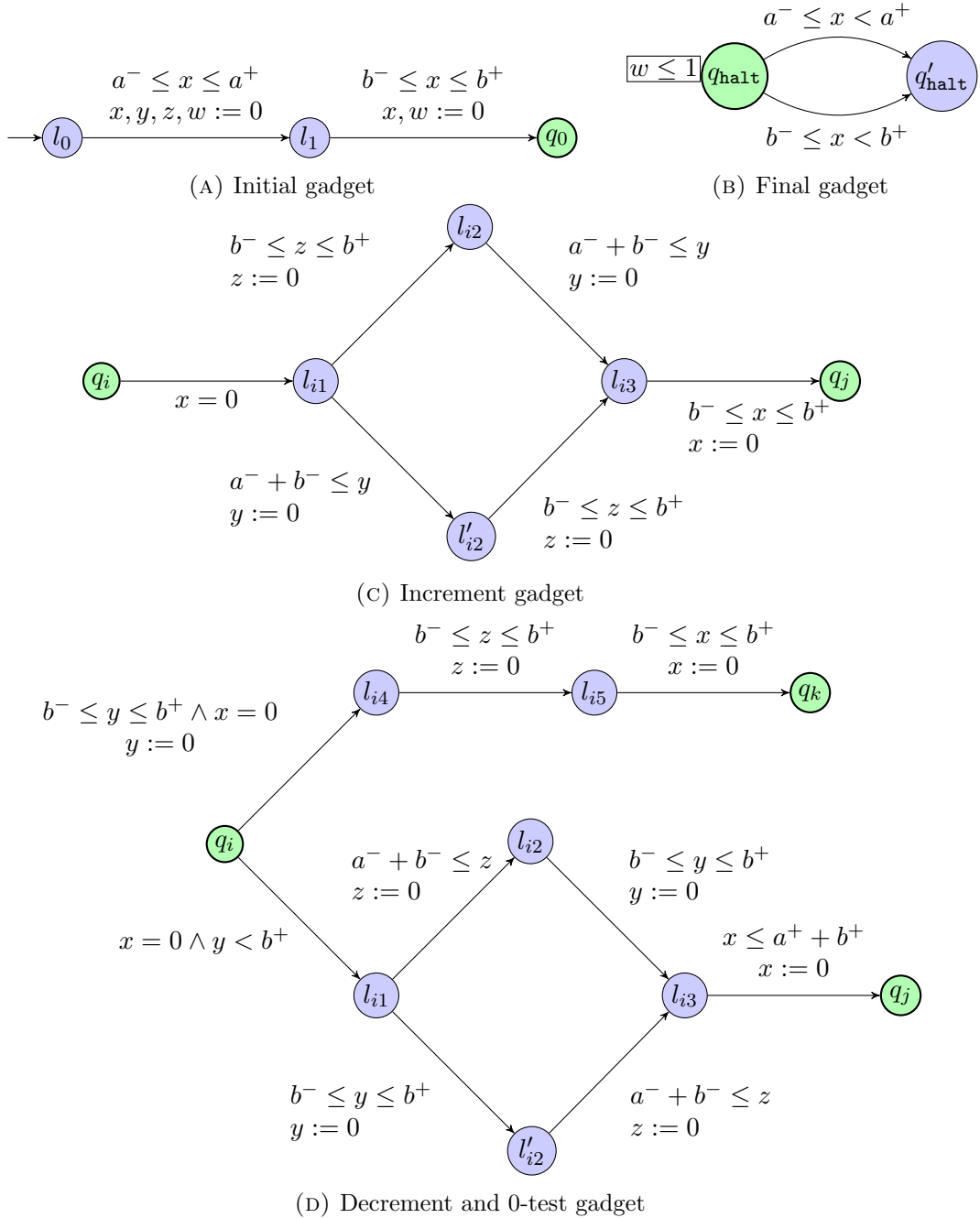
\begin{figure}

		\begin{subfigure}[b]{.62\textwidth}
	\centering
	\scalebox{\generalFigsScaleFactor}{
        \begin{tikzpicture}[PTA, node distance=3cm]
		\node[location, initial] (i0) {$l_0$};

		\node[location, right=of i0] (i1) {$l_1$};

		\node[location2CM, right=of i1] (l0) {$\cmspta_0$};

		\path (i0) edge node[above]{\begin{tabular}{c}$a^- \leq x \leq a^+$\\$x,y,z,w := 0$\end{tabular}} (i1);
		\path (i1) edge node[above]{\begin{tabular}{c}$b^- \leq x \leq b^+$\\$x,w := 0$\end{tabular}} (l0);

        \end{tikzpicture}
       }
       
        \caption{Initial gadget}
        \label{figure:EGemptiness:obLUPTA:initial}
	\end{subfigure}
		\begin{subfigure}[b]{.35\textwidth}
	\centering
	\scalebox{\generalFigsScaleFactor}{
        \begin{tikzpicture}[PTA, node distance=2cm]
		\node[location2CM] (lochalt) {\lochalt{}};
		\node [invariant,left] at (lochalt.west) {$w \leq 1$};
 
		\node[location, right=of lochalt] (lochaltprime) {\lochaltprime{}};

		\path (lochalt) edge[bend left] node[above]{$a^- \leq x < a^+$} (lochaltprime);
		\path (lochalt) edge[bend right] node[below]{$b^- \leq x < b^+$} (lochaltprime);
 
        \end{tikzpicture}
       }
       
        \caption{Final gadget}
        \label{figure:EGemptiness:obLUPTA:final}
	\end{subfigure}
		\begin{subfigure}[b]{\textwidth}
    \centering
	\scalebox{\generalFigsScaleFactor}{
        \begin{tikzpicture}[PTA, node distance=2.5cm]
            \node[location2CM] (si) {$\cmspta_i$};
            \node[location, right=of si] (l1) {$l_{i1}$};
            \node[location, above right=of l1] (l2) {$l_{i2}$};
            \node[location, below right=of l1] (l2p) {$l'_{i2}$};
            \node[location, below right=of l2] (l3) {$l_{i3}$};
            \node[location2CM, right=of l3] (sj) {$\cmspta_j$};

            \draw[->] (si) -- node[swap] {$x=0$} (l1);
            \draw[->] (l1) -- node[above left] {$\begin{array}{l}b^- \leq z \leq b^+\\ z:=0\end{array}$} (l2);
            \draw[->] (l1) -- node[swap] {$\begin{array}{l}a^- + b^-\leq y \\ y:=0\end{array}$} (l2p);
            \draw[->] (l2) -- node {$\begin{array}{l}a^- + b^-\leq y \\ y:=0\end{array}$} (l3);
            \draw[->] (l2p) -- node[below right, yshift=-1em, xshift=-1em] {$\begin{array}{l}b^- \leq z \leq b^+\\ z:=0\end{array}$} (l3);
            \draw[->] (l3) -- node[below, yshift=-.3em, xshift=1em] {$\begin{array}{l}b^- \leq x \leq b^+\\ x:=0\end{array}$} (sj);
        \end{tikzpicture}
       }
       
        \caption{Increment gadget}
        \label{figure:EGemptiness:obLUPTA:increment}
		\end{subfigure}
		\begin{subfigure}[b]{\textwidth}
	\centering
	\scalebox{\generalFigsScaleFactor}{
        \begin{tikzpicture}[PTA, node distance=2.5cm]
		\node[location2CM] (si) {$\cmspta_i$};

		\node[location, above right=of si] (si4) {$l_{i4}$};
		\node[location, right=of si4] (si5) {$l_{i5}$};
		\node[location2CM, right=of si5] (sk) {$\cmspta_k$};
 
				\node[location, below right=of si] (l1) {$l_{i1}$};
				\node[location, above right=of l1] (l2) {$l_{i2}$};
				\node[location, below right=of l1] (l2p) {$l'_{i2}$};
				\node[location, below right=of l2] (l3) {$l_{i3}$};
				\node[location2CM, right=of l3] (sj) {$\cmspta_j$};

				\draw[->] (si) -- node[below left, align=center]{$x=0 \land y<b^+$} (l1);
				\draw[->] (l1) -- node[xshift=25] {$\begin{array}{l}a^- + b^- \leq z\\ z:=0\end{array}$} (l2); %
				\draw[->] (l1) -- node[swap] {$\begin{array}{l}b^- \leq y \leq b^+\\ y:=0\end{array}$} (l2p); %
				\draw[->] (l2) -- node {$\begin{array}{l}b^- \leq y \leq b^+\\ y:=0\end{array}$} (l3);%
				\draw[->] (l2p) -- node[swap] {$\begin{array}{l}a^- + b^- \leq z\\ z:=0\end{array}$} (l3); %
				\draw[->] (l3) -- node {$\begin{array}{c}x \leq a^+ + b^+\\ x:=0\end{array}$} (sj); %

		\path (si) edge node{\begin{tabular}{c}$b^- \leq y \leq b^+ \land x = 0$\\$y := 0$\end{tabular}} (si4);
		\path (si4) edge node{\begin{tabular}{c}$b^- \leq z \leq b^+ $\\$z := 0$\end{tabular}} (si5);
		\path (si5) edge node{\begin{tabular}{c}$b^- \leq x \leq b^+$\\$x := 0$\end{tabular}} (sk);
 
        \end{tikzpicture}
       }
       
        \caption{Decrement and 0-test gadget}
        \label{figure:EGemptiness:obLUPTA:0-test}
	\end{subfigure}

    \caption{EG-emptiness for bounded L/U-PTAs}
    \label{figure:EGemptiness:obLUPTA}
    \end{figure}

	Let us consider the encoding used in the proof of \cref{theorem:EDemptiness:cbLU}, to which we will perform several modifications.

	First, we force the 2-counter machine to execute in a constant 1-time unit duration as follows:
	\begin{enumerate}
		\item We replace any occurrence of ``1'' in the encoding with a parameter, either $b^-$ or $b^+$ (depending on whether the occurrence of~1 occurs as a lower-bound or an upper-bound); hence the duration of an increment or decrement gadget is now at least~$b^-$ and at most~$b^+$.
			We give the increment gadget in \cref{figure:EGemptiness:obLUPTA:increment}.
			The encoding of a counter is as follows: when $x = 0$, then
				$y = b - a c_1$
				and
				$z = b - a c_2$,
				where $a = a^- = a^+$ and $b = b^- = b^+$ (for other parameter valuations, the machine is not properly simulated).
			Typically, $b$ will need to be sufficiently small compared to~1 to encode the required number of steps of the machine, and $a$ will need to be sufficiently small compared to~$b$ to encode the maximum value of the counters.
			The decrement part of the ``test and decrement'' instruction is modified similarly.
			
        \item We modify the zero-test part of the ``test and decrement'' instruction so that its duration is within $[b^- , b^+]$, as in \cref{figure:EGemptiness:obLUPTA:0-test}%
			: only the first transition from $\cmspta_i$ to $\loc_{i4}$ encodes the zero-test, the two other transitions from $\loc_{i4}$ to $\loc_{i5}$ and from~$\loc_{i5}$ to $\cmspta_k$ forcing $[b^- , b^+]$ time units to elapse while keeping the values of the clocks unchanged, assuming $a^- = a^+$ and $b^- = b^+$ (we will see later that other valuations do not matter).
			Let $a = a^- = a^+$ and $b = b^- = b^+$.
		The zero-test requires here that $b = y \land x = 0$; in addition, $z$ encodes $c_2$ as follows:
			$z = b - a c_2$.
			After reaching $l_{i4}$ and waiting enough time to take the transition to~$l_{i5}$ (\ie{} a duration in $a c_2$) we have:
			$ z = b$ and $x = y = a c_2$.
			After reaching $l_{i5}$ and waiting enough time to take the transition to~$\cmspta_{k}$ (\ie{} a duration in $b - a c_2$) we have:
			$ z = b - a c_2$ and $x = y = b$.
			Resetting~$x$ gives $x = 0$, $y = b$ and $z= b - a c_2$, which was the value when performing the 0-test.
			So the value of the clocks remains unchanged when $b^- = b^+$, and $[b^- , b^+]$ time units have elapsed in any case.
		\item We add to any location in the entire system an invariant $w \leq 1$, where~$w$ is a fresh clock that is never reset in the increment/decrement/zero-test gadgets.
			(These invariants are omitted in \cref{figure:EGemptiness:obLUPTA}.)
	\end{enumerate}
	Hence, the duration of any gadget is at least $b^-$ and therefore for any valuation $b^- > 0$ the number of operations the machine can perform is finite due to the global invariant $w \leq 1$.

	Then, before starting the 2-counter machine encoding, we add an initial gadget given in \cref{figure:EGemptiness:obLUPTA:initial}.
	This gadget constrains $a^- \leq a^+$, $b^- \leq b^+$, and is such that when leaving the gadget then $y,z \in [b^- \leq b^+]$ while $x,w$ are~0.
	When $b^- = b^+$, this correctly encodes that the value of both counters is~0.

    Then, we add a new \lochaltprime{} location (without any invariant, \ie{} not requiring $w \leq 1$), with two transitions from \lochalt{} as depicted in \cref{figure:EGemptiness:obLUPTA:final}.
    We then add a transition (with no guard) from any location of the encoding (except \lochalt{}) to~\lochaltprime{}.
	That is, for any increment gadget, if the value of the parameters is not small enough to correctly simulate the machine, then the system is not deadlocked, and can lead instead to \lochaltprime{}.
	(If the value is small enough, the system can either lead to \lochaltprime{} or continue in the 2-counter machine encoding.)
	We also add a transition to \lochaltprime{} (with no guard) from all locations in the initial gadget in \cref{figure:EGemptiness:obLUPTA:initial}.
	
	We assume the following bounds for the parameters: $a^-, a^+, b^+ \in [0,1]$ and $b^- \in (0, 1]$.

    Let us show that the 2-counter machine halts iff the set of valuations satisfying \EGtoutsauflochaltprime{} is not empty. %
	
	\begin{enumerate}
	 \item If $a^- > a^+$ or $b^- > b^+$, the initial gadget cannot be passed, and thanks to the transitions to \lochaltprime{}, all runs eventually reach \lochaltprime{}, hence \EGtoutsauflochaltprime{} does not hold.
	 
	 \item If $a^- < a^+$ and $b^- \leq b^+$, then the machine may not be correctly simulated: a given run will either reach \lochalt{}, in which case it will also reach \lochaltprime{} (as the guard from \lochalt{} to \lochaltprime{} does not forbid this run), or it will loop in the machine until it eventually gets ``blocked'' (since $b^- > 0$ and because of the invariant $w \leq 1$, for any  value of $b^-$, the maximal number of steps is $\frac{1}{b^-}$);
		when being blocked, it has no other option than going to \lochaltprime{}, thanks to the unguarded transitions from any location to \lochaltprime{}.
		Hence if $a^- < a^+$, \EGtoutsauflochaltprime{} does not hold.
	
	\item If $b^- < b^+$ (and $a^- \leq a^+$), again the machine may not be correctly simulated, and following a similar reasoning, \EGtoutsauflochaltprime{} again does not hold.

	\item If $a^- = a^+$ and $b^- = b^+ > 0$:
		\begin{enumerate}
			\item Either the machine does not halt: in this case, after a maximum number of steps (typically $\frac{1}{b^-}$), a gadget will be blocked due to the invariant $w \leq 1$, and the run will end in \lochaltprime{}.
			Hence if the 2-counter machine does not halt, \EGtoutsauflochaltprime{} does not hold.
			
        \item Or the machine halts: in this case, if $c$ is the maximum value of both $C_1$ and $C_2$ over the (necessarily finite) halting execution of the machine, and if $\runlength$ is the length of this execution, and if $c>0$, then for valuations such that
			$a^- = a^+ \leq \frac{b^-}{c}$
			and $b^- = b^+ \leq \frac{1}{\runlength}$, then there exists one run that correctly simulates the machine (beside plenty of runs that will go to \lochaltprime{} due to the unguarded transitions); this run that correctly simulates the machine eventually reaches \lochalt{}.
			From \lochalt{}, for such valuations, the system is deadlocked: indeed, the transitions from $\lochalt$ to $\lochaltprime$ can only be taken if $a^- < a^+$ or $b^- < b^+$.
			Hence \EGtoutsauflochaltprime{} holds.
			The set of such valuations %
            is certainly non-empty: $a^- = a^+ =\frac{1}{\runlength \times c}$ and $b^- = b^+ = \frac{1}{\runlength}$ belongs to it (if $c=0$ then we choose, \eg{}
				$b^-=b^+=1$ and $a^-=a^+=\frac{1}{2}$).
			Hence, if the 2-counter machine halts, there exist parameter valuations for which \EGtoutsauflochaltprime{} holds.
		\end{enumerate}
	\end{enumerate}
	Hence the 2-counter machine halts iff the set of valuations for which \EGtoutsauflochaltprime{} holds is not empty.
\end{proof}
\begin{rem}%
	The above construction works over 1 time unit (an invariant can be added to~$\lochaltprime{}$ too), so this gives an undecidability result over bounded time as well.
\end{rem}

We now prove that EG-emptiness is also undecidable for \emph{unbounded} L/U-PTAs.
When not considering L/U-PTAs, proving an undecidability result for bounded PTAs gives the undecidability for unbounded PTAs, as a bounded PTA can be simulated using a PTA (by, \eg{} adding the bounds as a guard between a fresh location prior to the initial location and the initial location, \eg{} $\param \in [\boundinf, \boundsup]$ becomes $ \boundinf \leq \clock \leq \boundsup \land \param = \clock$).
Recall that this is not true for L/U-PTAs, as such a construction requires to compare the clock and the parameter using an equality; in addition, L/U-PTAs are incomparable with bounded L/U-PTAs~\cite{ALR16FORMATS}.
In addition, our proof for unbounded L/U-PTAs uses one parameter less than for open bounded L/U-PTAs.

\begin{thm}\label{theorem:EGemptiness:LU}
	The EG-emptiness problem is undecidable for L/U-PTAs with at least 4 clocks and 3 parameters.
\end{thm}

\begin{proof}
	We will again use a reduction from the halting problem of a 2-counter machine.
	Our proof essentially relies on a mechanism similar to the proof of \cref{theorem:EGemptiness:obLU};
	however, we must use a different PTA encoding (the encoding used in the proof of \cref{theorem:EGemptiness:obLU} does not work for unbounded L/U-PTAs, as it strongly relies on the fact that $b^-$ be strictly positive), which prevents us to factor the proof as much as we would have wished.

	We propose here an encoding inspired by that of a 2-counter machine proposed in~\cite{BBLS15} to prove the undecidability of the EF-emptiness problem for PTAs with a single integer-valued parameter used to encode the maximum value of the two counters (although not considered in~\cite{BBLS15}, the proof also works identically with a rational-valued parameter).
	Starting from the initial configuration $(\cms_0, C_1 = 0, C_2=0)$ the machine either reaches \lochalt{} and halts, or loops forever.
	Knowing whether the machine halts is undecidable~\cite{Minsky67}.

	The encoding uses a single parameter~$a$.
	Two clocks $x$ and $y$ are used to encode the value of the counters, while a third clock~$z$ is used as an auxiliary clock.
    Whenever~$z = 0$, then $x= c_1$ and $y = c_2$.
		
	We modify this encoding by splitting the single parameter~$a$ into a lower-bound parameter~$a^-$ and an upper-bound parameter~$a^+$, in the spirit of previous undecidability results for L/U-PTAs in this paper (\cref{theorem:EDemptiness:cbLU,theorem:EGemptiness:obLU}).
	
	In addition, we request that the entire execution takes a time less than~$b^+$, where $b^+$ is a fresh upper-bound parameter; this is achieved by adding an invariant $w \leq b^+$ to all locations (with $w$ a fresh clock never reset after the initial gadget).
	
	We give the modified increment gadget for the first counter in \cref{figure:EGemptiness:LUPTA:increment} (invariants are omitted).
    Note that, if $z = 0$ when entering $\cmspta_i$ then the time to pass this gadget is in $[a^- + 1 , a^+ + 1]$.

	\begin{figure}

		\begin{subfigure}[b]{.55\textwidth}
	\centering
	\scalebox{\generalFigsScaleFactor}{
		\begin{tikzpicture}[PTA, node distance=3cm]
		\node[location, initial] (i0) {$l_0$};

		\node[location, right=of i0, xshift=-2em] (i1) {$l_1$};

		\node[location2CM, right=of i1] (l0) {$\cmspta_0$};

		\path (i0) edge node[above]{\begin{tabular}{c}$a^- \leq x \leq a^+$\\$x := 0$\end{tabular}} (i1);
		\path (i1) edge node[above]{\begin{tabular}{c}$0 < x \leq b^+$\\$x,y,z,w := 0$\end{tabular}} (l0);
		
		\end{tikzpicture}
	}
	
		\caption{Initial gadget}
		\label{figure:EGemptiness:LUPTA:initial}
		\end{subfigure}
	\hfill{}
		\begin{subfigure}[b]{.4\textwidth}
	\centering
	\scalebox{\generalFigsScaleFactor}{
		\begin{tikzpicture}[PTA, node distance=2.2cm]
		\node[location] (lochalt) {\lochalt{}};
		\node [invariant,left] at (lochalt.west) {$w \leq b^+$};

		\node[location, right=of lochalt] (lochaltprime) {\lochaltprime{}};

		\path (lochalt) edge[] node[above]{$a^- \leq x < a^+$} (lochaltprime);

		\end{tikzpicture}
	}
	
		\caption{Final gadget}
		\label{figure:EGemptiness:LUPTA:final}
		\end{subfigure}
		\begin{subfigure}[b]{\textwidth}
	\centering
	\scalebox{\generalFigsScaleFactor}{
		\begin{tikzpicture}[PTA, node distance=2.5cm]
		\node[location2CM] (si) {$\cmspta_i$};

		\node[location, right=of si] (li1) {$l_{i1}$};
		\node[location, right=of li1] (li4) {$l_{i4}$};
		\node[location, above=of li1, yshift=-.75cm] (li2) {$l_{i2}$};
		\node[location, below=of li1, yshift=+1cm] (li2') {$l_{i2'}$};
		\node[location, right=of li2] (li3) {$l_{i3}$};
		\node[location, below=of li4, yshift=+1cm] (li3') {$l_{i3'}$};
		\node[location2CM, right=of li4] (sj) {$\cmspta_j$};

		\path (si) edge node[above]{\begin{tabular}{c}$z = 1$\\$z := 0$\end{tabular}} (li1);
		\path (li1) edge node[left, yshift=+1em]{\begin{tabular}{c}$a^- \leq x \leq a^+$\\$x := 0$\end{tabular}} (li2);
		\path (li2) edge node[below]{\begin{tabular}{c}$a^- \leq y \leq a^+$\\$y := 0$\end{tabular}} (li3);
		\path (li3) edge node[right, yshift=+1em]{\begin{tabular}{c}$y = 1$\\$y := 0$\end{tabular}} (li4);

		\path (li1) edge node[left]{\begin{tabular}{c}$a^- \leq y \leq a^+$\\$y := 0$\end{tabular}} (li2');
		\path (li2') edge node[above]{\begin{tabular}{c}$y = 1$\\$y := 0$\end{tabular}} (li3');
		\path (li3') edge node[right]{\begin{tabular}{c}$a^- \leq x \leq a^+$\\$x := 0$\end{tabular}} (li4);

		\path (li4) edge node[above]{\begin{tabular}{c}$a^- \leq z \leq a^+$\\$z := 0$\end{tabular}} (sj);

		\end{tikzpicture}
	}
	
		\caption{Increment gadget}
		\label{figure:EGemptiness:LUPTA:increment}
		\end{subfigure}
		\begin{subfigure}[b]{\textwidth}
	\centering
	\scalebox{\generalFigsScaleFactor}{
		\begin{tikzpicture}[PTA, node distance=2.5cm]
		\node[location2CM] (si) {$\cmspta_i$};

		\node[location, right=of si] (li1) {$l_{i1}$};
		\node[location, right=of li1] (li4) {$l_{i4}$};
		\node[location, above=of li1, yshift=-.75cm] (li2) {$l_{i2}$};
		\node[location, below=of li1, yshift=+1cm] (li2') {$l_{i2'}$};
		\node[location, right=of li2] (li3) {$l_{i3}$};
		\node[location, below=of li4, yshift=+1cm] (li3') {$l_{i3'}$};
		\node[location2CM, right=of li4] (sj) {$\cmspta_j$};

		\node[location, below=of li2', yshift=+1cm] (li1'') {$l_{i1''}$};
		\node[location, right=of li1''] (li2'') {$l_{i2''}$};
		\node[location2CM, right=of li2''] (sk) {$\cmspta_{k}$};

		\path (si) edge node[above]{\begin{tabular}{c}$z = 0 \land x > 0$\end{tabular}} (li1);
		\path (li1) edge node[left, yshift=+1em]{\begin{tabular}{c}$a^- \leq x \leq a^+$\\$x := 0$\end{tabular}} (li2);
		\path (li2) edge node[below]{\begin{tabular}{c}$x = 1$\\$x := 0$\end{tabular}} (li3);
		\path (li3) edge node[right, yshift=+1em]{\begin{tabular}{c}$a^- \leq y \leq a^+$\\$y := 0$\end{tabular}} (li4);

		\path (li1) edge node[left]{\begin{tabular}{c}$a^- \leq y \leq a^+$\\$y := 0$\end{tabular}} (li2');
		\path (li2') edge node[above]{\begin{tabular}{c}$a^- \leq x \leq a^+$\\$x := 0$\end{tabular}} (li3');
		\path (li3') edge node[right]{\begin{tabular}{c}$x = 1$\\$x := 0$\end{tabular}} (li4);

		\path (li4) edge node[above]{\begin{tabular}{c}$a^- \leq z \leq a^+$\\$z := 0$\end{tabular}} (sj);

		\path (si) edge[bend right] node[below left]{\begin{tabular}{c}$z = 0 \land x = 0$\end{tabular}} (li1'');
		\path (li1'') edge node[above]{\begin{tabular}{c}$a^- + 1 \leq y \leq a^+ + 1 $\\$y := 0$\end{tabular}} (li2'');
		\path (li2'') edge node[above]{\begin{tabular}{c}$a^- + 1 \leq x \leq a^+ + 1 $\\$x,z := 0$\end{tabular}} (sk);
		\end{tikzpicture}
	}
	
		\caption{0-test and decrement gadget}
		\label{figure:EGemptiness:LUPTA:decrement}
		\end{subfigure}

	\caption{EG-emptiness for L/U-PTAs}
    \end{figure}
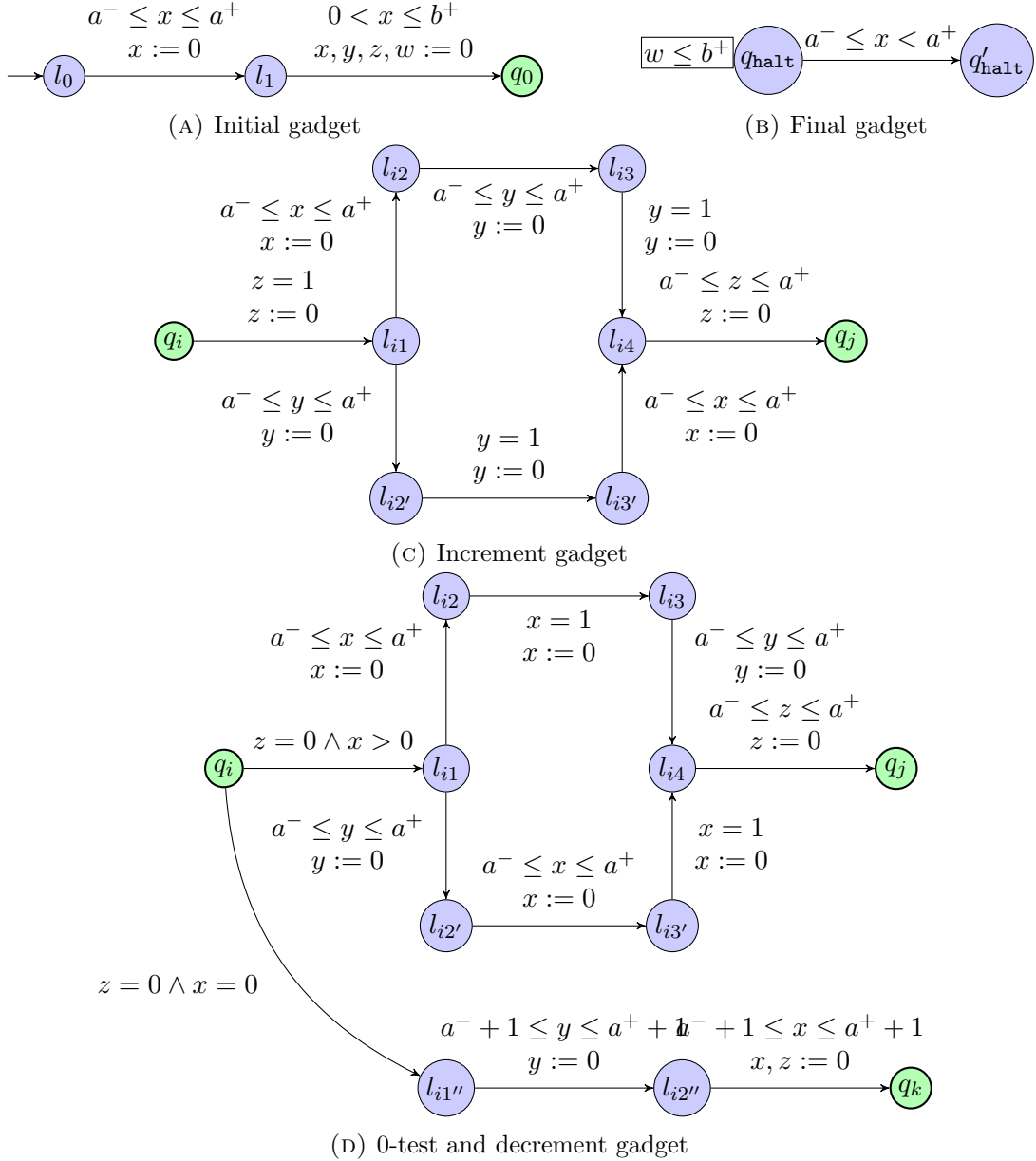

	The test and decrement gadget is similar, and given in \cref{figure:EGemptiness:LUPTA:decrement}.
	We performed a slight modification to the zero-test of~\cite{BBLS15}, that was executed in 0-time; we require in our construction that each gadget takes at least one time unit.
	Hence, we rewrote it in \cref{figure:EGemptiness:LUPTA:decrement} so as to force at least one time unit to elapse after the clocks are tested, and so that the final value of the clock is not changed, when $a^- = a^+$ (in the spirit of the same operation in the proof of \cref{theorem:EGemptiness:obLU}):
		when performing the zero-test, we have $x = z = 0$ and $y = c_2$.
		Then after $a - c_2 + 1$ time units (with $a = a^+ = a^-$), we have $x = z = a + 1 - c_2$ and $y = a + 1$,
		and we can take the transition to~$l_{i2''}$, resetting~$y$.
		Then after $c_2 $ time units, we have $x = z = a + 1$ and $y = c_2$
		and we can take the transition to~$l_{i2''}$, resetting~$x$ and~$z$.
		This gives finally $x = z = 0$ and $y = c_2$ and the time spent in the gadget is in $[a^- + 1, a^+ + 1]$, and therefore is more than one time unit.
	Gadgets for the second counter are symmetric.
	
	We add before the first instruction the initial gadget given in \cref{figure:EGemptiness:LUPTA:initial}, constraining $a^- \leq a^+$ and $b^+ > 0$, and resetting all clocks.

    In addition, just as in \cref{theorem:EGemptiness:obLU}, we add unguarded transitions from any location (including that of the initial gadget, but excluding~\lochalt{}) to a new location $\lochaltprime$.
	We finally add a transition from~\lochalt{} to \lochaltprime{} as shown in the final gadget in \cref{figure:EGemptiness:LUPTA:final}.
	
	Let us show that the 2-counter machine halts iff the set of valuations for which \EGtoutsauflochaltprime{} holds is not empty.
	We reason on the parameter valuations.
	\begin{enumerate}
		\item If $a^- > a^+$ or $b^+ = 0$, the initial gadget cannot be passed: any run is sent to~\lochaltprime{} because of the transitions to \lochaltprime{}, and therefore \EGtoutsauflochaltprime{} does not hold.

		\item If $a^- < a^+$ and $b^+ > 0$, 
			then the machine may not be correctly simulated: a given run will either reach \lochalt{}, in which case it will also reach \lochaltprime{} (as the guard from \lochalt{} to \lochaltprime{} in \cref{figure:EGemptiness:LUPTA:final} does not forbid this run), or it will loop in the machine until it eventually gets blocked: since $b^+ > 0$, since all gadgets require at least 1 time unit, for any  value of~$b^+$ the invariant $z \leq b^+$ will eventually block a transition after at most $b^+$~steps.
			When being blocked, a run has no other option than going to \lochaltprime{}, because of the unguarded transitions from any location to \lochaltprime{}.
			Hence if $a^- < a^+$ and $b^+ > 0$, \EGtoutsauflochaltprime{} does not hold.

		\item Now, assume $a^- = a^+$ and $b^+ > 0$.
		\begin{enumerate}
			\item Either the machine does not halt: in this case, after a maximum number of steps (typically at most $b^+$), a gadget will be blocked due to the invariant $z \leq b^+$, and the run will end in \lochaltprime{} because of the unguarded transitions from any location to \lochaltprime{}.
			Hence if the 2-counter machine does not halt, \EGtoutsauflochaltprime{} does not hold.

        \item Or the machine halts: in this case, if $c$ is the maximum value of both $C_1$ and $C_2$ over the (necessarily finite) halting execution of the machine, and if $\runlength$ is the length of this execution, and if $c>0$, then for valuations such that $a^- = a^+ \leq c$ and sufficiently large valuations of $b^+$ (typically $b^+ \geq \runlength \times (a^+ + 1)$ as a gadget can take up to $a^+ + 1$ time units%
        ), then there exists one run that correctly simulates the machine; %
				this run eventually reaches \lochalt{}.
			From \lochalt{}, for such values, the system is deadlocked. %
			Hence, if the 2-counter machine halts, there exist parameter valuations for which a run does not reach \lochaltprime{}, \ie{} for which \EGtoutsauflochaltprime{} holds.
		\end{enumerate}
	\end{enumerate}
	
    Hence the 2-counter machine halts iff the set of valuations for which \EGtoutsauflochaltprime{} holds is not empty.
\end{proof}

\ForLongVersion{
\begin{rem}
	The above construction works also for integer-valued parameters, so this gives an undecidability result for integer-valued parameters too.
	The proof also works over discrete time (with integer-valued parameters).
\end{rem}
}

\toutfaux{

\begin{thm}\label{theorem:bIPPTA-EG-undecidable}
    The AF-universality problem is undecidable for bounded IP-PTAs.
\end{thm}
\begin{proof}
	We reduce the problem of knowing whether the counters of a 2-counter machine grow unbounded along its execution, which is undecidable~\cite{Minsky67}, to the universality of the set of parameters for all runs eventually reach $\locerror{}$.
	
	Let us reuse the encoding of the 2-counter machine used in the proof of \cref{theorem:bip-EFu}, with the following modifications.
	First, let us remove parameter~$b$ (and the transition from the initial state to~$\locinit$ ensuring that $b \in [0, 1)$.
	Second, let us remove the guard from $l_{i2}$ to \locerror{}, \ie{} this transition can be taken any time.
	
	We now prove that the counters of the machine grow unbounded along its execution iff all valuations of~$a$ eventually reach~\locerror{}.
	
	First remark that for $a = 0$, any run eventually leads to \locerror{} (possibly through \lochalt{}), hence the property EG$(\Loc \setminus \{ \locerror{} \})$ does not hold.\ea{ah ben non ça c'est tout faux :=(}
	When $a > 0$, we have two cases:
	\begin{itemize}
		\item either the counters grow unbounded (say $C_1$ does), then whatever the value of $a>0$, at some point we have $ac_1 > 1$.
			More specifically, there is an increment of $C_1$ such that $ac_1 \leq 1$ and $a(c_1 + 1) > 1$.
			Then, when executing the corresponding increment gadget, \locerror{} can be reached from $l_{i2}$.
			Hence, the property EG$(\Loc \setminus \{ \locerror{} \})$ does not hold for any parameter valuation;
		\item or the counters remain bounded.
			Let $c$ be the maximal value of the counters.
			If $ca\leq 1$, then the transition to \locerror{} is never possible.
				Since the set of values of $a$ such that $a>0$ and $ca\leq 1$ is never empty, the set of valuations for which the automaton satisfies the property EG$(\Loc \setminus \{ \locerror{} \})$ is not empty.
	\end{itemize}
	
	Now, it remains to show that the constructed PTA is an IP-PTA; we reuse the reasoning from \cref{theorem:bip-EFu}.
	With the exception of $\locerror$, the result is clear: $a=0$ belongs to every reachable symbolic state, hence each symbolic state contains an integer parameter valuation, and hence from Lemma~\ref{lemma:Didier}, all symbolic states (except $\locerror$) contain at least one integer point.
	Now, concerning $\locerror$, despite the strict inequality ($x > 1$) entering the location, the fact that there is no invariant together with the fact that $a=0$ is an integer valuation gives that there exists at least one integer point.
	Hence this PTA is a (bounded) IP-PTA.
\end{proof}
\begin{cor}\label{corollary:EG-undecidable}
	The AF-universality problem is undecidable for IP-PTAs, for bounded PTAs and for PTAs.
\end{cor}
\begin{proof}
	From \cref{theorem:bIPPTA-EG-undecidable} and the fact that a bounded IP-PTA is also a bounded PTA, and a PTA.
\end{proof}

\ea{TODO one day: trouver le bug dans la preuve de décidabilité de EG-emptiness pour IP-PTA (et un contre-exemple) ; cf.\ \LaTeX{} code}\dl{Certes}

}

Finally, using a single parameter $a$ instead of both $a^-$ and $a^+$ in the above proof, we obtain a PTA (not L/U) with two parameters. Hence: 

\begin{cor}\label{corollary:EGemptiness:PTA}
	The EG-emptiness problem is undecidable for PTAs with at least 4 clocks and 2 parameters.
\end{cor}
\section{Summary}\label{section:summary}
\begin{table*}[tb!]
	\centering
	\setlength{\tabcolsep}{1pt} %
	\begin{tabular}{@{} | *{7}{c|}}
		\hline
		\cellHeader{Class} & \cellHeader{bL/U-PTAs} & \cellHeader{bIP-PTAs} & \cellHeader{L/U-PTAs} & \cellHeader{IP-PTAs} & \cellHeader{bPTAs} & \cellHeader{PTAs} \\
		\hline
		EF-empt. & \colCellDecNous{\crefabbr{theorem:bL/UPTA-EF-decidable}} & \colCellDecNous{\crefabbr{theorem:EF-emptiness:decidable:Didier}} & \colCellDec{\cite{HRSV02}} & \colCellUndecNous{\crefabbr{theorem:IPPTA-EF-undecidable}} & \colCellUndec{\cite{Miller00}} & \colCellUndec{\cite{AHV93}} \\
		\hline
		EF-univ. & \colCellDecNous{\crefabbr{theorem:bL/UPTA-EF-decidable}} & \colCellUndecNous{\crefabbr{theorem:bip-EFu}} & \colCellDec{\cite{BlT09}} & \colCellUndecNous{Cor.~\ref{corollary:EFu}} & \colCellUndecNous{Cor.~\ref{corollary:EFu}~} & \colCellUndecNous{Cor.~\ref{corollary:EFu}~} \\
		\hline
		AF-empt. & \colCellUndecNous{\crefabbr{theorem:bounded-LUPTA-AF-undecidable}} & \colCellUndecNous{Cor.~\ref{corollary:AF-undecidable}} & \colCellUndec{\cite{JLR15}} & \colCellUndecNous{Cor.~\ref{corollary:AF-undecidable}} & \colCellUndecNous{Cor.~\ref{corollary:AF-undecidable}} & \colCellUndec{\cite{JLR15}} \\
		\hline
	\end{tabular}

	\caption{Decidability of reachability (and AF) problems for PTAs and some subclasses}
    \label{table:summary:decidability}
\end{table*}

We give a summary of our results concerning EF-emptiness, EF-universality and AF-emptiness in \cref{table:summary:decidability}.
We give from left to right the (un)decidability for bounded L/U-PTAs, bounded IP-PTAs, L/U-PTAs, IP-PTAs, bounded PTAs, and PTAs.
Decidability is given in bold green preceded with~``$\surd$'', whereas undecidability is given in italic red preceded with~``$\times$''.
Our contributions are depicted using a plain background, whereas existing results are depicted using a light background.
\begin{figure}[tb]
	\newcommand{\figlength}{0.32\textwidth}
	\newcommand{\tikzscale}{.62}
	\newcommand{\scalefactor}{.72}
	
	\noindent\begin{subfigure}[b]{\figlength}
	\centering
		\scalebox{\scalefactor}{
		\begin{tikzpicture}[scale=\tikzscale,->, >=stealth', auto]
			\draw[decidable,contribution] (-1, 0) circle [radius=2.5];
			\node[legende, rotate=50] at (-2, 1) {bounded L/U};

			\draw[decidable] (1, 0) circle [radius=2.5];
			\node[legende] at (2.8, 0.8) {L/U};

			\draw[decidable,contribution] (3, -1.4)  arc[x radius = 3, y radius = 2.5, start angle=15, end angle=216]; %
			\draw[undecidable,contribution] (3, -1.4)  arc[x radius = 3, y radius = 2.5, start angle=15, end angle=-144]; %
			\node[legende] at (0, -4) {IP-PTA};

			\draw[decidable] (0, -1) ellipse (2.5 and .75);
			\node[legende] at (0, -1) {closed L/U};

			\draw[undecidable] (-1, 2.5) --++ (-5, 0);
			\draw[draw=none] (-1, 2.5) arc[radius = 2.5, start angle= 90, end angle= -65] node(arcfin){}; %
			\draw[decidable,contribution] (arcfin) --++ (-2.5, -1.35); %
			\draw[undecidable] (arcfin) --++ (-5, -2.7); %
			\node[legende,align=center] at (-4.5, -3) {bounded\\PTAs};

			\draw[undecidable] (-6, -5) rectangle (4.5, 3);
			\node[legende] at (3.75, -4) {PTAs};
			\end{tikzpicture}
		}
		\caption{EF-emptiness}
		\label{figure:summary:decidability:EF}
	\end{subfigure}
	\begin{subfigure}[b]{\figlength}
	\centering
		\scalebox{\scalefactor}{
		\begin{tikzpicture}[scale=\tikzscale,->, >=stealth', auto]
			\draw[decidable,contribution] (-1, 0) circle [radius=2.5];
			\node[legende, rotate=50] at (-2, 1) {bounded L/U};

			\draw[decidable] (1, 0) circle [radius=2.5];
			\node[legende] at (2.8, 0.8) {L/U};

			\draw[decidable,contribution] (3, -1.4)  arc[x radius = 3, y radius = 2.5, start angle=15, end angle=172]; %
			\draw[undecidable,contribution] (3, -1.4)  arc[x radius = 3, y radius = 2.5, start angle=15, end angle=-188]; %
			\node[legende] at (0, -4) {IP-PTA};

			\draw[decidable] (0, -1) ellipse (2.5 and .75);
			\node[legende] at (0, -1) {closed L/U};

			\draw[undecidable,contribution] (-1, 2.5) --++ (-5, 0);
			\draw[draw=none] (-1, 2.5) arc[radius = 2.5, start angle= 90, end angle= -65] node(arcfin){}; %
			\draw[undecidable,contribution] (arcfin) --++ (-5, -2.7); %
			\node[legende,align=center] at (-4.5, -3) {bounded\\PTAs};

			\draw[undecidable,contribution] (-6, -5) rectangle (4.5, 3);
			\node[legende] at (3.75, -4) {PTAs};
		\end{tikzpicture}
		}
		\caption{EF-universality}
		\label{figure:summary:decidability:AG}
	\end{subfigure}
	\begin{subfigure}[b]{\figlength}
	\centering
		\scalebox{\scalefactor}{
		\begin{tikzpicture}[scale=\tikzscale,->, >=stealth', auto]
			\draw[undecidable,contribution] (-1, 0) circle [radius=2.5];
			\node[legende, rotate=50] at (-2, 1) {bounded L/U};

			\draw[undecidable] (1, 0) circle [radius=2.5];
			\node[legende] at (2.8, 0.8) {L/U};

			\draw[undecidable,contribution] (0, -2) ellipse (3 and 2.5);
			\node[legende] at (0, -4) {IP-PTA};

			\draw[undecidable] (0, -1) ellipse (2.5 and .75);
			\node[legende] at (0, -1) {closed L/U};

			\draw[undecidable,contribution] (-1, 2.5) --++ (-5, 0);
			\draw[draw=none] (-1, 2.5) arc[radius = 2.5, start angle= 90, end angle= -65] node(arcfin){}; %
			\draw[undecidable,contribution] (arcfin) --++ (-5, -2.7); %
			\node[legende,align=center] at (-4.5, -3) {bounded\\PTAs};

			\draw[undecidable] (-6, -5) rectangle (4.5, 3);
			\node[legende] at (3.75, -4) {PTAs};
			\end{tikzpicture}
			}
		\caption{AF-emptiness}
		\label{figure:summary:decidability:AF}
	\end{subfigure}

	\caption{Decidability results for PTAs and subclasses}
    \label{figure:summary:decidability}
\end{figure}
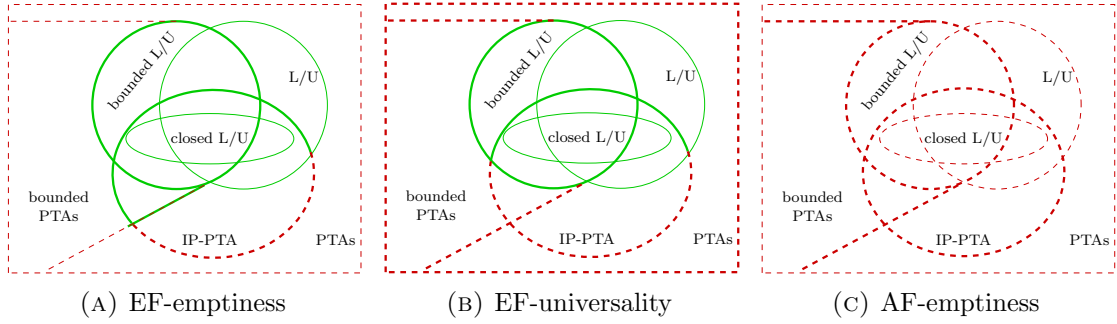

We give another summary in \cref{figure:summary:decidability}.
Note that bounded L/U-PTAs and L/U-PTAs are in fact incomparable of terms of expressiveness~\cite{ALR16FORMATS}; they are therefore not included into each other in the figures.
Decidability (resp.\ undecidability) is depicted in plain green (resp.\ dashed red); open problems are depicted in dotted black.
Our contributions are depicted with boldface font.

\begin{table*}[tb!]
	\centering
	\begin{tabular}{@{} | *{5}{c|}}
		\hline
		\cellHeader{Class} & \cellHeader{PTAs} & \cellHeader{L/U-PTAs} & \cellHeader{bo L/U-PTAs} & \cellHeader{bc L/U-PTAs} \\
		\hline
        EC-emptiness & \colCellUndecNous{\crefabbr{theorem:ECemptiness:PTA}} & \colCellDecNous{\crefabbr{theorem:ECemptiness:cbLU}} & \colCellOpen{}open & \colCellDecNous{\crefabbr{theorem:ECemptiness:LU}} \\
		\hline
		ED-emptiness & \colCellUndecNous{Cor.~\ref{corollary:EDemptiness:LU-PTA}} & \colCellUndecNous{Cor.~\ref{corollary:EDemptiness:LU-PTA}} & \colCellUndecNous{\crefabbr{theorem:EDemptiness:cbLU}} & \colCellUndecNous{Cor.~\ref{corollary:EDemptiness:LU-PTA}}\\
		\hline
		EG-emptiness & \colCellUndecNous{Cor.~\ref{corollary:EGemptiness:PTA}} & \colCellUndecNous{\crefabbr{theorem:EGemptiness:LU}} & \colCellUndecNous{\crefabbr{theorem:EGemptiness:obLU}} & \colCellDecNous{\crefabbr{theorem:EGemptiness:cbLU}} \\
		\hline
		AF-emptiness & \colCellUndec{\cite{JLR15}} & \colCellUndec{\cite{JLR15}} & \colCellUndecNous{Cor.~\ref{corollary:open-bounded-LUPTA-AF-undecidable}} & \colCellUndecNous{\crefabbr{theorem:bounded-LUPTA-AF-undecidable}} \\
		\hline
	\end{tabular}

	\caption{Decidability of liveness problems for PTAs and some subclasses}
    \label{table:summary:decidability:EG}
\end{table*}

We then review our results concerning EG-emptiness, ED-emptiness, EC-emptiness and AF-emptiness in \cref{table:summary:decidability:EG}, where ``bo L/U-PTAs'' and ``bc L/U-PTAs'' denote bounded open and bounded closed L/U-PTAs respectively.
Again, bold green preceded with~``$\surd$'' denotes decidability while red preceded with~``$\times$'' denotes undecidability.
Note that, with the exception of AF-emptiness for L/U-PTAs (and PTAs), all these contributions are new, \ie{} all these problems were open prior to our work.

\begin{table*}[tb!]
	\centering
	\setlength{\tabcolsep}{1pt} %
	\resizebox{\textwidth}{!}{
	\begin{tabular}{@{} | *{11}{c|}}
		\hline
		\cellHeader{Class} & \cellHeader{iL-PTA} & \cellHeader{L-PTA} & \cellHeader{bL/U-PTA} & \cellHeader{bIP-PTA} & \cellHeader{L/U-PTA} & \cellHeader{IP-PTA} & \cellHeader{bPTA} & \cellHeader{br-PTA} & \cellHeader{biPTA} & \cellHeader{PTA} \\
		\hline
		EF-synth. & \colCellDec{\cite{BlT09}} & \colCellOpen{}open & \colCellUndecNous{\crefabbr{theorem:intractability}} & \colCellUndecNous{Cor.~\ref{intractability-synthesis-IPPTA}} & \colCellUndec{\cite{JLR15}} & \colCellUndecNous{Cor.~\ref{intractability-synthesis-IPPTA}} & \colCellUndecNous{\crefabbr{theorem:intractability}} & \colCellDec{\cite{ALR21}} & \colCellDec{\cite{JLR15}} & \colCellUndec{\cite{AHV93}} \\ %
		\hline
	\end{tabular}
	}

	\caption{EF-synthesis for PTAs and some subclasses}
    \label{table:summary:synthesis}
\end{table*}

Finally, we review EF-synthesis, \ie{} for which subclasses of PTAs can the exact solution to the EF-synthesis problem be computed?
We review in \cref{table:summary:synthesis} from left to right the following subclasses:
	L-PTAs with integer-valued parameters,
	L-PTAs with rational-valued parameters,
	bounded L/U-PTAs,
	bounded IP-PTAs,
	L/U-PTAs,
	IP-PTAs,
	bounded PTAs,
	bounded reset-PTAs,
	bounded PTAs with integer-valued parameters,
	and
	PTAs.
Results for U-PTAs (resp.\ U-PTAs with integer-valued parameters) are identical to L-PTAs (resp.\ L-PTAs with integer-valued parameters)---and therefore not part of \cref{table:summary:synthesis}.
Clearly, beyond the obvious case of bounded PTAs with integer-valued parameters~\cite{JLR15}, only two subclasses allow for exact synthesis: integer-valued L-PTAs~\cite{BlT09} and bounded-reset PTAs~\cite{ALR21}.
The case of L-PTAs or U-PTAs over non-necessarily integer parameters remains open.

\section{Conclusion}\label{section:conclusion}

Despite the vast number of undecidability results linked to the formalism of parametric timed automata, and to which we also contribute in this paper, we exhibited a new subclass of PTAs (namely bounded IP-PTAs) for which the EF-emptiness problem is decidable.
By showing that bounded IP-PTAs are incomparable with L/U-PTAs, we strictly extend the set of PTAs for which this problem is decidable.
Although we showed that it cannot be decided whether a PTA is an IP-PTA, we introduced a new syntactic subclass of IP-PTAs, namely \emph{reset-PTAs}, for which, when bounded, the EF-emptiness problem is decidable.
It is worth noting that, to the best our knowledge, there is no other non-trivial set of syntactic restrictions making the reachability emptiness problem decidable for PTAs (aside from L/U-PTAs, of course).

We also considered several decision problems, and contributed in solving several open problems for PTAs and subclasses: this was achieved thanks to the results proved for IP-PTAs, and to (variations of) an original proof for the undecidability of the EF-emptiness problem for general PTAs with a single bounded rational-valued parameter and only non-strict constraints.%

We also have achieved some decidability for the existential parametric problem on the EG liveness property.
This could be done by imposing original constraints to the classical subclass of L/U-PTAs, pertaining to the topology of the domain of the parameter values.
This domain should be a closed and bounded hyperrectangle of the rational space.
The subclass together with the EG property really lies on the boundary of decidability: on the one hand, we have proved that considering unbounded, or bounded but open domains leads again to undecidability for~EG.
On the other hand, if we consider---instead of the EG property which asks for the existence of a maximal finite or infinite path staying in some locations---only infinite maximal paths (existence of discrete cycles), then we have proved that the problem becomes 
	decidable (for either closed bounded domains, or unbounded domains---the case of open bounded domains remains open).
And finally, if we consider only finite maximal paths (existence of deadlocks), then we have proved that the problem becomes consistently undecidable.

\subsection*{Future works}

Future work includes %
\begin{enumerate}
	\item studying the decidability of EC-emptiness for open bounded L/U-PTAs (possibly adapting techniques developed in~\cite{Sankur11}),
	\item extending the EG decidability result to shapes other than hyperrectangles, and
	\item studying actual synthesis, and implementing efficient algorithms.
\end{enumerate}
In addition, the decidability of problems we proved undecidable for L/U-PTAs should be studied for two subclasses of L/U-PTAs, where all parameters are upper bounds (U-PTAs) or all lower bounds (L-PTAs).

\section*{Acknowledgments}
This work was partially supported
by the ANR national research program ``PACS'' (ANR-14-CE28-0002)
and
by the ANR-NRF French-Singaporean research program \href{https://www.loria.science/ProMiS/}{ProMiS} (ANR-19-CE25-0015).

	\newcommand{\CCIS}{Communications in Computer and Information Science}
	\newcommand{\ENTCS}{Electronic Notes in Theoretical Computer Science}
	\newcommand{\FAC}{Formal Aspects of Computing}
	\newcommand{\FI}{Fundamenta Informaticae}
	\newcommand{\FMSD}{Formal Methods in System Design}
	\newcommand{\IJFCS}{International Journal of Foundations of Computer Science}
	\newcommand{\IJSSE}{International Journal of Secure Software Engineering}
	\newcommand{\IPL}{Information Processing Letters}
	\newcommand{\JAIR}{Journal of Artificial Intelligence Research}
	\newcommand{\JLAP}{Journal of Logic and Algebraic Programming}
	\newcommand{\JLAMP}{Journal of Logical and Algebraic Methods in Programming} %
	\newcommand{\JLC}{Journal of Logic and Computation}
	\newcommand{\LMCS}{Logical Methods in Computer Science}
	\newcommand{\LNCS}{Lecture Notes in Computer Science}
	\newcommand{\RESS}{Reliability Engineering \& System Safety}
	\newcommand{\STTT}{International Journal on Software Tools for Technology Transfer}
	\newcommand{\TCS}{Theoretical Computer Science}
	\newcommand{\ToPNoC}{Transactions on Petri Nets and Other Models of Concurrency}
	\newcommand{\TSE}{{IEEE} Transactions on Software Engineering}

	\newcommand{\IEEETSE}{{IEEE} Transactions on Software Engineering}

\bibliographystyle{alphaurl}
\bibliography{psynthesis}

\begin{thebibliography}{HKPV98}

\bibitem[ACEF09]{ACEF09}
{\'E}tienne Andr{\'e}, {\relax Th}omas Chatain, Emmanuelle Encrenaz, and
  Laurent Fribourg.
\newblock An inverse method for parametric timed automata.
\newblock {\em International Journal of Foundations of Computer Science},
  20(5):819--836, 10 2009.
\newblock \href {https://doi.org/10.1142/S0129054109006905}
  {\path{doi:10.1142/S0129054109006905}}.

\bibitem[AD94]{AD94}
Rajeev Alur and David~L. Dill.
\newblock A theory of timed automata.
\newblock {\em \TCS{}}, 126(2):183--235, April 1994.
\newblock \href {https://doi.org/10.1016/0304-3975(94)90010-8}
  {\path{doi:10.1016/0304-3975(94)90010-8}}.

\bibitem[AHV93]{AHV93}
Rajeev Alur, Thomas~A. Henzinger, and Moshe~Y. Vardi.
\newblock Parametric real-time reasoning.
\newblock In S.~Rao Kosaraju, David~S. Johnson, and Alok Aggarwal, editors,
  {\em Proceedings of the 25th annual {ACM} symposium on Theory of computing
  ({STOC} 1993)}, pages 592--601, New York, NY, USA, 1993. ACM.
\newblock \href {https://doi.org/10.1145/167088.167242}
  {\path{doi:10.1145/167088.167242}}.

\bibitem[AL17]{ALime17}
{\'E}tienne Andr{\'e} and Didier Lime.
\newblock Liveness in {L/U}-parametric timed automata.
\newblock In Alex Legay and Klaus Schneider, editors, {\em Proceedings of the
  17th International Conference on Application of Concurrency to System Design
  (ACSD 2017)}, pages 9--18. IEEE, 2017.
\newblock \href {https://doi.org/10.1109/ACSD.2017.19}
  {\path{doi:10.1109/ACSD.2017.19}}.

\bibitem[ALM20]{ALM20}
{\'E}tienne Andr{\'e}, Didier Lime, and Nicolas Markey.
\newblock Language preservation problems in parametric timed automata.
\newblock {\em \LMCS{}}, 16(1), January 2020.
\newblock \href {https://doi.org/10.23638/LMCS-16(1:5)2020}
  {\path{doi:10.23638/LMCS-16(1:5)2020}}.

\bibitem[ALR15]{ALR15}
{\'E}tienne Andr{\'e}, Didier Lime, and Olivier~H. Roux.
\newblock Integer-complete synthesis for bounded parametric timed automata.
\newblock In Miko\l{}aj Boja\'nczyk, S\l{}awomir Lasota, and Igor Potapov,
  editors, {\em Proceedings of the 9th International Workshop on Reachability
  Problems ({RP} 2015)}, volume 9328 of {\em \LNCS{}}, pages 7--19. Springer,
  September 2015.
\newblock \href {https://doi.org/10.1007/978-3-319-24537-9_2}
  {\path{doi:10.1007/978-3-319-24537-9_2}}.

\bibitem[ALR16a]{ALR16}
{\'E}tienne Andr{\'e}, Didier Lime, and Olivier~H. Roux.
\newblock Decision problems for parametric timed automata.
\newblock In Kazuhiro Ogata, Mark Lawford, and Shaoying Liu, editors, {\em
  Proceedings of the 18th International Conference on Formal Engineering
  Methods (ICFEM 2016)}, volume 10009 of {\em \LNCS{}}, pages 400--416.
  Springer, 2016.
\newblock \href {https://doi.org/10.1007/978-3-319-47846-3_25}
  {\path{doi:10.1007/978-3-319-47846-3_25}}.

\bibitem[ALR16b]{ALR16FORMATS}
{\'E}tienne Andr{\'e}, Didier Lime, and Olivier~H. Roux.
\newblock On the expressiveness of parametric timed automata.
\newblock In Martin Fränzle and Nicolas Markey, editors, {\em Proceedings of
  the 14th International Conference on Formal Modelling and Analysis of Timed
  Systems (FORMATS 2016)}, volume 9984 of {\em \LNCS{}}, pages 19--34.
  Springer, 2016.
\newblock \href {https://doi.org/10.1007/978-3-319-44878-7_2}
  {\path{doi:10.1007/978-3-319-44878-7_2}}.

\bibitem[ALR18]{ALR18FORMATS}
{\'E}tienne Andr{\'e}, Didier Lime, and Mathias Ramparison.
\newblock {TCTL} model checking lower/upper-bound parametric timed automata
  without invariants.
\newblock In David~N. Jansen and Pavithra Prabhakar, editors, {\em Proceedings
  of the 16th International Conference on Formal Modeling and Analysis of Timed
  Systems ({FORMATS} 2018)}, volume 11022 of {\em \LNCS{}}, pages 1--17.
  Springer, 2018.
\newblock \href {https://doi.org/10.1007/978-3-030-00151-3_3}
  {\path{doi:10.1007/978-3-030-00151-3_3}}.

\bibitem[ALR21]{ALR21}
{\'E}tienne Andr{\'e}, Didier Lime, and Mathias Ramparison.
\newblock Parametric updates in parametric timed automata.
\newblock {\em \LMCS{}}, 17(2):13:1--13:67, May 2021.
\newblock \href {https://doi.org/10.23638/LMCS-17(2:13)2021}
  {\path{doi:10.23638/LMCS-17(2:13)2021}}.

\bibitem[And16]{Andre16}
{\'E}tienne Andr{\'e}.
\newblock Parametric deadlock-freeness checking timed automata.
\newblock In Augusto Cesar~Alves Sampaio and Farn Wang, editors, {\em
  Proceedings of the 13th International Colloquium on Theoretical Aspects of
  Computing (ICTAC 2016)}, volume 9965 of {\em \LNCS{}}, pages 469--478.
  Springer, 2016.
\newblock \href {https://doi.org/10.1007/978-3-319-46750-4_27}
  {\path{doi:10.1007/978-3-319-46750-4_27}}.

\bibitem[And19]{Andre19STTT}
{\'E}tienne Andr{\'e}.
\newblock What's decidable about parametric timed automata?
\newblock {\em \STTT{}}, 21(2):203--219, 4 2019.
\newblock \href {https://doi.org/10.1007/s10009-017-0467-0}
  {\path{doi:10.1007/s10009-017-0467-0}}.

\bibitem[And21]{Andre21}
{\'E}tienne Andr{\'e}.
\newblock {IMITATOR} 3: Synthesis of timing parameters beyond decidability.
\newblock In Rustan Leino and Alexandra Silva, editors, {\em Proceedings of the
  33rd International Conference on Computer-Aided Verification ({CAV} 2021)},
  volume 12759 of {\em \LNCS{}}, pages 1--14. Springer, 2021.
\newblock \href {https://doi.org/10.1007/978-3-030-81685-8_26}
  {\path{doi:10.1007/978-3-030-81685-8_26}}.

\bibitem[BBLS15]{BBLS15}
Nikola Bene\v{s}, Peter Bezd\v{e}k, Kim~Gulstrand Larsen, and Ji\v{r}\'i Srba.
\newblock Language emptiness of continuous-time parametric timed automata.
\newblock In Magnús~M. Halldórsson, Kazuo Iwama, Naoki Kobayashi, and Bettina
  Speckmann, editors, {\em Proceedings of the 42nd International Colloquium on
  Automata, Languages, and Programming ({ICALP} 2015), Part {II}}, volume 9135
  of {\em \LNCS{}}, pages 69--81. Springer, July 2015.
\newblock \href {https://doi.org/10.1007/978-3-662-47666-6_6}
  {\path{doi:10.1007/978-3-662-47666-6_6}}.

\bibitem[BDR08]{BDR08}
Véronique Bruyère, Emmanuel Dall'Olio, and Jean-Francois Raskin.
\newblock Durations and parametric model-checking in timed automata.
\newblock {\em {ACM} Transactions on Computational Logic}, 9(2):12:1--12:23,
  2008.
\newblock \href {https://doi.org/10.1145/1342991.1342996}
  {\path{doi:10.1145/1342991.1342996}}.

\bibitem[BLT09]{BlT09}
Laura Bozzelli and Salvatore La~Torre.
\newblock Decision problems for lower/upper bound parametric timed automata.
\newblock {\em \FMSD{}}, 35(2):121--151, 2009.
\newblock \href {https://doi.org/10.1007/s10703-009-0074-0}
  {\path{doi:10.1007/s10703-009-0074-0}}.

\bibitem[BO14]{BO14}
Daniel Bundala and Joël Ouaknine.
\newblock Advances in parametric real-time reasoning.
\newblock In Erzs{\'{e}}bet Csuhaj{-}Varj{\'{u}}, Martin Dietzfelbinger, and
  Zolt{\'{a}}n {\'{E}}sik, editors, {\em Proceedings of the 39th International
  Symposium on Mathematical Foundations of Computer Science ({MFCS} 2014), Part
  {I}}, volume 8634 of {\em \LNCS{}}, pages 123--134. Springer, 2014.
\newblock \href {https://doi.org/10.1007/978-3-662-44522-8}
  {\path{doi:10.1007/978-3-662-44522-8}}.

\bibitem[BO17]{BO17}
Daniel Bundala and Joël Ouaknine.
\newblock On parametric timed automata and one-counter machines.
\newblock {\em Information and Computation}, 253:272--303, 2017.
\newblock \href {https://doi.org/10.1016/j.ic.2016.07.011}
  {\path{doi:10.1016/j.ic.2016.07.011}}.

\bibitem[BR07]{BR07}
V{\'e}ronique Bruy{\`e}re and Jean-Fran\c{c}ois Raskin.
\newblock Real-time model-checking: Parameters everywhere.
\newblock {\em Logical Methods in Computer Science}, 3(1:7):1--30, 2007.
\newblock \href {https://doi.org/10.2168/LMCS-3(1:7)2007}
  {\path{doi:10.2168/LMCS-3(1:7)2007}}.

\bibitem[BY03]{BY03}
Johan Bengtsson and Wang Yi.
\newblock Timed automata: Semantics, algorithms and tools.
\newblock In Jöörg Desel, Wolfgang Reisig, and Grzegorz Rozenberg, editors,
  {\em Lectures on Concurrency and {P}etri Nets, Advances in {P}etri Nets [This
  tutorial volume originates from the 4th Advanced Course on {P}etri Nets
  ({ACPN} 2003)]}, volume 3098 of {\em \LNCS{}}, pages 87--124. Springer, 2003.
\newblock \href {https://doi.org/10.1007/978-3-540-27755-2_3}
  {\path{doi:10.1007/978-3-540-27755-2_3}}.

\bibitem[CES86]{CES-86}
Edmund~M. Clarke, E.~Allen Emerson, and A.~Prasad Sistla.
\newblock Automatic verification of finite-state concurrent systems using
  temporal logic specifications.
\newblock {\em ACM Transactions on Programming Languages and Systems},
  8(2):244--263, 1986.
\newblock \href {https://doi.org/10.1145/5397.5399}
  {\path{doi:10.1145/5397.5399}}.

\bibitem[CL00]{CL00}
Franck Cassez and Kim~Guldstrand Larsen.
\newblock The impressive power of stopwatches.
\newblock In Catuscia Palamidessi, editor, {\em Proceedings of the 11th
  International Conference on Concurrency Theory ({CONCUR} 2000)}, volume 1877
  of {\em \LNCS{}}, pages 138--152. Springer, 2000.
\newblock \href {https://doi.org/10.1007/3-540-44618-4_12}
  {\path{doi:10.1007/3-540-44618-4_12}}.

\bibitem[Doy07]{Doyen07}
Laurent Doyen.
\newblock Robust parametric reachability for timed automata.
\newblock {\em \IPL{}}, 102(5):208--213, 2007.
\newblock \href {https://doi.org/10.1016/j.ipl.2006.11.018}
  {\path{doi:10.1016/j.ipl.2006.11.018}}.

\bibitem[GH21]{GH21}
Stefan Göller and Mathieu Hilaire.
\newblock Reachability in two-parametric timed automata with one parameter is
  {EXPSPACE}-complete.
\newblock In Markus Bläser and Benjamin Monmege, editors, {\em Proceedings of
  the 38th International Symposium on Theoretical Aspects of Computer Science
  ({STACS} 2021)}, volume 187 of {\em LIPIcs}, pages 36:1--36:18. Schloss
  Dagstuhl - Leibniz-Zentrum für Informatik, 2021.
\newblock \href {https://doi.org/10.4230/LIPIcs.STACS.2021.36}
  {\path{doi:10.4230/LIPIcs.STACS.2021.36}}.

\bibitem[HKPV98]{henzinger-JCSS-98}
Thomas~A. Henzinger, Peter~W. Kopke, Anuj Puri, and Pravin Varaiya.
\newblock What's decidable about hybrid automata?
\newblock {\em Journal of Computer and System Sciences}, 57(1):94--124, 1998.
\newblock \href {https://doi.org/10.1006/jcss.1998.1581}
  {\path{doi:10.1006/jcss.1998.1581}}.

\bibitem[HRSV02]{HRSV02}
Thomas Hune, Judi Romijn, Mari{\"e}lle Stoelinga, and Frits~W. Vaandrager.
\newblock Linear parametric model checking of timed automata.
\newblock {\em \JLAP{}}, 52-53:183--220, 2002.
\newblock \href {https://doi.org/10.1016/S1567-8326(02)00037-1}
  {\path{doi:10.1016/S1567-8326(02)00037-1}}.

\bibitem[JLR15]{JLR15}
Aleksandra Jovanovi{\'c}, Didier Lime, and Olivier~H. Roux.
\newblock Integer parameter synthesis for real-time systems.
\newblock {\em \TSE{}}, 41(5):445--461, 2015.
\newblock \href {https://doi.org/10.1109/TSE.2014.2357445}
  {\path{doi:10.1109/TSE.2014.2357445}}.

\bibitem[Lam77]{lamport-TSE-77}
Leslie Lamport.
\newblock Proving the correctness of multiprocess programs.
\newblock {\em \IEEETSE{}}, 3(2):125--143, 1977.

\bibitem[LRST09]{LRST09}
Didier Lime, Olivier~H. Roux, Charlotte Seidner, and Louis-Marie Traonouez.
\newblock {R}omeo: {A} parametric model-checker for {P}etri nets with
  stopwatches.
\newblock In Stefan Kowalewski and Anna Philippou, editors, {\em Proceedings of
  the 15th International Conference on Tools and Algorithms for the
  Construction and Analysis of Systems ({TACAS} 2009)}, volume 5505 of {\em
  \LNCS{}}, pages 54--57. Springer, March 2009.
\newblock \href {https://doi.org/10.1007/978-3-642-00768-2_6}
  {\path{doi:10.1007/978-3-642-00768-2_6}}.

\bibitem[Mil00]{Miller00}
Joseph~S. Miller.
\newblock Decidability and complexity results for timed automata and
  semi-linear hybrid automata.
\newblock In Nancy~A. Lynch and Bruce~H. Krogh, editors, {\em Proceedings of
  the Third International Workshop on Hybrid Systems: Computation and Control
  ({HSCC} 2000)}, volume 1790 of {\em \LNCS{}}, pages 296--309. Springer, 2000.
\newblock \href {https://doi.org/10.1007/3-540-46430-1_26}
  {\path{doi:10.1007/3-540-46430-1_26}}.

\bibitem[Min67]{Minsky67}
Marvin~L. Minsky.
\newblock {\em Computation: finite and infinite machines}.
\newblock Prentice-Hall, Inc., 1967.

\bibitem[Qua14]{Quaas14}
Karin Quaas.
\newblock {MTL}-model checking of one-clock parametric timed automata is
  undecidable.
\newblock In {\'{E}}tienne Andr{\'{e}} and Goran Frehse, editors, {\em
  Proceedings of the 1st International Workshop on Synthesis of Continuous
  Parameters ({SynCoP} 2014)}, volume 145 of {\em {EPTCS}}, pages 5--17, 2014.
\newblock \href {https://doi.org/10.4204/EPTCS.145.3}
  {\path{doi:10.4204/EPTCS.145.3}}.

\bibitem[San11]{Sankur11}
Ocan Sankur.
\newblock Untimed language preservation in timed systems.
\newblock In {\em Proceedings of the 36th International Symposium on
  Mathematical Foundations of Computer Science ({MFCS} 2011)}, volume 6907 of
  {\em \LNCS{}}, pages 556--567. Springer, August 2011.
\newblock \href {https://doi.org/10.1007/978-3-642-22993-0_50}
  {\path{doi:10.1007/978-3-642-22993-0_50}}.

\bibitem[Sch86]{schrijver-book-86}
Alexander Schrijver.
\newblock {\em Theory of linear and integer programming}.
\newblock John Wiley \& Sons, Inc., New York, NY, USA, 1986.

\bibitem[TLR09]{TLR09}
Louis-Marie Traonouez, Didier Lime, and Olivier~H. Roux.
\newblock Parametric model-checking of stopwatch {P}etri nets.
\newblock {\em Journal of Universal Computer Science}, 15(17):3273--3304, 2009.
\newblock \href {https://doi.org/10.3217/jucs-015-17-3273}
  {\path{doi:10.3217/jucs-015-17-3273}}.

\bibitem[TY01]{TY01}
Stavros Tripakis and Sergio Yovine.
\newblock Analysis of timed systems using time-abstracting bisimulations.
\newblock {\em \FMSD{}}, 18(1):25--68, 2001.
\newblock \href {https://doi.org/10.1023/A:1008734703554}
  {\path{doi:10.1023/A:1008734703554}}.

\bibitem[Wan96]{Wang96}
Farn Wang.
\newblock Parametric timing analysis for real-time systems.
\newblock {\em Information and Computation}, 130(2):131--150, 1996.
\newblock \href {https://doi.org/10.1006/inco.1996.0086}
  {\path{doi:10.1006/inco.1996.0086}}.

\end{thebibliography}

\end{document}